\documentclass[11pt,a4paper]{article}
\pdfoutput=1
\usepackage{jstylemarg}
\usepackage[utf8]{inputenc}
\usepackage{amsmath}
\usepackage{mathtools}
\usepackage{tikz}
\usepackage{empheq}
\usepackage{enumitem}
\usepackage{graphics}
\usepackage{wrapfig}
\usepackage{caption}
\usepackage{framed}
\usepackage{enumitem}

\newcommand{\bi}{\begin{itemize}}
\newcommand{\ei}{\end{itemize}}
\newcommand{\bea}{\begin{align}}
\newcommand{\eea}{\end{align}}
\newcommand{\be}{\begin{equation}}
\newcommand{\ee}{\end{equation}}

\newcommand{\pl}{{\partial}}

\newcommand{\tcb}{\textcolor{blue}}

\makeatletter
\renewcommand*\env@matrix[1][\arraystretch]{%
  \edef\arraystretch{#1}%
  \hskip -\arraycolsep
  \let\@ifnextchar\new@ifnextchar
  \array{*\c@MaxMatrixCols c}}
\makeatother

\author[a,b,c]{Charlotte SLEIGHT}

\author[c,d,e]{\quad Massimo TARONNA}

\affiliation[a]{Universit\'e Libre de Bruxelles
and International Solvay Institutes,\\
ULB-Campus Plaine CP231, 1050 Brussels, Belgium}

\affiliation[b]{School of Natural Sciences, Institute for Advanced Study,\\
1 Einstein Drive, Princeton, NJ 08540}

\affiliation[c]{Department of Physics, Princeton University,\\
Jadwin Hall, Princeton, NJ 08544}

\affiliation[d]{Dipartimento di Fisica ``Ettore Pancini", Universit\`a degli Studi di Napoli Federico II, \\Monte S. Angelo, Via Cintia, 80126 Napoli, Italy}

\affiliation[e]{INFN, Sezione di Napoli, Monte S. Angelo, Via Cintia, 80126 Napoli, Italy}

\emailAdd{charlotte.sleight@gmail.com, mtaronna@princeton.edu}

\title{\centering
\LARGE{Anomalous Dimensions from Crossing Kernels}}

\abstract{In this work we consider the problem of extracting the corrections to CFT data induced by the exchange of a primary operator and its descendents in the crossed channel. We show how those corrections which are analytic in spin can be systematically extracted from crossing kernels. To this end, we underline a connection between: Wilson polynomials (which naturally appear when considering the crossing kernels given recently in arXiv:1804.09334), the spectral integral in the conformal partial wave expansion, and Wilson functions. Using this connection, we determine closed form expressions for the OPE data when the external operators in 4pt correlation functions have spins $J_1$-$J_2$-$0$-$0$, in particular the anomalous dimensions of double-twist operators of the type $[\mathcal{O}_{J_1}\mathcal{O}_{J_2}]_{n,\ell}$ in $d$ dimensions and for both leading ($n=0$) and sub-leading ($n\ne0$) twist. The OPE data are expressed in terms of Wilson functions, which naturally appear as a spectral integral of a Wilson polynomial. As a consequence, our expressions are manifestly analytic in spin and are valid up to finite spin. We present some applications to CFTs with slightly broken higher-spin symmetry. The Mellin Barnes integral representation for $6j$ symbols of the conformal group in general $d$ and its relation with the crossing kernels are also discussed.}

\begin{document}
\begin{flushright}    
  {\texttt{PUPT-2567}}
\end{flushright}
\maketitle

\section{Introduction}

An important observation of the analytic bootstrap programme is that the inverse spin of primary operators, $1/\ell$, is a ``good'' perturbative expansion parameter. It was first observed in \cite{Komargodski:2012ek,Fitzpatrick:2012yx} how, under the assumption of unitary, the crossing equations simplify in the limit of large $\ell$ and are re-organised in terms of so called double-twist and generically multi-twist operators. Double-twist operators furthermore organise into analytic families which asymptote to free bound-states.\footnote{See also the earlier works \cite{PhysRevD.8.4383,Korchemsky:1992xv,Belitsky:2003ys,Alday:2007mf}.} The large spin bootstrap has explained many striking features of results in the numerical bootstrap programme \cite{Simmons-Duffin:2016wlq},\footnote{See e.g. the comprehensive reviews \cite{Simmons-Duffin:2016gjk,Poland:2018epd} and references therein.} which hold with remarkable accuracy down to very low spin $\ell \geq2$. Simply speaking, the large spin bootstrap captures the analytic data of CFTs which are highly constrained by causality and unitarity \cite{Hartman:2015lfa,Simmons-Duffin:2016wlq,Caron-Huot:2017vep}. From a dual AdS perspective these data also constrain Effective Field Theory in AdS, prescribing how higher derivative contact interactions organise into analytic families.
%(see fig.~\ref{fig:EFT}).

The aim of this note is to work out the explicit relation between crossing kernels of conformal partial waves and the above large spin bootstrap problem, extracting expressions for the corresponding OPE data that is analytic in spin. Such crossing kernels were recently given explicitly in terms of the hypergeometric function ${}_4F_3$ in \cite{Sleight:2018epi}, for both scalar and spinning external operators.\footnote{For related earlier work on crossing kernels and $6j$ symbols of the conformal group, see e.g. \cite{Krasnov:2005fu,Gadde:2017sjg,Hogervorst:2017sfd,Hogervorst:2017kbj,Gopakumar:2018xqi}.} We argue that Wilson polynomials provide a natural basis for such crossing kernels, which allows us to obtain closed formulas for the OPE data of double-twist operators in terms of Wilson functions. Explicit formulas in terms of Wilson functions are derived both for external scalar operators and for the case in which two of the external operators have arbitrary integer spin. We discuss some applications of our results to CFTs with slightly broken higher-spin symmetry. For the reader convenience we also detail in an appendix the Mellin Barnes integral representation for $6j$ symbols of the conformal group in general $d$ and its link with the crossing kernels used in this work.

\subsection{Anomalous dimensions from crossing kernels}
\label{subsec::approach}

For ease of presentation let us consider for now the simple case of four-point correlators of identical scalar primary operators ${\cal O}$ of scaling dimension $\Delta$,
\begin{equation}\label{OOOO}
    \langle {\cal O}\left(x_1\right){\cal O}\left(x_2\right){\cal O}\left(x_3\right){\cal O}\left(x_4\right) \rangle = \frac{{\cal A}\left(u,v\right)}{\left(x^2_{12}\right)^{\Delta}\left(x^2_{34}\right)^{\Delta}},
\end{equation}
with cross ratios
\begin{equation}
    u=\frac{x^2_{12}x^2_{34}}{x^2_{13}x^2_{24}}, \qquad  v=\frac{x^2_{14}x^2_{23}}{x^2_{13}x^2_{24}}.
\end{equation}
In \S \ref{subsec::extspinops} and \S \ref{sec::spinningcorrelators} we shall also consider four-point correlators in which two of the operators have arbitrary integer spin.
Associativity of the operator product expansion implies the crossing equations
\begin{equation}\label{gencrossing}
   u^{\Delta} \left(1+\sum_{\tau^\prime,\ell^\prime}a_{\tau^\prime,\ell^\prime} G_{\tau^\prime,\ell^\prime}\left(v,u\right)\right)=v^{\Delta} \left(1+\sum_{\tau,\ell}a_{\tau,\ell} G_{\tau,\ell}\left(u,v\right)\right),
\end{equation}
where we have separated the contribution of the identity operator. The $G_{\tau,\ell}$ are conformal blocks encoding the exchange of a conformal multiplet with lowest weight primary operator of twist $\tau$ and spin $\ell$. The RHS is the ${\sf s}$-channel conformal block expansion of the correlator \eqref{OOOO} and the LHS is the conformal block expansion in the ${\sf t}$-channel.

As shown in \cite{Fitzpatrick:2012yx,Komargodski:2012ek} the identity contribution in the ${\sf t}$-channel entails the existence of ``double-twist'' operators $\left[{\cal O}{\cal O}\right]_{n,\ell}$, whose scaling dimensions and OPE coefficients approach the mean field theory values in the limit of large $\ell$:
\begin{subequations}
\begin{align}
\tau_{n,\ell} &\rightarrow \tau^{(0)}_{n,\ell},\\
a_{n,\ell} &\rightarrow a^{(0)}_{n,\ell},
\end{align}
\end{subequations}
where $\tau^{(0)}_{n,\ell} = 2\Delta+2n$ in this example of equal external operators and the $a^{(0)}_{n,\ell}$ have been computed in \cite{Dolan:2000ut,Heemskerk:2009pn,Fitzpatrick:2011dm}.\footnote{For the results on the mean field theory OPE coefficients of operators $\left[{\cal O}_{J}{\cal O}\right]$ with primary operator ${\cal O}_{J}$ of arbitrary spin $J$, which we employ in \S \ref{subsec::extspinops} and \S\ref{sec::spinningcorrelators} of this work, see \cite{Sleight:2018epi}.}
\begin{equation}\label{OPEnl}
    a_{n,\ell}^{(0)}=\frac{2^{\ell} (-1)^{n} (\Delta)^2_n\left(-\frac{d}{2}+\Delta+1\right)^2_n (n+\Delta)^2_{\ell}}{\ell!\, n! \left(\frac{d}{2}+\ell\right)_n (d-2 n-2\Delta)_n (\ell+2 n+2\Delta-1)_{\ell} \left(-\frac{d}{2}+\ell+n+2\Delta\right)_n}\,.
\end{equation}

Corrections to the above are induced by operators in the OPE of ${\cal O}$ with itself exchanged in the crossed channels. In particular, for an operator of twist $\tau^\prime$ we have the large spin expansion \cite{Alday:2015eya}  
\begin{subequations}\label{intlsexpgamma}
\begin{align}
\tau_{n,\ell}&=\tau^{(0)}_{n,\ell}+\gamma_{n,\ell}\\
\gamma_{n,\ell}&=- \frac{c^{(0)}_n}{\mathfrak{J}^{\tau^\prime}}\left(1+\sum^{\infty}_{k=1}\frac{c^{(k)}_n}{\mathfrak{J}^{2k}}\right)\,,   
\end{align}
\end{subequations}
which naturally organises itself in terms of the conformal spin $\mathfrak{J}$, whose dependence on $\ell$ is given by 
\begin{equation}
    \mathfrak{J}^2=\left(\ell+\tfrac{\tau_{n,\ell}}2\right)\left(\ell+\tfrac{\tau_{n,\ell}}2-1\right)\,.
\end{equation}
For external scalar operators, the leading contributions $c^{(0)}_n$ were determined in \cite{Fitzpatrick:2012yx,Komargodski:2012ek,Vos:2014pqa}. The corrections $c^{(k)}_n$ for leading double-twist operators (i.e. $n=0$) were considered in \cite{Alday:2015ewa,Dey:2017fab} in general dimensions $d$, while for specific dimensions there are results available for general $n$ \cite{Heemskerk:2009pn,Kaviraj:2015cxa,Alday:2017gde,Aharony:2018npf}.\footnote{For related work see \cite{Kaviraj:2015xsa,Dey:2016zbg}.} There has been some progress for external operators of low spin, where in \cite{Li:2015itl} leading contributions $c^{(0)}_0$ for mixed correlators involving external spin one currents and the stress tensor were extracted in $d=3$. Leading contributions $c^{(0)}_n$ for external Fermions and general $n$ have been determined in $d=4$ \cite{Elkhidir:2017iov}.

It was recently clarified in \cite{Caron-Huot:2017vep} that the CFT data $\left\{\tau_{n,\ell},\gamma_{n,\ell}\right\}$ above is analytic in spin, and thus the large spin expansion \eqref{intlsexpgamma} is an asymptotic expansion of a function that is analytic in the conformal spin $\mathfrak{J}$ (see also \cite{Alday:2017vkk}). In this work we provide the latter analytic expressions for the OPE data using recent results \cite{Sleight:2018epi} for the crossing kernels of conformal partial waves in general dimensions $d$, which include external spinning operators and kernels for double-twist operators with general $n$.\footnote{For external scalar operators, earlier works have obtained re-summations for certain scaling dimensions and dimensions $d$ by considering an explicit re-summation of the series \eqref{intlsexpgamma} \cite{Alday:2015ewa,Alday:2017gde}, and also in \cite{Caron-Huot:2017vep,Aharony:2018npf} using the inversion formula \cite{Caron-Huot:2017vep}. We reproduce these results and moreover extend them to more general cases. See also \cite{Giombi:2018vtc} \cite{Sleight:2018epi} for anomalous dimensions of finite spin double-trace operators in large $N$ CFTs induced by double-trace flows.}${}^{,}$\footnote{It is important to keep in mind that, from the perspective of crossing symmetry, the analytic results which we shall present have to be supplemented with solutions that have finite support in spin \cite{Heemskerk:2009pn}. These contributions however can be further constrained by causality and can be reduced to a finite number \cite{Caron-Huot:2017vep}. In this work we shall not consider these finite spin contributions.} At large spin, the $1/\mathfrak{J}$ expansion of our results gives the extension of the corrections $c^{(k)}_n$ listed in the above paragraph to these more general cases (external spins, sub-leading twists $n\ne0$ and general $d$). 

\begin{figure}[t]
    \centering
    \captionsetup{width=0.95\textwidth}
    \includegraphics[width=0.75\textwidth]{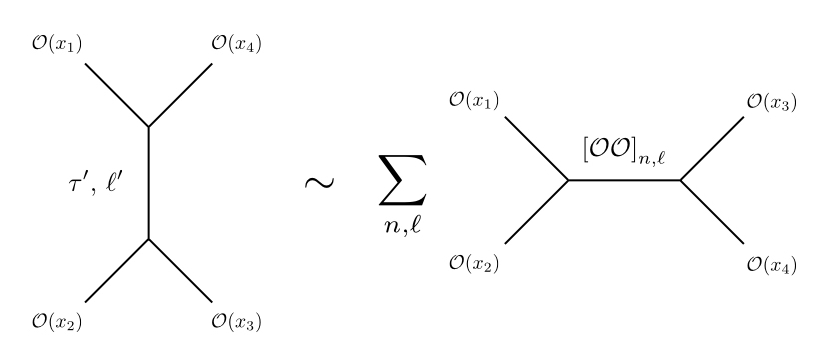}
    \caption{${\sf s}$-channel decomposition of the exchange of an operator of twist $\tau^\prime$ and spin $\ell^\prime$ (+descendents) in the crossed channel.}
    \label{fig:crossing}
\end{figure}

\paragraph{Approach.} We are considering the following decomposition problem\footnote{In equation \eqref{ttosblock} we use the weak equality ``$\sim$" instead of "=" to emphasise that technically this identity can only be used when considering single-valued sums of conformal blocks \cite{ElShowk:2011ag}, like within a 4pt correlator.}
\begin{align}\label{ttosblock}
\left(\frac{u}{v}\right)^{\Delta}\left( a_{\tau^\prime,\ell^\prime}  G_{\tau^\prime,\ell^\prime}\left(v,u\right)\right) \sim   \sum_{n,\ell}a_{n,\ell}\, G_{\tau_{n,\ell},\ell}\left(u,v\right),
\end{align}
where above the LHS gives the contribution of a conformal block in the $\sf{t}$-channel and the RHS gives its expansion in terms of primary operators in the $\sf{s}$-channel.

Considering now small corrections to the conformal data with respect to their mean-field theory values, we can write:
\begin{align}\label{expsmu}
    &a_{n,\ell}\, G_{\tau_{n,\ell},\ell}\left(u,v\right)\\ \nonumber & \hspace*{0.3cm}=u^{\Delta+n}\left(\frac{\gamma_{n,\ell}}2\,a_{n,\ell}^{(0)}\,f_{2\Delta+2n,\ell}(v)\,\log u +\gamma_{n,\ell}\,a_{n,\ell}^{(0)}\,\frac{1}{2}\pl_{n}\,(f_{2\Delta+2n,\ell}(v))+a_{n,\ell}^{(1)}f_{2\Delta+2n,\ell}(v)+{\cal O}\left(u\right)\right).
\end{align}
The collinear conformal block $f_{\tau,\ell}(v)$ is defined by the small $u$ limit of the ${\sf s}$-channel conformal block $G_{\tau,\ell}\left(u,v\right)$ \cite{Alday:2010zy}:
\begin{equation}
f_{\tau,\ell}(v)=\lim_{u\rightarrow0} u^{-\tau/2} G_{\tau,\ell}\left(u,v\right)=\left(1-v\right)^{\ell}{}_2F_1\left(\frac{\tau+2\ell}{2},\frac{\tau+2\ell}{2},\tau+2\ell;1-v\right),
\end{equation}
so that \eqref{expsmu} gives the contribution from the double-twist primary operators $\left[{\cal O}{\cal O}\right]_{n,\ell}$. The ${\cal O}(u)$ terms in \eqref{expsmu} correspond to contributions of its descendents.\footnote{Recall that the contribution of a twist $\tau$ operator in the ${\sf s}$-channel is proportional to $u^{\tau/2}$.}

For simplicity, in this introductory section we focus explicitly on the procedure for extracting the OPE data $\{\gamma_{0,\ell},a^{(1)}_{0,\ell}\}$ of the leading double-twist operators $\left[{\cal O}{\cal O}\right]_{n,\ell}$ with $n=0$ from the LHS of equation \eqref{ttosblock}. The case of general $n>0$ (subleading twists) is considered in \S \ref{ExchSub} and involves the additional technical step of projecting away the contributions from conformal multiplets of each lower twist $n^\prime<n$.\footnote{A way to do this was given in \cite{Sleight:2018epi}, which entails acting with so-called twist block operators whose kernels contain any conformal block of a given twist.} The point is that contributions from sub-leading twists mix with the contributions from the descendents of the lower twist conformal multiplets and so they have to be disentangled. After this projection, the procedure follows in the same way as for the $n=0$ case described here.

From equation \eqref{expsmu} we see that anomalous dimension $\gamma_{0,\ell}$ can be read off from the $u^\Delta \log u$ term in the ${\sf s}$-channel expansion of the ${\sf t}$-channel conformal block on the LHS of \eqref{ttosblock}, while the corrections $a_{0,\ell}^{(1)}$ to the OPE coefficients are encoded in the terms proportional to $u^{\Delta}$. To extract them the contribution from $\pl_{n}\,(f_{2\Delta+2n,\ell}(v))$ has to be disentangled, for which one employs the identity\footnote{This identity can be obtained considering the conformal blocks as functions of conformal spin.} \cite{Alday:2015ewa,Dey:2017fab}
\begin{align}
  \gamma_{0,\ell}\,a_{0,\ell}^{(0)}\,\frac{1}{2}\pl_{n}\,(f_{2\Delta+2n,\ell}(v))\Big|_{n=0}=-\frac{ \gamma_{0,\ell}}2\,a_{0,\ell}^{(0)}\,f_{2\Delta,\ell}(v)\,\log(1-v).
\end{align}
The $\{\gamma_{0,\ell},a^{(1)}_{0,\ell}\}$ contributions can then be disentangled in the crossing equation \eqref{ttosblock} as
\begin{subequations}\label{sumgammaope}
\begin{align}
  \sum_{\ell}\frac{\gamma_{0,\ell}}2\,a_{0,\ell}^{(0)}\,f_{2\Delta+2n,\ell}(v)\,&= A_{0}\left(v\right), \\
  \sum_{\ell}\,a_{0,\ell}^{(1)}\,f_{2\Delta+2n,\ell}(v)\,&= B_{0}\left(v\right)-A_0\left(v\right)\log\left(1-v\right),
\end{align}
\end{subequations}
where
\begin{subequations}\label{A0B0}
\begin{align}
   A_{0}\left(v\right) & = \left[\left(\frac{u}{v}\right)^{\Delta}\left( a_{\tau^\prime,\ell^\prime}  G_{\tau^\prime,\ell^\prime}\left(v,u\right)\right)\right]_{u^{\Delta}\log u},\\
   B_{0}\left(v\right) & = \left[\left(\frac{u}{v}\right)^{\Delta}\left( a_{\tau^\prime,\ell^\prime}  G_{\tau^\prime,\ell^\prime}\left(v,u\right)\right)\right]_{u^{\Delta}}.
\end{align}
\end{subequations}

At this point it is instructive to turn to the Mellin representation of CFT correlators \cite{Mack:2009mi}, which has proven to be an invaluable tool in the conformal bootstrap \cite{Sen:2015doa,Gopakumar:2016wkt,Gopakumar:2016cpb,Gopakumar:2018xqi}. See \cite{Sleight:2018epi} for the notations and conventions we employ in this note. In Mellin space the collinear conformal blocks $f_{\tau,\ell}(v)$ are represented by orthogonal polynomials ${\cal Q}_{\tau,\ell}\left(s\right)$ \cite{Costa:2012cb} (see also \cite{Korchemsky:1994um}),\footnote{In particular 
\begin{equation}
    f_{\tau,\ell}(v)=\int^{i\infty}_{-i\infty} \frac{ds}{4\pi i} v^{-(s+\tau)/2} {\tilde \rho}_{\{\Delta\}}\left(s,\tau\right){\cal Q}_{\tau,\ell}\left(s\right).
\end{equation}
The polynomial ${\cal Q}_{\tau,\ell}\left(s\right)$ can be expressed in terms of a continuous Hahn polynomial $Q_\ell^{(a,b,c,d)}(s)$ \cite{andrews_askey_roy_1999}, 
\begin{equation}\label{mackpoly}
    {\cal Q}_{\tau,\ell}\left(s\right)=\left(-1\right)^\ell \ell! \left(\mathfrak{N}_\ell^{(\tau,\tau+\tau_1-\tau_2-\tau_3+\tau_4,-\tau_1+\tau-2,\tau_3-\tau_4)}\right)^{-1}\,Q_\ell^{(\tau,\tau+\tau_1-\tau_2-\tau_3+\tau_4,-\tau_1+\tau_2,\tau_3-\tau_4)}(s),
\end{equation}
where $\tau_i$ are the twists of the external operators and $\mathfrak{N}_\ell^{(a,b,c,d)}$ is the normalisation of their bi-linear form. See appendix D of \cite{Sleight:2018epi} for the relevant properties, notations and definitions used in this note.} in terms of which equations \eqref{sumgammaope} for $\{\gamma_{0,\ell},a^{(1)}_{0,\ell}\}$ read\footnote{The sum in \eqref{sumgammaopeb} arises from the Mellin representation of the Taylor expansion of $\log(1-v)$.} 
\begin{subequations}\label{sumgammaope2}
\begin{align}
  \sum_{\ell}\frac{\gamma_{0,\ell}}2\,a_{0,\ell}^{(0)}\,{\cal Q}_{2\Delta,\ell}(s)\,&= {\cal A}_{0}\left(s\right), \\
  \sum_{\ell}\,a_{0,\ell}^{(1)}\,{\cal Q}_{2\Delta,\ell}(s)\,&= {\cal B}_{0}\left(s\right)+\sum^{\infty}_{k=1}\frac{1}{k}{\cal A}_0\left(s+2k\right)\frac{\left(\frac{s}{2}+\Delta\right)_k}{\left(-\frac{s}{2}-\Delta\right)_k},\label{sumgammaopeb}
\end{align}
\end{subequations}
where ${\cal A}_{0}\left(s\right)$ and ${\cal B}_{0}\left(s\right)$ are the Mellin representations of ${\cal A}_{0}\left(v\right)$ and ${\cal B}_{0}\left(v\right)$:
\begin{subequations}\label{ABmellin}
\begin{align}
    A_{0}(v)&=\int^{i\infty}_{-i\infty} \frac{ds}{4\pi i} v^{-(s+2\Delta)/2} {\tilde \rho}_{\{\Delta\}}\left(s,2\Delta\right){\cal A}_0\left(s\right), \\
    B_{0}(v)&=\int^{i\infty}_{-i\infty} \frac{ds}{4\pi i} v^{-(s+2\Delta)/2} {\tilde \rho}_{\{\Delta\}}\left(s,2\Delta\right){\cal B}_0\left(s\right),
\end{align}
\end{subequations}
with the reduced Mellin measure 
\begin{equation}
    {\tilde \rho}_{\{\tau_i\}}\left(s,t\right)=\Gamma\left(\tfrac{s+t}{2}\right)\Gamma\left(\tfrac{s+t+\tau_1-\tau_2-\tau_2+\tau_4}{2}\right)\Gamma\left(\tfrac{-s-\tau_1+\tau_2}{2}\right)\Gamma\left(\tfrac{-s-\tau_4+\tau_3}{2}\right).
\end{equation}
The $\tau_i$ are the twists of the external operators (in this discussion $\tau_i=\Delta$).

Using the orthogonality of the continuous Hahn polynomials, the anomalous dimensions $\gamma_{0,\ell}$ are thus given by the Mellin integral
\begin{equation}\label{anomdtcb}
    \frac{\gamma_{0,\ell}}{2}a^{(0)}_{0,\ell} = \frac{(-1)^{\ell}}{\ell!}\int^{i\infty}_{-i \infty} \frac{ds}{4\pi i} {\tilde \rho}_{\{\Delta\}}\left(s,2\Delta\right){\cal A}_{0}\left(s\right)Q^{(2\Delta,2\Delta,0,0)}_{\ell}\left(s\right),
\end{equation}
which is the projection of the ${\sf t}$-channel conformal block \eqref{ABmellin} onto the $u^{\Delta}\log u$ contribution from leading double-twist operators $\left[{\cal O}{\cal O}\right]_{0,\ell}$ in the ${\sf s}$-channel.

Instead of conformal blocks, it is often useful to expand conformal four-point functions in terms of an orthogonal basis of single-valued functions known as conformal partial waves (CPWs) \cite{Mack1974,Mack:1974sa,Dobrev:1975ru} 
\begin{equation}\label{cpweA}
    {\cal A}\left(u,v\right)=\sum_{\ell} \int^{\frac{d}{2}+i \infty}_{\frac{d}{2}-i\infty} \frac{d{\tilde \Delta}}{2\pi i}\,a_\ell({\tilde \Delta}){\cal F}_{{\tilde \Delta},\ell}\left(u,v\right),
\end{equation}
where the spectral integral in the exchanged dimension ${\tilde \Delta}$ is over the principal series. The spectral function $a_\ell({\tilde \Delta})$ is meromorphic, whose poles in ${\tilde \Delta}$ correspond to the physical exchanged operators with the OPE data encoded in the residues. See e.g. \cite{Caron-Huot:2017vep} for a more recent discussion. Each conformal partial wave is a linear combination of a conformal block and its shadow \cite{Dolan:2011dv}
\begin{equation}\label{cpwdefint}
    {\cal F}_{{\tilde \Delta},\ell} = \,G_{{\tilde \Delta},\ell}+\# \, G_{d-{\tilde \Delta},\ell}.
\end{equation}
The ${\sf s}$-channel expansion of a ${\sf t}$-channel conformal partial wave ${\cal F}_{{\tilde \Delta},\ell^\prime}\left(v,u\right)$ takes the same form as in equation \eqref{ttosblock}, though extracting the OPE data \eqref{sumgammaope2} now entails evaluating a spectral integral. For example, for the anomalous dimensions \eqref{anomdtcb} we have to evaluate:
\begin{equation}\label{anomdint}
      \frac{\gamma_{0,\ell}}{2}a^{(0)}_{0,\ell} = \int^{\frac{d}{2}+i \infty}_{\frac{d}{2}-i\infty} \frac{d{\tilde \Delta}}{2\pi i}\,a_{\ell^\prime}({\tilde \Delta})\, {}^{({\sf t})}\mathfrak{I}_{{\tilde \Delta},\ell^\prime|\ell}\left(t=2\Delta\right),
\end{equation}
where $\mathfrak{I}^{({\sf t})}_{{\tilde \Delta},\ell^\prime|\ell}\left(t\right)$ is the crossing kernel of a ${\sf t}$-channel conformal partial wave ${\cal F}_{{\tilde \Delta},\ell^\prime}$ onto the contribution from a spin-$\ell$ exchange of dimension $t$ in the ${\sf s}$-channel (see \cite{Sleight:2018epi}):\footnote{It should be emphasised that the crossing kernels \eqref{cpwccint} are functions of $t$, where $t$ is not fixed to any particular value. The poles in $t$ are encoded in the full Mellin measure \eqref{rhoM}, which in the illustrative example above are located at $t=2\Delta+2n$ with $n \in \mathbb{Z}_{\geq 0}$. The example anomalous dimension given in \eqref{anomdint} is for double-twist operators of leading twist ($n=0$).}
\begin{align}\label{cpwccint}
    {}^{(\sf{t})}\mathfrak{I}_{{\tilde \Delta},\ell^\prime|\ell}\left(t\right)&=\frac{(-1)^{\ell}}{\ell!}\int^{i\infty}_{-i \infty} \frac{ds}{4\pi i}\, {\tilde \rho}_{\{\Delta\}}\left(s,t\right){\cal F}_{{\tilde \Delta},\ell^\prime}\left(s,t\right)Q^{(t,t,0,0)}_{\ell}\left(s\right).
\end{align}
The contributions from a CPW in the ${\sf u}$-channel differ from \eqref{cpwccint} by a factor of $\left(-1\right)^{\ell+\ell^\prime}$. The crossing kernels \eqref{cpwccint} were computed explicitly in \cite{Sleight:2018epi} for general $t$, where they were given in terms of the hypergeometric function ${}_4F_3$. For example, for the simple case of $\ell^\prime=0$ and equal external scalars of dimension $\Delta$ we have 
\begin{multline}\label{simpleinv}
    {}^{(\sf{t})}\mathfrak{I}_{{\tilde \Delta},0|\ell}\left(t\right)= \frac{(-2)^{\ell}\Gamma\left(\tfrac{d+t-2\Delta-{\tilde \Delta}}2\right)^2\Gamma\left(\tfrac{t-2\Delta+{\tilde \Delta}}2\right)^2\left(\tfrac{t}{2}\right)_{\ell}^2\Gamma({\tilde \Delta})}{\ell!\,\Gamma\left(\tfrac{d}2+t-2\Delta\right)(t+\ell-1)_{\ell}\Gamma\left(\frac{d}{2}-{\tilde \Delta}\right)\Gamma\left(\frac{{\tilde \Delta}}{2}\right)^4}\\\times\,{}_4 F_3\left(\begin{matrix}-\ell,t+\ell-1,\tfrac{d-{\tilde \Delta}}2+\frac{t}{2}-\Delta,\tfrac{{\tilde \Delta}}2+\frac{t}{2}-\Delta\\\tfrac{t}{2},\tfrac{t}2,\tfrac{d}{2}+t-2\Delta\end{matrix};1\right),
\end{multline}
which, as noted in \cite{Sleight:2018epi}, is proportional to a Wilson polynomial. Analogous expressions for crossing kernels with general exchanged spin $\ell^\prime$ in the crossed channel and for external spinning operators are reviewed in \S \ref{sec::crossingkernels}. In appendix \ref{6jsymbols} we explicate how these kernels are related to $6j$ symbols of the conformal group.

Given a crossing kernel, the task of extracting the OPE data thus boils down to evaluating its spectral integral. One of the main results of this work is the evaluation of the spectral integral of a generic crossing kernel, which has the form
\begin{equation}\label{genspecint}
     {}^{(\sf{t})}I_{ \Delta^\prime,\ell^\prime|\ell}\left(t\right) = n_{\Delta^\prime,\ell^\prime} \int^{ \infty}_{-\infty} \frac{d\nu}{2\pi}\,a_{\ell^\prime}(\nu)\, {}^{({\sf t})}\widehat{\mathfrak{I}}_{\frac{d}{2}+i\nu,\ell^\prime|\ell}\left(t\right),
\end{equation}
where $n_{\Delta^\prime,\ell^\prime}$ and the hat on the crossing kernel denotes a convenient choice of normalisation which, is defined in \S \ref{spectral}. Note that the spectral parameter $\nu$ is related to ${\tilde \Delta}$ in \eqref{cpweA} via ${\tilde \Delta}=\frac{d}{2}+i\nu$, $\nu \in \mathbb{R}$. The spectral function $a_{\ell}(\nu)$ has the form
\begin{equation}
    a_{\ell^\prime}(\nu) = a_{\tau^\prime, \ell^\prime}\, w_{a_i}\left(\nu\right) \frac{1}{\nu^2+(\Delta^\prime-\frac{d}{2})^2} \frac{1}{\left(\frac{d}{2}+\pm i \nu-1\right)_{\ell^\prime}},\label{anu}
\end{equation}
where $a_{\tau^\prime, \ell^\prime}$ is the OPE coefficient of the physical operator of spin $\ell^\prime$ and twist $\tau^\prime=\Delta^\prime-\ell^\prime$ exchanged in the crossed-channel \eqref{ttosblock}, and we have identified a spectral weight function 
\begin{align}\label{wa}
    w_{a_i}(\nu)&=\frac{\Gamma\left(a_1\pm \tfrac{i\nu}2\right)\Gamma\left(a_2\pm \tfrac{i\nu}2\right)\Gamma\left(a_3\pm \tfrac{i\nu}2\right)\Gamma\left(a_4\pm \tfrac{i\nu}2\right)}{\Gamma(\pm i\nu)}\,,
\end{align}
whose significance will become apparent shortly and we have employed the usual notation $\Gamma(a\pm b)\equiv \Gamma(a+b)\Gamma(a-b)$. The form \eqref{anu} of the spectral function was fixed by Polyakov in \cite{Polyakov:1974gs} by requiring that it decays exponentially as $\nu \rightarrow \pm i \infty$ so that the integral in $\nu$ along the real line is well-defined.\footnote{\label{foo::Witten} Remarkably, though in hindsight perhaps to be expected, this spectral function coincides with the spectral function arising from a Witten diagram for the exchange of a particle of spin-$\ell^\prime$ and mass $(m R_{\text{AdS}})^2=\Delta^\prime(\Delta^\prime-d)-\ell^\prime$ in AdS$_{d+1}$, as noted in \cite{Gopakumar:2016wkt,Gopakumar:2016cpb}. Such spectral functions have been computed for exchange Witten diagrams involving scalar external legs in \cite{Costa:2014kfa} and spinning external legs in \cite{Sleight:2017fpc}.}

To evaluate the spectral integral \eqref{genspecint}, a key observation is that the spectral weight function we singled out in \eqref{anu} is the measure with respect to which Wilson polynomials \cite{Wilson1980} are orthogonal. Wilson polynomials of degree $\ell$ are defined as 
\begin{align}\label{wilson}
    \mathcal{W}_{\ell}(\nu^2;a_i)&={}_4F_3\left(\begin{matrix}-\ell,a_1+a_2+a_3+a_4+\ell-1,a_1+\tfrac{i\nu}2,a_1-\tfrac{i\nu}2\\a_1+a_2,a_1+a_3,a_1+a_4\end{matrix};1\right),
\end{align}
and in our normalisation the precise orthogonality relation reads: 
\begin{multline}\label{OrtoN}
    \int_{-\infty}^{+\infty}\frac{d\nu}{2\pi}\,w_{a_i}(\nu)\,\mathcal{W}_\ell(\nu^2;a_i)\,\mathcal{W}_{\ell^\prime}(\nu^2;a_i)\\=\delta_{\ell,\ell^\prime}\,\frac{\ell! (a_2+a_3)_{\ell} (a_2+a_4)_{\ell} (a_3+a_4)_{\ell} (a_1+a_2+a_3+a_4+{\ell}-1)_{\ell}}{(a_1+a_2)_{\ell} (a_1+a_3)_{\ell} (a_1+a_4)_{\ell}(a_1+a_2+a_3+a_4)_{2 \ell}}\\\times\,\frac{4\,\Gamma (a_1+a_2) \Gamma (a_1+a_3) \Gamma (a_1+a_4) \Gamma (a_2+a_3) \Gamma (a_2+a_4) \Gamma (a_3+a_4)}{\Gamma (a_1+a_2+a_3+a_4)}.
\end{multline}
As we shall demonstrate, all crossing kernels determined in \cite{Sleight:2018epi} are naturally expressed in terms of finite sums of Wilson polynomials in the form
\begin{equation}
    {}^{({\sf t})}\widehat{\mathfrak{I}}_{\frac{d}{2}+i\nu,\ell^\prime|\ell}\left(t\right) = \sum_{j}\beta_j\left(t\right)\,\mathcal{W}_{j}(\nu^2;a_i),
\end{equation}
where the parameters $a_i$ match those of the measure \eqref{wa} and the number of terms in the sum depends only on $\ell^\prime$, ensuring analyticity in $\ell$. An example of this type of decomposition that we have already seen is the crossing kernel \eqref{simpleinv}. This observation reduces the spectral integral \eqref{genspecint} of the crossing kernel into a finite sum of spectral integrals of the Wilson polynomial with respect to the measure \eqref{wa}. The latter spectral integrals can be evaluated in closed form, which we carry out with full generality in \S \ref{EvaluationSp}. In all cases such integrals are finite sums of Wilson functions, which are the analytic continuation of Wilson polynomials \eqref{wilson} to non-integer $\ell$.\footnote{As we shall see in \S \ref{EvaluationSp}, Wilson functions are moreover a particular case of the $\psi$-function defined on page 127 of Lucy Slater's \cite{Slater}, which decomposes into a sum of two 1-balanced ${}_4F_3$ Hypergeometric functions and has many other interesting and useful properties.} 

The above perspective could give a further understanding behind the appearance of Wilson functions \cite{Groenevelt2006,doi:10.1063/1.524047,Hogervorst:2016hal,Hogervorst:2017sfd,Gopakumar:2018xqi} and Wilson polynomials \cite{Giombi:2018vtc,Sleight:2018epi} in various expressions available in the literature for crossing kernels/$6j$ symbols. We emphasise that, as a consequence of choosing Wilson polynomials as a basis for the crossing kernels, our results provide explicit analytic in spin expressions in general $d$ for the spectral integral \eqref{genspecint} and corresponding OPE data that are in particular valid (i.e. finite) for all values of scaling dimensions and spins. 

\paragraph{Large Spin Expansion}

In section \S \ref{sec::dtanom} of this work we consider above results for the spectral integral \eqref{genspecint} of crossing kernels at large spin. We highlight an alternative way to obtain the large spin expansion from the crossing kernels \cite{Sleight:2018epi}, which does not involve evaluating a spectral integral. In particular, at large spin the additional contributions generated by the shadow conformal multiplet in the conformal partial wave in the definition \eqref{cpwccint} can be projected away by hand.\footnote{Such contributions also become manifest in the following asymptotic behaviour of the crossing kernel \eqref{cpwccint} at large $\ell$
\begin{multline}
     {}^{({\sf t})}\mathfrak{I}_{{\tilde \Delta},\ell^\prime|\ell}\left(2\Delta\right) \sim \left(\frac1{\ell}\right)^{{\tilde \Delta}-\ell^\prime} \frac{\Gamma (\Delta )^2 \Gamma ({\tilde \Delta}+\ell^\prime)}{2^{\ell^\prime}\Gamma \left(\frac{{\tilde \Delta}+\ell^\prime}{2}\right)^2 \Gamma \left(\Delta -\frac{{\tilde \Delta}-\ell^\prime }{2}\right)^2}\\-\left(\frac1{\ell}\right)^{d-{\tilde \Delta}-\ell^\prime}\frac{\Gamma (\Delta )^2 \Gamma (d-{\tilde \Delta}+\ell^\prime)}{2^{\ell^\prime}\Gamma \left(\frac{d-{\tilde \Delta}+\ell^\prime}{2}\right)^2 \Gamma \left(\Delta -\frac{d-{\tilde \Delta}-\ell^\prime }{2}\right)^2}\,a^{\text{sh.}}_{0,0,\ell^\prime}+...\,,
\end{multline}
from which one clearly identifies the shadow contribution on the second line (cf. equation \eqref{intlsexpgamma}). The coefficient $a^{\text{sh.}}_{0,0,\ell^\prime}$ is the shadow OPE coefficient which is not important for this discussion and is defined in \cite{Sleight:2018epi}, section 4.5.} This moreover provides a re-summation of the large spin expansion which is analytic in spin for a large range of parameters (though not all, in which case it is only valid asymptotically). To project away the shadow contributions at large spin, we employ the Mellin representation of the hypergeometric function ${}_4F_3$ appearing in the crossing kernels:
\begin{multline}\label{mellinhyper}
    {}_4F_3\left(\begin{matrix}
    a_1,a_2,a_3,a_4\\
    b_1,b_2,b_3
    \end{matrix};z\right)=\frac{\Gamma (b_1) \Gamma (b_2) \Gamma (b_3)}{\Gamma (a_1) \Gamma (a_2) \Gamma (a_3) \Gamma (a_4)}\\\times\int \frac{ds}{2\pi i}\,\frac{ \Gamma (s) \Gamma (a_1-s) \Gamma (a_2-s) \Gamma (a_3-s) \Gamma (a_4-s)}{\Gamma (b_1-s) \Gamma (b_2-s) \Gamma (b_3-s)}\,(-z)^{-s}\,.
\end{multline}
E.g. for the $\ell^\prime=0$ example crossing kernel \eqref{simpleinv} above we have (for $t=2\Delta$):
\begin{subequations}\label{intabex}
\begin{align}
    a_1&=-\ell, && b_1=\Delta\\
    a_2&=2\Delta+\ell-1, && b_2=\Delta\\
    a_3&=\tfrac{d-{\tilde \Delta}}{2}, && b_3=\tfrac{d}{2}\\
    a_4&=\tfrac{{\tilde \Delta}}{2}
\end{align}
\end{subequations}
The representation \eqref{mellinhyper} has two useful features:
\begin{itemize}
   
   \item It is manifest how the crossing kernel decomposes into the individual contributions from the physical conformal block and its shadow appearing in the conformal partial wave \eqref{cpwdefint}, whose respective families of primary and descendants are captured by the poles of individual and distinct $\Gamma$-functions in the Mellin variable $s$. Closing the contour in the positive $s$ plane, such contributions are encoded in the residues of $\Gamma(a_3-s)$ and $\Gamma(a_4-s)$, respectively. For instance, for the simple case of a scalar exchange \eqref{simpleinv}, such poles are:
\begin{subequations}
\begin{align}
    \Gamma \left(\frac{{\tilde \Delta}}{2}-s\right) &\rightarrow \text{physical block},\\
    \Gamma \left(\frac{d-{\tilde \Delta}}{2}-s\right)  &\rightarrow \text{shadow block}.
\end{align}
\end{subequations}
In this way we recover contribution of a single conformal block by simply dropping the shadow residues. It is also useful to note that in a large spin expansion the poles of $\Gamma(a_2-s)$ can be dropped. We have moreover checked that they give contributions which vanish for integer $\ell$.\footnote{\label{foo::1}In the simplest example of the scalar exchange these poles are encoded in:
\begin{equation}
    \Gamma (\ell+2\Delta-1-s)\,,
\end{equation}
and their re-summation gives terms proportional to $1/\Gamma(-\ell)$ which vanish for integer $\ell$. We shall drop these contributions from the Mellin integral keeping only those of the physical conformal block.}  This will also be discussed in the next bullet point. 

 \item The Mellin representation allows to systematically determine the asymptotic expansion in $1/\mathfrak{J}^2$ from the known \cite{Fields} asymptotic behaviour \eqref{asymratgamma} of the following simple ratios of $\Gamma$-functions: 
\begin{equation}\label{gammaratio}
    \frac{\Gamma(a_1-s)\Gamma(a_2-s)}{\Gamma(a_1)\Gamma(a_2)}\sim (-1)^s\frac{\Gamma (1-a_1)}{\Gamma (1-a_1+s)}\frac{\Gamma(a_2-s)}{\Gamma(a_2)},
\end{equation}
where (schematically) $a_1\sim -\ell$ and $a_2\sim \ell$. In all examples $a_1$ and $a_2$ are the only entries of the hypergeometric function that depend on $\ell$. On the right hand side we have conveniently removed the essential singularity around $\ell\to\infty$ with a replacement that does not affect the residues at $s=-n$ for $n\in \mathbb{N}$. In \S\ref{sec:LargeSpin} we review how to obtain the expansion of \eqref{gammaratio} in $1/\mathfrak{J^2}$, which is given in terms of generalised Bernoulli polynomials.\footnote{The utility of the asymptotic behaviour of simple ratios of Gamma functions was noted in \cite{Dey:2017fab} where it was used to determine the large spin expansion of the Mack polynomials \eqref{mackpoly}, which can be expressed explicitly in terms of the hypergeometric functon ${}_3F_{2}$.}

\end{itemize}

In terms of coefficients $a_i$ and $b_i$, with (schematically) $a_1\sim -\ell$, $a_2\sim \ell$, $a_3\sim\tfrac{d-{\tilde \Delta}}2$ and $a_4\sim\tfrac{{\tilde \Delta}}2$ (which is the case for all crossing kernels), the projection onto the contribution of a single physical conformal block is simply the replacement:\footnote{Equation \eqref{largespinsub} is obtained by re-summing the following series associated to the non-shadow poles 
\begin{equation}
    -\tfrac{\Gamma (b_1) \Gamma (b_2) \Gamma (b_3)}{\Gamma (a_1) \Gamma (a_2) \Gamma (a_3) \Gamma (a_4)}\sum_n \tfrac{(-1)^{2-n} \Gamma (1-a_1) \Gamma (a_1) \Gamma (a_4+n) \Gamma (a_2-a_4-n) \Gamma (a_3-a_4-n)}{n! \Gamma (-a_1+a_4+n+1) \Gamma (-a_4+b_1-n) \Gamma (-a_4+b_2-n) \Gamma (-a_4+b_3-n)}\,,
\end{equation}
after removing the essential singularities as explained in footnote \ref{foo::1}.} 
\begin{multline}
    {}_4F_3\left(\begin{matrix}
    a_1,a_2,a_3,a_4\\
    b_1,b_2,b_3
    \end{matrix};1\right)\rightarrow\\-\frac{\Gamma (1-a_1) \Gamma (b_1) \Gamma (b_2) \Gamma (b_3) \Gamma (a_2-a_4) \Gamma (a_3-a_4)}{\Gamma (a_2) \Gamma (a_3) \Gamma (-a_1+a_4+1) \Gamma (b_1-a_4) \Gamma (b_2-a_4) \Gamma (b_3-a_4)}\\\times \, _4F_3\left(\begin{matrix}a_4,a_4-b_1+1,a_4-b_2+1,a_4-b_3+1\\-a_1+a_4+1,-a_2+a_4+1,-a_3+a_4+1\end{matrix};1\right)\,.\label{largespinsub}
\end{multline}
The above projection operation, with due care about the analytic continuation of the hypergeometric function when varying its arguments, can be straightforwardly performed on the crossing kernels expressed in terms of ${}_4F_3$ to obtain a re-summation of the $1/\mathfrak{J}^2$ expansion \eqref{intlsexpgamma} for the double-twist operator anomalous dimensions induced the exchange of a physical conformal block \eqref{ttosblock} in the crossed channel. We stress that, although it is not analytic in spin for all values of the parameters (in which case the resulting expression is only valid asymptotically), in many cases this re-summation matches the analytic result in spin which we obtain by evaluating the spectral integral as in \eqref{anomdint}.\footnote{The mismatch can be ascribed in general to the poor behaviour of conformal blocks at infinity. This can generate boundary terms in the Mellin or spectral plane, which are cured by including the double-twist poles in the spectral measure as in \eqref{wa}.} For all applications of our results considered in \S \ref{slightlyB}, the re-summation of the large spin expansion obtained in the way described above coincides with the analytic results obtained instead by evaluating the spectral integral \eqref{genspecint}.

\subsection{Outline and summary of results}

\begin{itemize}[leftmargin=*]
    \item In \S \ref{spectral} we discuss the spectral integral of generic crossing kernels, underlining the connection between crossing kernels and Wilson polynomials. In particular, we highlight that the spectral integral can be regarded as an inner product of the crossing kernel with the physical poles in $\nu$, where the spectral measure is given by the Wilson measure \eqref{wa}
\begin{subequations}
\begin{align}
    {}^{({\sf t})}I_{\Delta^\prime,\ell^\prime|\ell}(t) &\;\sim\; \left\langle \frac{1}{\nu^2+\left(\Delta^\prime-\tfrac{d}{2}\right)^2}\,\Bigg|\,{}^{({\sf t})}\widehat{\mathfrak{I}}_{\frac{d}{2}+i\nu,\ell^\prime|\ell}\left(t\right)\right\rangle\,,\\
    \left\langle p(\nu^2)|q(\nu^2)\right\rangle&\;=\;\int_{-\infty}^{+\infty}\,\frac{d\nu}{2\pi}\, w_{a_i}(\nu)\,p(\nu^2)\,q(\nu^2).
\end{align}
\end{subequations}

\item In \S \ref{EvaluationSp} we detail the general approach to evaluating such spectral integrals. We show how crossing kernels are naturally decomposed as a finite sum of Wilson polynomials. This reduces the task of evaluating the spectral integral of crossing kernels to the evaluation of the spectral integrals of Wilson polynomials, which we refer to as seed integrals $ \phi_\ell(a_i)$:
\begin{align}
  \phi_\ell(a_i) = \int_{-\infty}^{+\infty}\,\frac{d\nu}{2\pi}\, w_{a_i}(\nu) \frac{1}{\nu^2+\left(\Delta^\prime-\tfrac{d}{2}\right)^2} {\cal W}_\ell
(a_i).
\end{align}
We show that such spectral integrals of Wilson polynomials are given by Wilson functions (see e.g. eq.~(\tcb{3.2}) of \cite{Groenevelt2003}), which provide an analytic continuation of the Wilson polynomial in its degree $\ell$. These features ensure that the spectral integral of the corresponding crossing kernel is analytic in spin $\ell$. Wilson functions admit various convenient explicit forms \cite{Groenevelt2003}, for example: as a (``very well poised") hypergeometric function ${}_7F_6$, a combination of $1$-balanced (Saalsch\"utzian) hypergeometric functions ${}_4F_3$ or, less explicitly, in terms of integrated products of hypergeometric functions ${}_2F_1$.

\item In \S \ref{ExchScalar} we consider the case of equal external scalar operators. Using the result for the spectral integral of the corresponding crossing kernel, we extract the anomalous dimensions of leading double-twist operators $\left[{\cal O}{\cal O}\right]_{0,\ell}$ induced by the exchange of a scalar with arbitrary twist $\tau^\prime$ in the crossed channel
\begin{multline}
    \gamma_{0,\ell}=-\frac{2\,\Gamma (\tau^\prime) \Gamma \left(\tau^\prime+1-\frac{d}{2}\right)}{\Gamma(\tfrac{d}{2}) \Gamma \left(\frac{\tau^\prime}{2}\right)^4 \Gamma \left(\tfrac{2\Delta-\tau^\prime}{2}\right)^2 \Gamma \left(\tfrac{2 \Delta +\tau^\prime-d}{2}\right)^2}\,\\ \times \phi_\ell\left(\frac{d}{4},\frac{d}{4},\Delta -\frac{d}{4},\Delta -\frac{d}{4},\frac{2\tau^\prime-d}{4},\frac{2\tau^\prime-d}{4}\right)
\end{multline}
Here $\Delta$ is the scaling dimension of the external operators ${\cal O}$ and the result is for arbitrary $d$. This expression is manifestly analytic in spin-$\ell$ from the definition of $\phi_{\ell}$, which is a Wilson function.
\item In \S \ref{ExchSpin} the results of section  \S \ref{ExchScalar} are generalised to the case where the exchanged operator of twist  $\tau^\prime$  in the crossed channel has arbitrary spin $\ell^\prime$. This induces the following expression for the anomalous dimensions of leading double-twist operators $\left[{\cal O}{\cal O}\right]_{0,\ell}$: 
\begin{equation}
    \gamma_{0,\ell}=2\,n_{\tau^\prime,\ell^\prime}\sum_{j=0}^{2\ell^\prime}\beta_{\ell,j}^{(\ell^\prime)}\,\phi_{\ell-j}\left(\tfrac{d+2\ell^\prime}{4}, \tfrac{d+2\ell^\prime}{4}, \Delta -\tfrac{d-2\ell^\prime}{4},\Delta -\tfrac{d-2\ell^\prime}{4},{\tfrac{\tau^\prime+\ell^\prime}2-\tfrac{d}{4}},{\tfrac{\tau^\prime+\ell^\prime}2-\tfrac{d}{4}}\right)\,,
\end{equation}
where the coefficients $\beta_{\ell,j}^{(\ell^\prime)}$ are the expansion coefficients of the crossing kernel in terms of Wilson polynomials, which are derived in appendix \S \ref{CrossingToWilson}. The coefficient $n_{\tau^\prime,\ell^\prime}$ is a normalisation and is given explicitly in \eqref{normgenellprime}. The result is a finite linear combination of Wilson functions $\phi_{\ell-j}$ and therefore manifestly analytic in spin $\ell$.

\item In \S \ref{ExchSub} we consider the generalisation of the results in section \S \ref{ExchScalar} to double-twist operators $\left[{\cal O}{\cal O}\right]_{n,\ell}$ of sub-leading twist. I.e. for all $n\ne0$. The anomalous dimensions of these operators induced by the exchange of a scalar of twist $\tau^\prime$ in the crossed channel are given by: 
\begin{equation}
    \gamma_{n,\ell}=2\,n_{\tau^\prime,0}\sum_{i=0}^{2n}\beta_i^{(\ell,n)}\phi_{\ell+i}\left(\frac{d}{4},\frac{d}{4},\Delta -\frac{d}{4},\Delta -\frac{d}{4},\frac{2\tau^\prime-d}{4},\frac{2\tau^\prime-d}{4}\right),
\end{equation}
where the normalisation coefficient $n_{\tau^\prime,0}$ is given explicitly in equation \eqref{ccoeff}.

\item In \S \ref{spinninWilson} we consider external spinning operators, focusing on the case where there are two operators of (totally symmetric) spin $J_1$ and $J_2$ together with two other scalar operators. Focusing on the contributions to leading twist operators in the ${\sf s}$-channel induced by the exchange of a scalar of twist $\tau^\prime$ in the crossed channel, we evaluate the spectral integral of the corresponding crossing kernel. When extracting OPE data in this case, there is an operator mixing problem which we discuss in detail (see also \cite{Aharony:2018npf} for a recent related discussion). Because of operator mixing, the result for the spectral integral only gives access to ``averages" of the anomalous dimensions of double-twist operators $\left[{\cal O}_{J_1}{\cal O}_{J_2}\right]_\ell$ at that order, which we find are proportional to a single Wilson function: 
\begin{multline}
    c^{(0)}_{\mathcal{O}_{J_1}\mathcal{O}_{J_2}[\mathcal{O}_{J_1}\mathcal{O}_{J_2}]_{(\ell)}}c^{(0)}_{\mathcal{O}\mathcal{O}[\mathcal{O}\mathcal{O}]_{(\ell)}}\gamma_{0,\ell} \\ = 2n_{\tau^\prime,0}\,\beta_{\ell}\,\phi_{\ell}\left(\frac{d}{4}+J_1,\frac{d}{4}+J_2,-\frac{d}{4}+J_1+\Delta,-\frac{d}{4}+J_2+\Delta,\frac{2\tau^\prime-d}{4},\frac{2\tau^\prime-d}{4}\right),
\end{multline}
where we considered the case of equal external twists $\tau_i=\Delta$. The coefficients $c^{(0)}_{\mathcal{O}_{J_1}\mathcal{O}_{J_2}[\mathcal{O}_{J_1}\mathcal{O}_{J_2}]_{(\ell)}}$ are the mean field theory OPE coefficients and the normalisation coefficient $n_{\tau^\prime,0}$ in this spinning case is given explicitly in equation \eqref{normalj1j2}

\item In \S \ref{sec::dtanom} we consider the above results at large spin. We moreover highlight an alternative way to obtain the large spin expansion from the crossing kernels \cite{Sleight:2018epi}, which does not involve evaluating a spectral integral. In particular, at large spin the additional contributions generated by the shadow conformal multiplet in the conformal partial wave in the definition \eqref{cpwccint} can be projected away by hand using the Mellin representation of the hypergeometric functions in terms of which the crossing kernels \cite{Sleight:2018epi} are given. This alternative approach moreover provides a re-summation of the large spin expansion which is analytic in spin (and thus coincides with the results obtained in \S \ref{SpectralFull}) for a large range of parameters. When this is not the case this re-summation is only valid asymptotically.

\item In \S \ref{slightlyB} we consider applications of our results to CFTs with slightly broken higher-spin symmetry in the large $N$ limit. In \S \ref{subsubsec::scalarcrr} we first extract ${\cal O}\left(1/N\right)$ anomalous dimensions of $\left[{\cal O}{\cal O}\right]_{n,\ell}$ double-trace operators for the critical Boson/Fermion and quasi Fermion/Boson theories. In \S \ref{sec::spinningcorrelators} we consider external spinning operators, using the results of \S \ref{spinninWilson} to extract anomalous dimensions of leading twist $\left[{\cal O}_J{\cal O}\right]_{0,\ell}$ double-trace operators in the quasi Boson theory at ${\cal O}\left(1/N\right)$, where ${\cal O}$ is an operator of arbitrary spin $J$. In each subsection we confirm existing results for these theories and also obtain new ones.

\end{itemize}

Various complicated formulas and technicalities are collected in the appendices. 

In appendix \ref{6jsymbols} we discuss the relation of our approach and the corresponding results to $6j$ symbols of the conformal group.
\newpage

\section{Spectral integrals and double-twist anomalous dimensions}\label{SpectralFull}

In section \S \ref{spectral} we write down the spectral integral \eqref{genspecint} for a generic crossing kernel, and provide a method to evaluate it in \S \ref{EvaluationSp} which is based on the decomposition of crossing kernels in terms of a finite sum of Wilson polynomials detailed in appendix \S \ref{CrossingToWilson}. In \S \ref{ExchScalar} we consider explicitly the spectral integral of crossing kernels corresponding to the contribution of a scalar exchange in the crossed channel to the exchange of leading twist operators of general spin $\ell$ in the ${\sf s}$-channel. The generalisation to exchanged operators of spin-$\ell^\prime$ in the crossed channel is presented in \S \ref{ExchSpin}. The generalisation to contributions to operators of subleading twist is considered in \S \ref{ExchSub}, and to spinning external operators in \S \ref{spinninWilson}.

\subsection{The spectral integral}\label{spectral}

Our goal is to evaluate the spectral integral
\begin{equation}\label{int2}
    {}^{({\sf t})}I_{\Delta^\prime,\ell^\prime|\ell}\left(t\right) = n_{\Delta^\prime,\ell^\prime} \int^{ \infty}_{-\infty} \frac{d\nu}{2\pi}\,a_{\ell^\prime}(\nu)\, {}^{({\sf t})}\widehat{\mathfrak{I}}_{\frac{d}{2}+i\nu,\ell^\prime|\ell}\left(t\right),
\end{equation}
which we consider for both external scalar operators and when two of the external operators have non-zero spin. The normalised crossing kernel in \eqref{int2} is related to the original un-hatted crossing kernel as obtained in \cite{Sleight:2018epi} via
\begin{equation}\label{CrossingNorm}
    {}^{({\sf t})}\widehat{\mathfrak{I}}_{\frac{d}{2}+i\nu,\ell^\prime|\ell}\left(t\right)=\frac{\pi^{d/2}}{\kappa_{\frac{d}{2}-i\nu,\ell^\prime}\alpha_{\ell_3,\ell_4,\ell^\prime;\tau_3,\tau_4,\frac{d}{2}+i\nu-\ell^\prime}}{}^{({\sf t})}{\mathfrak{I}}_{\frac{d}{2}+i\nu,\ell^\prime|\ell}\left(t\right),
\end{equation}
where the functions $\kappa$ and $\alpha$ are defined for instance in eq.~(\tcb{2.30}) and  eq.~(\tcb{A.13}) of \cite{Sleight:2018epi}, and arise from the shadow transform appearing in the integral representation of CPWs. This normalisation is convenient as the crossing kernel is then a polynomial in $\nu$ (in particular with respect to the Wilson measure \eqref{wa}), which facilitates its decomposition in terms of Wilson polynomials. 

The Polyakov spectral function $a_{\ell^\prime}(\nu)$ is given by the spectral function arising from a Witten diagram (\emph{c.f.} footnote \ref{foo::Witten}) for the exchange of a spin-$\ell^\prime$ particle of mass {\footnotesize $(m R_{\text{AdS}})^2=\Delta^\prime(\Delta^\prime-d)-\ell^\prime$} in AdS$_{d+1}$, which for external operators of twist $\tau_i$ and spin $J_i$ is given by equation (3.29) of \cite{Sleight:2017fpc}\footnote{For generic spins $J_i$ one should also specify the three-point conformal structures of the conformal partial wave in \eqref{cpwccint}, which are parametrised by the $n_i$ in equation (3.29) of \cite{Sleight:2017fpc}. The crossing kernels we consider in this work consist of three-point conformal structures that involve at most one spinning operator, in which case the conformal structure is unique and corresponds to $n_i=0$.}
\begin{equation}\label{anuwitten}
     {a}_{\ell^\prime}(\nu)=\frac{\nu^2}{\pi}\,\frac{1}{\nu^2+\left(\Delta^\prime-\tfrac{d}{2}\right)^2}\,{\sf B}_{J_1,J_2,\ell^\prime;\tau_1,\tau_2,\tfrac{d}{2}+i\nu-\ell^\prime}\,{\sf B}_{\ell^\prime,J_3,J_4;\tfrac{d}{2}-i\nu-\ell^\prime,\tau_3,\tau_4},
\end{equation}
where we set for simplicity the bulk coupling constant to one. There is a non-trivial OPE coefficient $a_{\tau^\prime,\ell^\prime}$ generated by the integration over the volume of AdS
\begin{subequations}
\begin{align}
    a_{\tau^\prime,\ell^\prime}&=\frac{{\sf B}_{J_1,J_2,\ell^\prime;\tau_1,\tau_2,\tau^\prime}{\sf B}_{\ell^\prime,J_3,J_4;\tau^\prime,\tau_3,\tau_4}}{C_{\Delta^\prime,\ell^\prime}},\\
    C_{\Delta^\prime,\ell^\prime}&=\frac{(\ell^\prime+\Delta^\prime-1)\Gamma(\Delta^\prime)}{2\pi^{d/2}(\Delta^\prime-1)\Gamma\left(\Delta^\prime+1-\tfrac{d}2\right)},
\end{align}
\end{subequations}
which one should factor out as in \eqref{anu}. The coefficient on the second line is the two-point function normalisation. 

It is important to keep in mind that, when the exchanged spin $\ell^\prime$ in the crossed channel is non-zero, the spectral function $a_{\ell^\prime}\left(\nu\right)$ contains contributions from a finite number of integer space spurious poles. We displayed these poles explicitly in \eqref{anu}, where they encoded by the Pochhammer factor
\begin{equation}
    \frac1{\left(\frac{d}{2}\pm i \nu -1\right)_{\ell^\prime}}\,.
\end{equation}
Such poles are non-physical when they do not overlap with the physical poles at $\frac{d}{2}\pm i\nu=\tau^\prime+\ell^\prime$, however their contribution cancels when considering the full Witten exchange diagram.\footnote{Technically speaking, the spectral function \eqref{anuwitten} is generated by the traceless and transverse part of the AdS propagator in the Witten exchange diagram, which encodes the physical exchange of the single-particle state. This is accompanied by a tail of contact terms, which are generated by the off-shell terms in the AdS propagator. These contact terms cancel the spurious poles, see e.g. \cite{Costa:2014kfa}.} We therefore need not consider the contributions from these poles when evaluating the spectral integral. Since these poles are finite in number at fixed spin $\ell^\prime$, we can separate them from the physical poles using the following expansion: 
\begin{equation}\label{spurious}
    \frac{1}{\nu^2+\left(\Delta^\prime-\tfrac{d}{2}\right)^2}\frac{1}{\left(\frac{d}{2}\pm i \nu -1\right)_{\ell^\prime}}=\frac{A_{\ell^\prime,\tau^\prime}}{\nu^2+\left(\Delta^\prime-\tfrac{d}{2}\right)^2}+\sum_{n=0}^{\ell^\prime-1}\frac{B_{\ell^\prime,\tau^\prime}^{(n)}}{\nu^2+\left(\tfrac{d-2}2+n\right)^2}
\end{equation}
where
\begin{subequations}
\begin{align}
    A_{\ell^\prime,\tau^\prime}&=\frac{1}{(\Delta^\prime-1)_{\ell^\prime} (d-\Delta^\prime-1)_{\ell^\prime}}\,,\\
    B_{\ell^\prime,\tau^\prime}^{(n)}&=-\frac{4 (-1)^n (d+2 n-2) \Gamma (d+n-2)}{n! (\Delta^\prime+n-1) (d-\Delta^\prime+n-1)\Gamma (\ell^\prime-n) \Gamma (d+\ell^\prime+n-2)}\,.
\end{align}
\end{subequations}
The expansion \eqref{spurious} makes sense whenever $\Delta^\prime-\tfrac{d}2\neq \tfrac{d}2-1+n$ with $0\leq n\leq\ell^\prime-1$. Otherwise one of the spurious poles collides with a physical pole, which generates a double-pole. This corresponds to the appearance of (partially-)massless representations \cite{Joung:2012rv}. In the latter case the result can be defined as a limit of the sum of the two colliding poles, which remains non-singular in the limit (see e.g. \S(4.3) of \cite{Sleight:2018epi}).

With the above observation in mind, the spectral integral \eqref{genspecint} of the crossing kernel can be expressed in the form
\begin{align}\label{SpectralInt}
 \frac{1}{n_{\Delta^\prime,\ell^\prime}a_{\tau^\prime,\ell^\prime}} {}^{({\sf t})}{I_{\tau^\prime,\ell^\prime|\ell}}(t)&=A_{\ell^\prime,\tau^\prime}\,{}^{({\sf t})}{I_{\tau^\prime,\ell^\prime|\ell}^{(\text{phys.})}}(t)+\sum_{n=0}^{\ell^\prime-1}\,B_{\ell^\prime,\tau^\prime}^{(n)}\,{}^{({\sf t})}{I_{\tau^\prime,\ell^\prime|\ell}^{(n)}}(t)\,,
\end{align}
where 
\begin{equation}\label{physspectint}
    {}^{({\sf t})}{I_{\Delta^\prime,\ell^\prime|\ell}^{(\text{phys.})}}(t) =  \int^{+\infty}_{-\infty}\frac{d\nu}{2\pi}\,w_{a_i}\left(\nu\right)\, \frac{1}{\nu^2+\left(\Delta^\prime-\tfrac{d}{2}\right)^2}\,{}^{({\sf t})}\widehat{\mathfrak{I}}_{\frac{d}{2}+i\nu,\ell^\prime|\ell}\left(t\right),
\end{equation}
is the contribution from the physical pole that we need to evaluate, and
\begin{equation}
    {}^{({\sf t})}{I_{\ell^\prime|\ell}^{(n)}}(t) = \int^{+\infty}_{-\infty}\frac{d\nu}{2\pi}\,w_{a_i}\left(\nu\right)\, \frac{1}{\nu^2+\left(\tfrac{d-2}2+n\right)^2}\,{}^{({\sf t})}\widehat{\mathfrak{I}}_{\frac{d}{2}+i\nu,\ell^\prime|\ell}\left(t\right),
\end{equation}
are the contributions from the spurious poles which we can neglect, though the spectral integral can be evaluated in exactly the same way!

In order to perform the above type of spectral integral, we express the hatted crossing kernels \eqref{CrossingNorm} in terms of Wilson polynomials. In fact, for $\ell^\prime=0$ it has already been observed \cite{Sleight:2018epi} (and in \cite{Giombi:2018vtc} for $t=2\Delta$ and external scalars) that the corresponding crossing kernels are proportional to a single Wilson polynomial. In appendix \ref{CrossingToWilson} we show that for non-zero $\ell^\prime$ the crossing kernels can be expressed as a finite sum of Wilson polynomials of the form
\begin{equation}
    {}^{({\sf t})}\widehat{\mathfrak{I}}_{\frac{d}{2}+i\nu,\ell^\prime|\ell}\left(t\right)=\sum_{j}\beta_j(t)\,\mathcal{W}_{j}(\nu^2;a_1,a_2,a_3,a_4)\,,
\end{equation}
where the number of terms in the sum is a function of $\ell^\prime$, and not of $\ell$. In this way, the spectral integral \eqref{int2} of the crossing kernel acquires a very natural meaning as an inner product with respect to the Wilson measure \eqref{wa}:
\begin{equation}
    \left\langle p(\nu^2)|q(\nu^2)\right\rangle=\int_{-\infty}^{+\infty}\,\frac{d\nu}{2\pi}\, w_{a_i}(\nu)\,p(\nu^2)\,q(\nu^2).
\end{equation}
In particular 
\begin{equation}
    {}^{({\sf t})}{I_{\tau^\prime,\ell^\prime|\ell}^{(\text{phys.})}}(t)=\left\langle \frac{1}{\nu^2+\left(\Delta^\prime-\tfrac{d}{2}\right)^2}\,\Bigg|\,{}^{({\sf t})}\widehat{\mathfrak{I}}_{\frac{d}{2}+i\nu,\ell^\prime|\ell}\left(t\right)\right\rangle\,.
\end{equation}
which appears to select Wilson polynomials as a natural basis for crossing kernels. From this perspective, crossing can be rephrased as the orthogonal projection of the \emph{physical} spectral poles onto the polynomial basis of crossing kernels! In hindsight, this property would appear to give further clarity behind the appearance of Wilson polynomials in expressions for crossing kernels given so far in the literature \cite{Giombi:2018vtc,Sleight:2018epi}.

\subsection{Evaluating the spectral integral}\label{EvaluationSp}

Since we decompose crossing kernels into a finite sum of Wilson polynomials, to evaluate their spectral integral \eqref{int2} we simply need to know how to evaluate a seed integral of the general form:
\begin{subequations}\label{Wintegral}
\begin{align}
    {{\phi}}_\ell(a_i)&= \left\langle \frac{1}{4}\frac{\Gamma\left(a_5 \pm \tfrac{i \nu}2\right)}{\Gamma\left(1+a_6 \pm \tfrac{i \nu}2\right)}\,\Bigg|\,\mathcal{W}_\ell(\nu^2;a_i)\right\rangle\\
    &=\int_{-\infty}^{+\infty}\frac{d\nu}{2\pi}\,w_{a_i}(\nu)\,\frac{1}{4}\frac{\Gamma\left(a_5 \pm \tfrac{i \nu}2\right)}{\Gamma\left(1+a_6 \pm \tfrac{i \nu}2\right)}\,\mathcal{W}_\ell(\nu^2;a_i)\,,
\end{align}
\end{subequations}
where $\mathcal{W}_\ell(\nu^2;a_i)$ is a degree $\ell$ Wilson polynomial with parameters $a_i$ matching those of the spectral weight $w_{a_i}(\nu)$. This integral is slightly more general for our purposes, where for the specific type of spectral integral \eqref{physspectint} under consideration we have $a_5=a_6=\frac{1}{2}\left(\Delta^\prime-\frac{d}{2}\right)$. In particular:
\begin{equation}
    \frac{1}{4}\frac{\Gamma\left(a_5 \pm \tfrac{i \nu}2\right)}{\Gamma\left(1+a_6 \pm \tfrac{i \nu}2\right)}\Big|_{a_5=a_6=\frac{1}{2}\left(\Delta^\prime-\frac{d}{2}\right)} = \frac{1}{\nu^2+\left(\Delta^\prime-\tfrac{d}{2}\right)^2}.
\end{equation}
We shall keep $a_5$ and $a_6$ arbitrary in the following, in order to be as general as possible.

The integral \eqref{Wintegral} may appear to be formidable, however we can evaluate it explicitly using the following trick. The basic idea is to reduce the integral to a finite sum of simpler integrals of the same form as in \eqref{Wintegral} but with $\ell=0$,
\begin{equation}\label{Wintegral2}
    \phi_0(a_i)=\int_{-\infty}^{+\infty} \frac{d\nu}{2\pi} \,w_{a_i}(\nu)\,\frac{1}{4}\frac{\Gamma\left(a_5 \pm \frac{i\nu}{2}\right)}{\Gamma\left(1+a_6 \pm \frac{i \nu}{2}\right)}.
\end{equation}
This can be achieved by first expanding the Wilson polynomial as\footnote{Note that our choice of normalisation for Wilson polynomials differs from the one usually adopted in the literature \cite{Wilson1980}.}
\begin{align}\label{expwilson}
    \mathcal{W}_\ell(\nu^2;a_i)=\sum_{k=0}^\ell\frac{(-\ell)_k \left(a_1\pm\frac{i \nu }{2}\right)_k (a_1+a_2+a_3+a_4+\ell-1)_k}{k! (a_1+a_2)_k (a_1+a_3)_k (a_1+a_4)_k}\,,
\end{align}
so that
\begin{equation}\label{ksum}
    \phi_\ell(a_i)=\sum_{k=0}^{\ell}\frac{(-\ell)_k (a_1+a_2+a_3+a_4+\ell-1)_k}{k! (a_1+a_2)_k (a_1+a_3)_k (a_1+a_4)_k}\phi_0(a_1+k,a_{i>1})\,,
\end{equation}
where we absorbed the $\nu$-dependent Pochhammer symbols in \eqref{expwilson} into the Wilson measure \eqref{wa}. 

The spectral integral \eqref{Wintegral2} may still appear rather complicated. However, it can be simplified using the Barnes' 2nd Lemma, which provides a useful transformation formula:\footnote{It may be useful to note that equation \eqref{Idnu} plays a key role in performing various spectral integrals that arise in Witten diagram computations, which generally involve products of $\Gamma(a_1\pm\tfrac{i\nu}2)$.}
\begin{equation}\label{Idnu}
    \frac{\Gamma \left(a_1\pm\frac{i\nu }{2}\right)  \Gamma \left(a_3\pm\frac{i\nu }{2}\right)  \Gamma \left(a_4\pm\frac{i\nu }{2}\right) }{\Gamma (a_1+a_3) \Gamma (a_1+a_4) \Gamma (a_3+a_4)}=\int\frac{ds}{2\pi i}\,\frac{\Gamma (-s\pm\frac{i\nu}2)  \Gamma \left(a_1+s\right) \Gamma \left(a_3+s\right) \Gamma \left(a_4+s\right)}{\Gamma \left(a_1+a_3+a_4+s\right)}\,.
\end{equation}
Note that the above transformation formula breaks the manifest symmetry of the integrand \eqref{Wintegral2} under permutations of $a_i$, $i=1,\ldots,5$. Exchanging the order of integration we arrive to the following simpler double-integral
\begin{multline}
    \phi_0(a_i)=\Gamma (a_1+a_3) \Gamma (a_1+a_4)\Gamma (a_3+a_4)\int_{-i\infty}^{+i\infty}\frac{ds}{2\pi i}\frac{ \Gamma (a_1+s)   \Gamma (a_3+s) \Gamma (a_4+s) }{4 \Gamma (a_1+a_3+a_4+s)}\\\times\int_{-\infty}^{+\infty}\frac{d\nu}{2\pi}\frac{\Gamma \left(a_2\pm\frac{i\nu }{2}\right)\Gamma \left(a_5\pm\frac{i\nu }{2}\right) \Gamma \left(-s\pm\frac{i\nu }{2}\right)}{\Gamma (\pm i\nu )  \Gamma \left(a_6\pm\frac{i\nu }{2}+1\right)}\,.
\end{multline}
In this form the spectral integral is more manageable due to the reduction in Gamma function factors, and can be evaluated in the usual way by evaluating the residues of each pole in the three series of poles associated to three of the six $\Gamma$-functions in the numerator. The contribution from each series of poles can be further re-summed in terms of a ``very well poised'' ${}_5F_4$ hypergeometric function.\footnote{In general it might be useful to keep in mind that that spectral integrals involving products of $\Gamma(a_i\pm\tfrac{i\nu}2)$ can often be expressed in terms of sums of very well poised Hypergeometric functions.} For instance, the series of poles for $\nu=i (2a_2+2n)$ gives the contribution:
\begin{multline}
    \frac{2\, \Gamma (a_5-a_2) \Gamma (a_2+a_5) \Gamma (-a_2-s) \Gamma (a_2-s)}{\Gamma (-2 a_2) \Gamma (-a_2+a_6+1) \Gamma (a_2+a_6+1)}\\\times\, _5F_4\left(\begin{matrix}
    2 a_2,&a_2+1,&a_2+a_5,&a_2-a_6,&a_2-s\\&a_2,&a_2-a_5+1,&a_2+a_6+1,&a_2+s+1\end{matrix};1\right)\,.
\end{multline}
One can then use equation (\tcb{1}) in \S \tcb{4.4} of \cite{Bailey} to re-sum such hypergeometric functions as a ratio of Gamma functions:
\begin{multline}
    {}_5F_4\left(\begin{matrix}
    a,&1+\tfrac{a}2,&c,&d,&e\\
    &\tfrac{a}2,&1+a-c,&1+a-d,&1+a-e
    \end{matrix},1\right)\\=\frac{\Gamma (a-c+1) \Gamma (a-d+1) \Gamma (a-e+1) \Gamma (a-c-d-e+1)}{\Gamma (a+1) \Gamma (a-c-d+1) \Gamma (a-c-e+1) \Gamma (a-d-e+1)}\,.
\end{multline}
The contribution of each series of poles can then be simplified and combined so that we end up with the following Mellin integral:
\begin{multline}
    \phi_0(a_i)=\frac{\Gamma (a_1+a_3) \Gamma (a_1+a_4) \Gamma (a_2+a_5) \Gamma (a_3+a_4)}{\Gamma (1-a_2+a_6) \Gamma (1-a_5+a_6)}\\\times\int_{-i \infty}^{+i\infty}\frac{ds}{2\pi i}\,\frac{\Gamma (-s) \Gamma (a_1+a_2+s) \Gamma (a_2+a_3+s) \Gamma (a_2+a_4+s) \Gamma (-a_2+a_5-s) \Gamma (a_6-a_5+s+1)}{\Gamma (a_2+a_6+s+1) \Gamma (a_1+a_2+a_3+a_4+s)}\,.
\end{multline}
Before evaluating the above Mellin integral it is first convenient to perform the sum over $k$ in \eqref{ksum}, which very nicely can be re-summed to the same type of Mellin integral but with different parameters:
\begin{multline}\label{phil}
    \phi_{\ell}(a_i)=\,\frac{\Gamma (a_1+a_2) \Gamma (a_1+a_3) \Gamma (a_1+a_4) \Gamma (a_2+a_5)\Gamma(a_3+a_4+\ell)}{\Gamma(a_1+a_2+\ell)\Gamma (1-a_2+a_6) \Gamma (1-a_5+a_6)}\\\times\int_{-i \infty}^{+i\infty}\frac{ds}{2\pi i}\,\frac{\Gamma (-s) \Gamma (a_1+a_5+s) \Gamma (a_3+a_5+s) \Gamma (a_4+a_5+s) \Gamma (1-a_2+a_6+s) \Gamma (a_2-a_5+\ell-s)}{\Gamma (1+a_5+a_6+s) \Gamma (a_1+a_3+a_4+a_5+\ell+s)}\,.
\end{multline}
We can evaluate this Mellin integral either by picking the residues on the positive real axis, which re-sum to a pair of ${}_4F_3$ 1-balanced hypergeometric functions, or directly in terms of a well poised $_7F_6$ hypergeometric function using eq. (\tcb{4.7.1.3}) of \cite{Slater}, in which case we obtain:
\begin{framed}
\paragraph{Result for the seed integral \eqref{Wintegral}:}
\begin{align}\label{SpectralRes}
    \phi_\ell(a_i)&=\Gamma (a_1+a_2) \Gamma (a_1+a_3) \Gamma (a_1+a_4) \Gamma (a_1+a_5) \Gamma (a_2+a_5) \Gamma (a_3+a_5) \Gamma (a_4+a_5) \\\nonumber&\times\frac{\Gamma (a_2+a_3+\ell) \Gamma (a_2+a_4+\ell) \Gamma (a_3+a_4+\ell) \Gamma (a_6-a_5+\ell+1)}{\Gamma (1-a_5+a_6)}\, \psi(a;b,c,d,e,f),\\\nonumber
    a&=a_1+a_2+a_3+a_4+2 a_5+\ell-1\,,\\\nonumber
    b&=a_1+a_5,\,\\\nonumber
    c&=a_2+a_5,\,\\\nonumber
    d&=a_3+a_5,\,\\\nonumber
    e&=a_4+a_5,\,\\\nonumber
    f&=a_1+a_2+a_3+a_4+a_5-a_6+\ell-1\,,
\end{align}
in terms of the $\psi$ function defined in \cite{Slater} (page 127):
\begin{multline}
    \psi(a;b,c,d,e,f)=\tfrac{\Gamma(a+1)}{\Gamma(1+a-b)\Gamma(1+a-c)\Gamma(1+a-d)\Gamma(1+a-e)\Gamma(1+a-f)\Gamma(2+2a-b-c-d-e-f)}\\\times\,{}_7F_6\left(\begin{matrix}
    a,& 1+\tfrac{a}2, &b, &c, &d,& e,& f\\
    &\tfrac{a}{2},&1+a-b,&1+a-c,&1+a-d,&1+a-e,&1+a-f
    \end{matrix};1\right).
\end{multline}
\end{framed}
In the mathematical literature, the $\psi$-function is also referred to as the Wilson function (see e.g. eq.~(\tcb{3.2}) of \cite{Groenevelt2003}), which provides an analytic continuation of Wilson polynomial for non-integer degree. The analytic continuation is obtained by considering the Wilson polynomials as Eigenfunctions of certain difference operators.
It is interesting to note that our result \eqref{SpectralRes} for the spectral integral \eqref{Wintegral} is well defined for all physical range of parameters, since:\footnote{Note that this property does not hold for the explicit form of the crossing kernels derived in \cite{Gopakumar:2018xqi} in terms of ${}_7F_{6}$ which, as the authors of the paper also noted, requires an analytic continuation to be applicable in general.} 
\begin{equation}\label{psiconv}
    2 + 2 a - b - c - d - e - f=1 - a_5 + a_6 + \ell > 0.
\end{equation}

In the following sections it will also prove convenient to employ the identity (\tcb{4}) in \S \tcb{4.4} of \cite{Bailey}, which allows to decompose a well poised ${}_7F_6$ into a sum of two $1$-balanced (Saalsch\"utzian) ${}_4F_3$. The general identity for the $\psi$-function reads:
\begin{align}
    \psi(a;b,c,d,e,f)&=\tfrac{\Gamma (a-d-e-f+1)}{\Gamma (a-b+1) \Gamma (a-c+1) \Gamma (a-d-e+1) \Gamma (a-d-f+1) \Gamma (a-e-f+1) \Gamma (2 a-b-c-d-e-f+2)}\\\nonumber
    &\times\,_4F_3\left(\begin{matrix}a-b-c+1,d,e,f\\a-b+1,a-c+1,-a+d+e+f\end{matrix};1\right)\\\nonumber
    &+\tfrac{\Gamma (-a+d+e+f-1)}{\Gamma (d) \Gamma (e) \Gamma (f) \Gamma (a-b-c+1) \Gamma (2 a-b-d-e-f+2) \Gamma (2 a-c-d-e-f+2)}\\
    &\times\, _4F_3\left(\begin{matrix}a-d-e+1,a-d-f+1,a-e-f+1,2 a-b-c-d-e-f+2\\a-d-e-f+2,2 a-b-d-e-f+2,2 a-c-d-e-f+2\end{matrix};1\right),\nonumber
\end{align}
which is valid if the $\psi$-function is convergent, which is when: $\text{Re}(2+2a-b-c-d-e-f)>0$. It is interesting to notice how the permutation symmetry in the parameters $b$, $c$, $d$, $e$ and $f$ is manifest on the left hand side of the equation, but highly non-trivial on the right hand side. This allows to obtain various different transformations of the corresponding ${}_4F_3$ hypergeometric functions which yield the same $\psi$-function.\footnote{Note that there are various such decompositions and not all of them are manifestly analytic in spin! Analyticity in spin is however manifest in the $\psi$-function.}

Another convenient representation of the $\psi$-function was given in \cite{Groenevelt2003} eq. \tcb{(6.5)}.\footnote{See also \cite{Gopakumar:2018xqi} where a similar representation was used to perform the $\epsilon$-expansion.} This representation after changing variables and using a transformation for the ${}_2F_1$ hypergeometric function reads:
\begin{align}\label{2F1rep}
    \psi(a;b,c,d,e,f)&=\tfrac{1}{\Gamma (f) \Gamma (a-f+1)^2 \Gamma (a-b-c+1) \Gamma (a-d-e+1) \Gamma (2 a-b-c-d-e-f+2)}\\\nonumber&\times\,\int_0^1\, dy \, y^{a-f}(1-y)^{2 a-b-c-d-e-f+1}\\\nonumber&\times\, _2F_1\left(\begin{matrix}a-b-f+1,a-c-f+1\\a-f+1\end{matrix};y\right) \, _2F_1\left(\begin{matrix}a-d-f+1,a-e-f+1\\a-f+1\end{matrix};y\right)\,.
\end{align}

With the result \eqref{SpectralRes} for the seed integral \eqref{Wintegral} at hand, one can now evaluate the spectral integrals of the type \eqref{anomdint} for any crossing kernel in \cite{Sleight:2018epi} by simply expanding the crossing kernel in terms of Wilson polynomials as outlined in appendix \S \ref{CrossingToWilson}. This is a sometimes cumbersome but straightforward procedure. In many cases this can be done explicitly or implemented with a computer algebra program. Therefore, the results presented in this section solve the crossing problem up to finite spin for generic operator exchanges, reducing the crossing problem to simply determining the form the of the crossing kernel of interest which, if it is not already known, can be worked out following the method of \cite{Sleight:2018epi}.

\subsection{Exchanged scalar operators}\label{ExchScalar}

In this section we give the result for the spectral integral \eqref{Wintegral} for a scalar of twist $\tau^\prime$ exchanged in the crossed channel 
\begin{equation}\label{int2lprimezero}
    {}^{({\sf t})}I_{\tau^\prime,0|\ell}\left(t\right) = n_{\tau^\prime,0} \int^{\infty}_{-\infty} \frac{d\nu}{2\pi}\,a_{0}(\nu)\, {}^{({\sf t})}\widehat{\mathfrak{I}}_{\frac{d}{2}+i\nu,0|\ell}\left(t\right),
\end{equation}
where the normalisation $n_{\tau^\prime,0}$ reads
\begin{align}
    n_{\tau^\prime,0}&=\frac{\pi ^{-\frac{d}{2}} \Gamma (\tau^\prime) \Gamma \left(\tau^\prime+1-\frac{d}{2}\right)}{\Gamma \left(\frac{\tau^\prime}{2}\right)^4 \Gamma \left(\tfrac{2\Delta-\tau^\prime}{2}\right)^2 \Gamma \left(\tfrac{2\Delta +\tau^\prime-d}{2}\right)^2}\,,\label{ccoeff}
\end{align}
and the spectral function $a_{0}(\nu)$ is given by \eqref{anuwitten} with $\ell^\prime=0$, $J_i=0$ and $\tau_i=\Delta$.

The expression for the crossing kernel in terms of Wilson polynomials can in this case simply be read off from equation \eqref{simpleinv}
\begin{equation}
    {}^{({\sf t})}\widehat{\mathfrak{I}}_{\frac{d}{2}+i\nu,0|\ell}\left(t\right)=a_{0,\ell}^{(0)}\,\beta(t)\,\mathcal{W}_{\ell}(\nu^2;a_1,a_2,a_3,a_4)\,,
\end{equation}
where 
\begin{equation}
    \beta(t)= \frac{\pi ^{d/2}2^{2 \Delta -t}  (-1)^\ell \Gamma (\Delta )^2 \Gamma \left(\ell+\Delta -\frac{1}{2}\right) \Gamma \left(\ell+\frac{t}{2}\right) \Gamma (\ell+t-1)}{\Gamma \left(\frac{t}{2}\right)^2 \Gamma (\ell+\Delta ) \Gamma (\ell+2 \Delta -1) \Gamma \left(\ell+\frac{t}{2}-\frac{1}{2}\right) \Gamma \left(\frac{d}{2}+t-2 \Delta \right)}\,,
\end{equation}
and
\begin{align}
    a_1&= \frac{d}{4}-\Delta +\frac{t}{2},& a_2&= \frac{d}{4}-\Delta +\frac{t}{2},& a_3&= \Delta -\frac{d}{4},& a_4&= \Delta -\frac{d}{4}\,.
\end{align}
The evaluation of \eqref{int2lprimezero} is then a simple application of the result \eqref{SpectralRes}, which gives 
\begin{empheq}[box=\fbox]{align}
 \label{ScalarAnalyticT}
    {}^{({\sf t})}I_{\tau^\prime,0|\ell}\left(t\right)&=-a_{0,\ell}^{(0)}\,\tfrac{(-1)^\ell 2^{2 \Delta -t}\, \Gamma (\Delta )^2  \Gamma (\tau^\prime) \Gamma \left(-\frac{d}{2}+\tau^\prime+1\right) \Gamma \left(\ell+\Delta -\frac{1}{2}\right) \Gamma \left(\ell+\frac{t}{2}\right) \Gamma (\ell+t-1)}{\Gamma \left(\frac{t}{2}\right)^2 \Gamma \left(\frac{\tau^\prime}{2}\right)^4 \Gamma \left(\Delta -\frac{\tau^\prime}{2}\right)^2 \Gamma (\ell+\Delta ) \Gamma (\ell+2 \Delta -1) \Gamma \left(\ell+\frac{t}{2}-\frac{1}{2}\right) \Gamma \left(-\frac{d}{2}+\Delta +\frac{\tau^\prime}{2}\right)^2 \Gamma \left(\frac{d}{2}+t-2 \Delta \right)}\\& \hspace*{0.5cm}\times\,\phi_\ell\left(\tfrac{d}{4}-\Delta+\tfrac{t}{2},\tfrac{d}{4}-\Delta+\tfrac{t}{2},\Delta -\tfrac{d}{4},\Delta -\tfrac{d}{4},\tfrac{2 \tau^\prime-d}{4},\tfrac{2\tau^\prime-d}{4}\right).\nonumber
\end{empheq}
The above result admits a straightforward generalisation to external operators of arbitrary twist upon re-instating the dependence on the external twists $\tau_i$ in the pre-factor and in the coefficients $a_i$, as shown in \S \ref{spinninWilson}.

Inserting $t=2\Delta$, this gives the following result for the anomalous dimension \eqref{anomdint} of leading double-twist operators $\left[{\cal O}{\cal O}\right]_{0,\ell}$ of spin $\ell$, where $\Delta$ is the scaling dimension of the scalar operator ${\cal O}$:
\begin{empheq}[box=\fbox]{equation}
\label{ScalarAnalytic}
    \gamma_{0,\ell}=-\frac{2\,\Gamma (\tau^\prime) \Gamma \left(\tau^\prime+1-\frac{d}{2}\right)}{\Gamma(\tfrac{d}{2}) \Gamma \left(\frac{\tau^\prime}{2}\right)^4 \Gamma \left(\tfrac{2\Delta-\tau^\prime}{2}\right)^2 \Gamma \left(\tfrac{2 \Delta +\tau^\prime-d}{2}\right)^2}\,\phi_\ell\left(\frac{d}{4},\frac{d}{4},\Delta -\frac{d}{4},\Delta -\frac{d}{4},\frac{2\tau^\prime-d}{4},\frac{2\tau^\prime-d}{4}\right).
\end{empheq}
We stress that analyticity in spin is manifest just from the definition of the function $\phi_\ell$. For $d=4$ this result coincides with the result given in \cite{Liu:2018jhs} which was obtained using a different approach specific to the $d=4$ case. The above result straightforwardly generalises to the case of unequal external operators using eq.~(\tcb{4.45}) of \cite{Sleight:2018epi}.

\subsection{Exchanged spinning operators}\label{ExchSpin}

In this section we extend the results of the previous section to the spectral integral of crossing kernels for CPWs with exchanged spin-$\ell^\prime$ in the crossed channel. We take $t=2\Delta$, appropriate for the corrections to the OPE data of leading double-twist operators. The result for the spectral integral thus gives the analytic in spin anomalous dimensions \eqref{anomdint} of leading twist double-trace operators $\left[{\cal O}{\cal O}\right]_{0,\ell}$ induced by the exchange of spin-$\ell^\prime$ operator of twist $\tau^\prime$ in the crossed channel. 

There is more than one way to perform the spectral integral \eqref{int2} for $\ell^\prime\ne0$, leading to different transformation formulas for the final result. In all cases the result can be expressed as a sum of Wilson functions. All of them differ by how we fix the spectral weight and from the decomposition of the crossing kernels in terms of the corresponding orthogonal Wilson polynomials. 

Following the method outlined in \S \ref{EvaluationSp}, to evaluate the spectral integral we need only decompose crossing kernel in terms of Wilson polynomials. We derive the following decomposition in appendix \ref{CrossingToWilson}:
\begin{equation}
    {}^{({\sf t})}\widehat{\mathfrak{I}}_{\frac{d}{2}+i\nu,\ell^\prime|\ell}\left(t=2\Delta\right)=a_{0,\ell}^{(0)}\sum_{j=0}^{2\ell^\prime}\beta_{\ell,j}^{(\ell^\prime)}\,\mathcal{W}_{\ell-j}(\nu^2;a_1,a_2,a_3,a_4)\,.
\end{equation}
with
\begin{align}\label{aiSpinning}
    a_1&= \frac{d+2\ell^\prime}{4},& a_2&= \frac{d+2\ell^\prime}{4},&a_3&= \Delta -\frac{d-2\ell^\prime}{4},&a_4= \Delta -\frac{d-2\ell^\prime}{4}\,.
\end{align}

Using the result for the seed integral \eqref{SpectralRes}, this decomposition of the crossing kernel immediately gives the following contribution to the anomalous dimension of leading twist double-trace operators $\left[{\cal O}{\cal O}\right]_{0,\ell}$ of spin-$\ell$  
\begin{empheq}[box=\fbox]{equation}
    \gamma_{0,\ell}=2\,n_{\tau^\prime,\ell^\prime}\sum_{j=0}^{2\ell^\prime}\beta_{\ell,j}^{(\ell^\prime)}\,\phi_{\ell-j}\left(\tfrac{d+2\ell^\prime}{4}, \tfrac{d+2\ell^\prime}{4}, \Delta -\tfrac{d-2\ell^\prime}{4},\Delta -\tfrac{d-2\ell^\prime}{4},\tfrac{\tau^\prime+\ell^\prime-2d}2,\tfrac{\tau^\prime+\ell^\prime-2d}2\right)\,.
\end{empheq}
The normalisation $n_{\tau^\prime,\ell^\prime}$ reads in this case: 
\begin{align}\label{normgenellprime}
    n_{\tau^\prime,\ell^\prime}&=\tfrac{2^{4 \ell^\prime+2\tau^\prime-4}\pi ^{-\frac{d}{2}-1} (2 \ell^\prime+\tau^\prime-1) \Gamma (d-\ell^\prime-\tau^\prime -1) \Gamma \left(\ell^\prime+\tfrac{\tau^\prime}{2}-\tfrac{1}{2}\right)^2 \Gamma \left(-\tfrac{d}{2}+\ell^\prime+\tau^\prime+1\right)}{\Gamma (2\ell^\prime+\tau^\prime -1)\Gamma (d-\tau^\prime -1)\Gamma \left(\Delta -\tfrac{\tau^\prime}{2}\right)^2  \Gamma \left(\ell^\prime+\frac{\tau^\prime}{2}\right)^2  \Gamma \left(-\frac{d}{2}+\ell^\prime+\Delta+\frac{\tau^\prime}{2}\right)^2}\,,
\end{align}

In $d=4$ this expression matches the result obtained in \cite{Liu:2018jhs} which uses a different approach tailored to the $d=4$ case. 

It might be useful to note that extra care has to be taken when using the above expression for $\ell<2\ell^\prime$ due to some zeros in the coefficients $\beta_{\ell,j}$ which are compensated by singularities in $\phi_{\ell-j<0}$. This form of the result was introduced in order to be able to write down the final result for arbitrary $\ell$ in its simplest form. For such values of $\ell^\prime$ one can use the representation \eqref{2F1rep} and integrate by parts in $y$ so as to remove singularities at $y\sim1$ of the type $\frac1{(1-y)^\#}$ which arise for $\mathbb{N}\ni2a-b-c-d-e-f+1<0$. As anticipated, the price to pay is a much more cumbersome expression for arbitrary $\ell$, but it is anyway straightforward to obtain by adding a finite number of boundary terms in the $y$ integral \eqref{2F1rep} which only contribute for $\ell<2\ell^\prime$.

\subsection{Subleading double-twist operators}\label{ExchSub}

Since the result \eqref{ScalarAnalytic} for the spectral integral holds for arbitrary $t$, naively one might expect that setting $t=2\Delta+2n$ would give the anomalous dimension induced for subleading double-twist operators $\left[{\cal O}{\cal O}\right]_{n,\ell}$ by a scalar of twist $\tau^\prime$ in the crossed channel, just as for the leading twist case \eqref{anomdint}. I.e.:
\begin{equation}\label{anomdintsub}
      \frac{\gamma_{n,\ell}}{2}a^{(0)}_{n,\ell} \; \overset{?}{=} \; \int^{ \infty}_{-\infty} \frac{d\nu}{2\pi}\,a_{0}(\nu)\, {}^{({\sf t})}\mathfrak{I}_{\frac{d}{2}+i\nu,0|\ell}\left(t=2\Delta+2n\right).
\end{equation}
The above expression is however not quite correct, since the crossing kernel ${}^{({\sf t})}\mathfrak{I}_{\frac{d}{2}+i\nu,0|\ell}\left(t=2\Delta+2n\right)$ also contains contributions to conformal multiplets of double-twist operators $\left[{\cal O}{\cal O}\right]_{n^\prime,\ell}$ of lower twist $n^\prime < n$. To extract the anomalous dimensions $\gamma_{n,\ell}$ we first have to subtract the contributions to all double-twist operators $\left[{\cal O}{\cal O}\right]_{n^\prime,\ell}$ with $n^\prime < n$. A general method to do this was given in \cite{Sleight:2018epi}, where it was also carried out explicitly for the specific crossing kernel we are considering here. The projected crossing kernel reads (see \cite{Sleight:2018epi}, section 5.1.2):\footnote{For ease of presentation we focus on the case with equal external operators of twists $\tau_i=\Delta$. The general case with generic external twists $\tau_i$ follows in exactly the same way starting from the more general crossing kernels studied in \cite{Sleight:2018epi}.}
\begin{align}\label{gamma0000nl}
    \alpha_n( 2\Delta+2n) {}^{({\sf t})}\mathfrak{I}^{\text{projected}}_{\frac{d}{2}+i\nu,0|\ell}\left(t=2\Delta+2n\right) = \frac{2\, \Gamma (\frac{d}{2}+i\nu)}{\Gamma \left(\frac{d}{4}+\frac{i\nu}{2}\right)^4 \Gamma \left(-i\nu\right)}\sum_{j=0}^n\,D_{j}T^n_{n-j,j},
\end{align}
where
\begin{equation}\label{alphaCoeff}
    \alpha_n(x)=(-2)^{3 n}\, n! \left(\tfrac{d}{2}+\ell\right)_n (d-2 n-x )_n \left(-\tfrac{d}{2}+\ell+n+x \right)_n\,,
\end{equation}
and 
\begin{multline}
    T^n_{ij}=\,\int_{-i\infty}^{i\infty} \frac{ds}{4\pi i}\,\Gamma(-\tfrac{s}{2})^2\Gamma\left(\tfrac{s+\frac{d}{2}-i\nu}{2}+i\right)\Gamma\left(\tfrac{s+\frac{d}{2}-i\nu}2+j\right)\,Q_{\ell}^{2\Delta+2n,2\Delta+2n,0,0}(s)\\=\underbrace{\frac{2^{\ell} \Gamma \left(\tfrac{2 j+\frac{d}{2}+i\nu}{2}\right)^2 \Gamma \left(\tfrac{2 i+\frac{d}{2}-i\nu }{2}\right)^2\left(\tfrac{2 \Delta +2n}{2}\right)_{\ell}^2}{(\ell+2 \Delta+2n -1)_{\ell} \Gamma \left(\tfrac{d+2 i+2 j}{2} \right)}}_{t_{ij}^{n}}\, {}_4F_3\left(\begin{matrix}-\ell,2 \Delta +2n +\ell-1,i+\frac{d}{4}-\frac{i\nu}{2},j+\frac{d}{4}+\frac{i\nu}{2}\\\frac{d}{2}+i+j,\Delta +n,\Delta +n\end{matrix};1\right)\,.
\end{multline}
We give the explicit form of the coefficients $D_j$ up to $n=3$ in appendix \ref{Dcoeffs}. All such coefficients can be systematically worked out by solving a linear system for any $n$ \cite{Sleight:2018epi}. 

We can now write down a spectral integral of the projected crossing kernel \eqref{gamma0000nl} that gives the anomalous dimensions $\gamma_{n,\ell}$: 
\begin{equation}\label{anomdintsub2}
      \frac{\gamma_{n,\ell}}{2}a^{(0)}_{n,\ell}=n_{\Delta^\prime,0}\int^{\infty}_{-\infty} \frac{d\nu}{2\pi}\,a_{0}(\nu)\, {}^{({\sf t})}\widehat{\mathfrak{I}}^{\text{projected}}_{\frac{d}{2}+i\nu,0|\ell}\left(t=2\Delta+2n\right),
\end{equation}
where the normalisation $n_{\Delta^\prime,0}$ is given by \eqref{ccoeff} as before. We evaluate this integral in the same way as before by decomposing the projected crossing kernel in terms of Wilson polynomials. The latter decomposition takes the form 
\begin{equation}\label{projecteddecomp}
    {}^{({\sf t})}\widehat{\mathfrak{I}}^{\text{projected}}_{\frac{d}{2}+i\nu,0|\ell}\left(t=2\Delta+2n\right) = \sum_{i=0}^{2n}\beta_i^{(\ell,n)}\mathcal{W}_{\ell+i}(\nu^2;a_i).
\end{equation}
We have not yet been able to obtain a general closed form expression for all coefficients $\beta_i^{(\ell,n)}$ though they satisfy the following property:
\begin{equation}\label{SumBeta}
    \sum_{i=0}^{2n}\beta_i^{(\ell,n)}=\frac{(-1)^\ell\,\pi^{d/2}}{\Gamma(\tfrac{d}{2})}\,,
\end{equation}
and we have managed to obtained a closed form expression for the following values of $i$ in the sum \eqref{SumBeta}:
\begin{subequations}\label{betas}
\begin{align}
    \beta_0^{(\ell,n)}&=\frac{\pi^{d/2}}{\Gamma(\tfrac{d}{2})}\,\frac{(-1)^{\ell+n} (-d+n+2 \Delta +1)_n \left(\tfrac{d+2 \ell}{2}\right)_n \left(\tfrac{d-2 n-4 \Delta +2}{2}\right)_n (\ell+\Delta )_n}{\left(-\tfrac{d-2 \Delta -2}{2}\right)_n{}^2 4^n n! \left(\tfrac{2 \ell+2 \Delta +1}{2}\right)_n}\,,\\
    \beta_1^{(\ell,n)}&=-\frac{n (2 \Delta +2 \ell+1) (d-4 \Delta -2 \ell-2 n)}{(\Delta +\ell) (d-4 \Delta -2 \ell)}\beta_0^{(\ell,n)}\,,\\
    \beta_{2n}^{(\ell,n)}&=\frac{\pi^{d/2}}{\Gamma(\tfrac{d}{2})}\,\frac{(-1)^\ell (-d+n+2 \Delta +1)_n \left(\tfrac{d+2 \ell+2 n}{2}\right)_n (\ell+n+\Delta )_n \left(-\tfrac{d-2 \ell-2 n-4 \Delta}{2}\right)_n}{\left(-\tfrac{d-2 \Delta -2}{2}\right)_n{}^2 4^n n! \left(\tfrac{2 \ell +2 n+2 \Delta -1}{2}\right)_n}\,,\\
    \beta_{2n-1}^{(\ell,n)}&=-\frac{n (d+2 \ell+2 n-2) (2 \Delta +2 \ell+4 n-3)}{(d+2 \ell+4 n-2) (\Delta +\ell+2 n-1)}\,\beta_{2n}^{(\ell,n)}\,.
\end{align}
\end{subequations}

With the expansion \eqref{projecteddecomp} of the projected crossing kernel in terms of Wilson polynomials, as before the result for the spectral integral \eqref{anomdintsub2} can be immediately written down using the result \eqref{Wintegral} for the seed spectral integrals. This gives the following expression for the anomalous dimensions induced by a scalar of twist $\tau^\prime$ in the crossed channel: 
\begin{empheq}[box=\fbox]{equation}
\label{N>0}
    \gamma_{n,\ell}=2\,n_{\tau^\prime,0}\sum_{i=0}^{2n}\beta_i^{(\ell,n)}\phi_{\ell+i}\left(\frac{d}{4},\frac{d}{4},\Delta -\frac{d}{4},\Delta -\frac{d}{4},\frac{2\tau^\prime-d}{4},\frac{2\tau^\prime-d}{4}\right)\,.
\end{empheq}

The closed form expressions \eqref{betas} for the coefficients $\beta_i^{(\ell,n)}$, together with \eqref{SumBeta} give a relatively simple form for anomalous dimensions \eqref{N>0} up to $n=2$. For $n>2$ the coefficients $\beta_i^{(\ell,n)}$ do not factorise and we have not yet managed to obtain a closed form expression for them.

\subsection{Spinning external operators}\label{spinninWilson}

The decomposition of the crossing kernels obtained in \cite{Sleight:2018epi} in terms of Wilson polynomials orthogonal with respect to the spectral measure is very general and can be seamlessly applied also to the cases with spinning external operators. In this way the spectral integral of crossing kernels with spinning external legs can also be evaluated using the seed spectral integrals in \S \ref{EvaluationSp}.

In this section we shall consider spinning crossing kernels of CPWs for four-point correlators involving two spinning operators of spins $J_1$ and $J_2$,
\begin{equation}\label{0O1OOO2}
    \langle {\cal O}_{J_1}\left(x_1\right){\cal O}_{J_2}\left(x_2\right){\cal O}_3\left(x_3\right){\cal O}_4\left(x_4\right) \rangle.
\end{equation}
For simplicity we shall consider exchanged scalars of twist $\tau^\prime$ in the ${\sf t}$-channel ($\ell^\prime=0$) and their contribution to double-trace operators of leading twist in the ${\sf s}$-channel. In addition to $\left[{\cal O}_3{\cal O}_4\right]$ composed of scalar operators ${\cal O}_3$ and ${\cal O}_4$ of twists $\tau_3$ and $\tau_4$, in the ${\sf s}$-channel there includes contributions to double-twist operators of the form $\left[{\cal O}_{J_1}{\cal O}_{J_2}\right]_\ell$ involving operators ${\cal O}_{J_1}$ and ${\cal O}_{J_2}$ of spins $J_1$ and $J_2$, and twists $\tau_1$ and $\tau_2$, respectively. The relevant crossing kernels are reviewed in \S \ref{subsec::cccbextspin} which, as was already noted in \cite{Sleight:2018epi}, are proportional to a single Wilson polynomial:\footnote{Recall that, since we are considering contributions to leading twist operators in the ${\sf s}$-channel we can fix $t=\tau_1+\tau_2$.}
\begin{equation}\label{0spinningdecomp}
    {}^{(\sf{t})}\widehat{\mathfrak{J}}_{\tfrac{d}2+i\nu,0|\ell}(t=\tau_1+\tau_2)=\beta_{\ell}\,\mathcal{W}_{\ell}(\nu^2;a_i)\,,
\end{equation}
with\footnote{Contrary to the previous cases we included the mean field theory OPE in the definition of $\beta_l$.}
\begin{equation}
    \beta_\ell=\tfrac{\pi ^{d/2} (-1)^\ell 2^{-J_1-J_2+\ell} \Gamma (J_1-J_2+\ell+\tau_1) \Gamma (J_1+J_2+\ell+\tau_1+\tau_2-1) \Gamma \left(\ell+\frac{\tau_1}{2}+\frac{\tau_2}{2}+\frac{\tau_3}{2}-\frac{\tau_4}{2}\right)}{\Gamma (2 J_1+\tau_1) \Gamma (-J_1-J_2+\ell+1) \Gamma (2 \ell+\tau_1+\tau_2-1) \Gamma \left(J_1+J_2+\frac{\tau_1}{2}+\frac{\tau_2}{2}+\frac{\tau_3}{2}-\frac{\tau_4}{2}\right) \Gamma \left(\frac{d}{2}+J_1+J_2+\frac{\tau_1}{2}+\frac{\tau_2}{2}-\frac{\tau_3}{2}-\frac{\tau_4}{2}\right)}\,,
\end{equation}
and 
\begin{align}
    a_1&=\frac{1}{4} (d+4 J_1+2\tau_1-2\tau_4)\,,& a_2&=\frac{1}{4} (d+4 J_2+2\tau_2-2\tau_3)\,,\\
    a_3&=\frac{1}{4} (-d+4 J_1+2\tau_1+2\tau_4)\,,&
    a_4&=\frac{1}{4} (-d+4 J_2+2\tau_2+2\tau_3)\,.
\end{align}
For brevity, here we only present explicitly the $\sf{t}$-channel crossing kernel. The $\sf{u}$-channel kernels, which are given in \S \ref{subsec::cccbextspin}, will differ from the $\sf{t}$-channel kernels by a sign which only contributes when $J_1+J_2+\ell$ is odd. 

For the spectral integral \eqref{int2} of the crossing kernel, the spectral function \eqref{anu} is given by \eqref{anuwitten} with $J_3=J_4=0$ and $\ell^\prime=0$. Using the result \eqref{Wintegral} for the seed integral, the result for the spectral integral is then immediately given by \begin{empheq}[box=\fbox]{align}\label{spinninspecint}
    {}^{(\sf{t})}I_{\tau^\prime,0|\ell}(t=\tau_1+\tau_2)&=n_{\tau^\prime,0}\,\beta_{\ell}\,\phi_{\ell}\left(a_1,a_2,a_3,a_4,\frac{2\tau^\prime-d}{4},\frac{2\tau^\prime-d}{4}\right)\,,
\end{empheq}
where the normalisation $n_{\tau^\prime,0}$ for general $J_1$ and $J_2$ reads 
\begin{multline}\label{normalj1j2}
    n_{\tau^\prime,0}=\tfrac{ \pi^{-d/2}\Gamma(\tau^\prime)\Gamma \left(-\frac{d}{2}+\tau^\prime+1\right)}{\Gamma \left(\tfrac{\tau^\prime-\tau_1+\tau_4}{2}\right) \Gamma \left(\tfrac{\tau^\prime-\tau_2+\tau_3}{2}\right) \Gamma \left(\tfrac{2 J_1+\tau^\prime+\tau_1-\tau_4}{2}\right) \Gamma \left(\tfrac{2 J_1-\tau^\prime+\tau_1+\tau_4}{2}\right) \Gamma \left(\tfrac{2 J_2+\tau^\prime+\tau_2-\tau_3}{2}\right) \Gamma \left(\tfrac{2 J_2-\tau^\prime+\tau_2+\tau_3}{2}\right) }\\\times\tfrac{1}{\Gamma \left(\tfrac{-d+2 J_1+\tau^\prime+\tau_1+\tau_4}{2}\right) \Gamma \left(\tfrac{-d+2 J_2+\tau^\prime+\tau_2+\tau_3}{2}\right)}\,.
\end{multline}
A similar expression also holds for the crossing kernel of the $\sf{u}$-channel CPWs \eqref{CKj1j2u}.

\paragraph{Double-trace anomalous dimensions} Let us now discuss the relation of the result \eqref{spinninspecint} to the anomalous dimensions of double-twist operators. In the following we consider the case of equal external twists $\tau_i=\Delta$.\footnote{Note that anomalous dimensions are only generated for $\tau_1+\tau_2=\tau_3+\tau_4$.} 

For general $J_1\neq J_2$, there is no mean field theory part to correlators of the type \eqref{0O1OOO2}. In such a case, the only way to generate corrections to double-trace anomalous dimensions  at (i.e. $\log u$ terms) is to have an ${\cal O}\left(1\right)$ mixing between degenerate double twist operators:
\begin{equation}\label{0mix}
    \Sigma_i^{(\ell)}=\sum_{J_1,J_2}b_{[\mathcal{O}_{J_1}\mathcal{O}_{J_2}]_{\ell},\Sigma_i^{(\ell)}}\,[\mathcal{O}_{J_1}\mathcal{O}_{J_2}]_\ell\,,
\end{equation}
where the coefficients $b_{[\mathcal{O}_{J_1}\mathcal{O}_{J_2}]_{\ell},\Sigma_i^{(\ell)}}$ express the Eigenfunction of the dilatation operator in terms of the standard free theory double-twist operators.
In this way, the mean field theory OPE coefficients $c^{(0)}_{\mathcal{O}_{J_1}\mathcal{O}_{J_2}\Sigma^{(\ell)}_i}$ and $c^{(0)}_{\mathcal{O}\mathcal{O}\Sigma^{(\ell)}_i}$ can both give a non-trivial contribution to the four-point function whenever the Eigenvectors $\Sigma^{(\ell)}_i$ of the dilatation operator are a linear combination of double-twist operators. We denote by $\gamma_{0,\ell}^{(i)}$ the corresponding Eigenvalues at this order. 

Considering small corrections to the mean field theory values, we arrive to: 
\begin{align}
    c^{(0)}_{\mathcal{O}_{J_1}\mathcal{O}_{J_2}[\mathcal{O}_{J_1}\mathcal{O}_{J_2}]_{(\ell)}}c^{(0)}_{\mathcal{O}\mathcal{O}[\mathcal{O}\mathcal{O}]_{(\ell)}}\gamma_{0,\ell}&=\sum_{i}c^{(0)}_{\mathcal{O}_{J_1}\mathcal{O}_{J_2}\Sigma_i^{(\ell)}}c^{(0)}_{\mathcal{O}\mathcal{O}\Sigma_i^{(\ell)}}\,\gamma_{0,\ell}^{(i)}\\\nonumber&\equiv c^{(0)}_{\mathcal{O}_{J_1}\mathcal{O}_{J_2}[\mathcal{O}_{J_1}\mathcal{O}_{J_2}]_{\ell}}c^{(0)}_{\mathcal{O}\mathcal{O}[\mathcal{O}\mathcal{O}]_{\ell}}\sum_{i}b_{[\mathcal{O}_{J_1}\mathcal{O}_{J_2}]_\ell,\Sigma_i^{(\ell)}}\,b_{[\mathcal{O}\mathcal{O}],\Sigma_i^{(\ell)}}\gamma_{0,\ell}^{(i)}\,,
\end{align}
where we have used \eqref{0mix} together with the mean-field theory result to factor out the mean-field theory OPE coefficients. In the end the average $\gamma_{0,\ell}$ is weighted by the coefficients $b_{[\mathcal{O}_{J_1}\mathcal{O}_{J_2}]_{\ell},\Sigma_i^{(\ell)}}$ only. Finally, simplifying the mean-field theory OPE, we recover the following average of the anomalous dimensions $\gamma_{0,\ell}^{(i)}$ of the double-trace operators $\Sigma_i$:
\begin{equation}\label{0avgamma}
    \gamma_{0,\ell}=\sum_{i}b_{[\mathcal{O}_{J_1}\mathcal{O}_{J_2}]_\ell,\Sigma_i^{(\ell)}}\,b_{[\mathcal{O}\mathcal{O}],\Sigma_i^{(\ell)}}\gamma_{0,\ell}^{(i)}\,,
\end{equation}
which we can extract from the result \eqref{spinninspecint} for the spectral integral of the crossing kernel \eqref{0spinningdecomp}. In particular:
\begin{equation}\label{DTANOMspin}
   \frac{1}{2}  c^{(0)}_{\mathcal{O}_{J_1}\mathcal{O}_{J_2}[\mathcal{O}_{J_1}\mathcal{O}_{J_2}]_{(\ell)}}c^{(0)}_{\mathcal{O}\mathcal{O}[\mathcal{O}\mathcal{O}]_{(\ell)}}\gamma_{0,\ell}= {}^{(\sf{t})}I_{\tau^\prime,0|\ell}(t=2\Delta).
\end{equation}
The mean field theory OPE coefficients $ c^{(0)}_{\mathcal{O}\mathcal{O}[\mathcal{O}\mathcal{O}]_{(\ell)}}$ are given by setting $n=0$ in equation \eqref{OPEnl}. On the other hand, the coefficients $c^{(0)}_{\mathcal{O}_{J_1}\mathcal{O}_{J_2}[\mathcal{O}_{J_1}\mathcal{O}_{J_2}]_{(\ell)}}$ so far are only known explicitly for $J_2=0$ or $J_1=0$. Setting $J_1=J$ and $J_2=0$, we have \cite{Sleight:2018epi}
\begin{equation}\label{0spinningMean}
    \left(c^{(0)}_{\mathcal{O}_J\mathcal{O}[\mathcal{O}_J\mathcal{O}]_{\ell}}\right)^2=\frac{2^{\ell-J} (2 J+\tau_1)_{\ell-J} (\tau_2)_{\ell-J}}{(\ell-J)! (\ell+J+\tau_1+\tau_2-1)_{\ell-J}}\,,
\end{equation}
where we recall that above we are considering the case $\tau_1=\tau_2=\Delta$.

Disentangling the degeneracy in \eqref{0avgamma} to obtain the anomalous dimensions $\gamma_{0,\ell}^{(i)}$ is a difficult problem in general, which we don't attempt to solve here.\footnote{In order to fully solve this problem we would need to apply the techniques of \cite{Sleight:2017fpc,Sleight:2018epi} to obtain full crossing kernels for more general spinning correlators, which we postpone for now.}

\section{Comparison with large spin double-twist anomalous dimensions}
\label{sec::dtanom}

In this section we consider a different way to obtain double-trace anomalous dimensions by projecting away the shadow contribution directly from the crossing kernel at large spin. This approach was outlined at the end of \S \ref{subsec::approach} and does not involve evaluating spectral integrals. We then compare with the results obtained in the previous section.

\subsection{Exchanged scalar operators}
\label{subsec::exchscalarop}

We begin starting with the simplest case of scalar external operators of equal scaling dimension $\Delta$, with a scalar operator of twist $\tau^\prime$ exchanged in the ${\sf t}$-channel (i.e. $\ell^\prime=0$ in \eqref{ttosblock}). In this section we restrict ourselves to the anomalous dimensions of leading double-twist operators ($n=0$), which implies $t=2\Delta$ in the corresponding CPW crossing kernel \eqref{simpleinv}. 

Before projecting away the shadow contribution, the anomalous dimension induced by the crossing kernel is given by (for more details see section 5 of \cite{Sleight:2018epi})\footnote{This is just the analogue of equation \eqref{anomdtcb} but for the CPW \eqref{cpwdefint} with ${\tilde \Delta}=\tau^\prime$ instead of the conformal block \eqref{A0B0}.}
\begin{align}\label{cpwanom0ell}
  a^{(0)}_{0,\ell} \frac{\gamma^{\text{CPW}}_{0,\ell}}{2}&=a_{\tau^\prime,0}\,\mathfrak{I}^{({\sf t})}_{\tau^\prime,0|\ell}\left(t=2\Delta\right)\\
   &= a^{(0)}_{0,\ell} \frac{\gamma^{\text{CPW}}_{0,0}}{2} \, _4F_3\left(\begin{matrix}-\ell,\ell+2 \Delta -1,\frac{d-\tau^\prime}{2},\frac{\tau^\prime }{2}\\\frac{d}{2},\Delta ,\Delta \end{matrix};1\right)\,,
\end{align}
where
\begin{equation}
    \gamma^{\text{CPW}}_{0,0}=\frac{2\,\Gamma (\tau^\prime) \Gamma \left(\frac{d-\tau^\prime}{2}\right)^2}{\Gamma \left(\frac{d}{2}\right) \Gamma \left(\frac{\tau^\prime}{2}\right)^2 \Gamma \left(\frac{d}{2}-\tau^\prime\right)}\,a_{\tau^\prime,0},
\end{equation}
proportional to the OPE coefficient $a_{\tau^\prime,0}$ in the ${\sf t}$-channel. 

To obtain the double-twist operator anomalous dimensions $\gamma_{0,\ell}$ induced by the physical conformal block (i.e. equation \eqref{anomdtcb}), following the prescription outlined at the end of \S \ref{subsec::approach} we project away the contributions from the shadow conformal multiplet in \eqref{cpwanom0ell} by employing the Mellin representation \eqref{mellinhyper} of the ${}_4F_3$ hypergeometric function and closing the Mellin contour on the physical block poles. This gives
\begin{multline}
    \gamma_{0,\ell}=-\gamma^{\text{CPW}}_{0,0}\,\frac{\Gamma (\Delta )^2 \Gamma (\ell+1) \Gamma (\tau^\prime) \Gamma \left(\ell+2 \Delta -\frac{\tau^\prime}{2}-1\right)}{\Gamma \left(\frac{\tau^\prime}{2}\right)^2 \Gamma \left(\Delta-\frac{\tau^\prime}{2}\right)^2 \Gamma (\ell+2\Delta-1) \Gamma \left(\ell+\frac{\tau^\prime}{2}+1\right)}\\\times \, _4F_3\left(\begin{matrix}\frac{\tau^\prime-d}{2}+1,\frac{\tau^\prime-2\Delta }{2}+1,\frac{\tau^\prime-2\Delta}{2}+1,\frac{\tau^\prime}{2}\\\frac{\tau^\prime}{2}+\ell+1,\frac{\tau^\prime}{2}-2 \Delta -\ell+2,\tau^\prime-\frac{d}{2} +1\end{matrix};1\right)\,\label{ScalarExch}
\end{multline}
which can be immediately obtained using the replacement \eqref{largespinsub}. It might be useful for the reader to note that the ${}_4F_3$ hypergeometric functions in \eqref{cpwanom0ell} and \eqref{ScalarExch} are $1$-balanced\footnote{A hypergeometric function ${}_{n+1}F_n\left(\begin{matrix}a_1,\ldots,a_{n+1}\\b_1,\ldots , b_n\end{matrix};z\right)$ is said to be $n$-balanced if $\sum_i b_i-\sum_i a_i=n$. If $\text{Re}(n)>0$ then the hypergeometric function converges for $z=1$.} and for this reason they are both well defined at argument $z=1$.\footnote{Integer balanced hypergeometric functions have in general logarithmic singularities at $z=1$. See e.g. \cite{HypZ=1}.} At the end of this subsection we give the expansion of \eqref{ScalarExch} in $1/\mathfrak{J^2}$, which we checked to agree with known expressions. 

\begin{figure}[t]
    \centering
    \captionsetup{width=0.95\textwidth}
    \includegraphics[width=\textwidth]{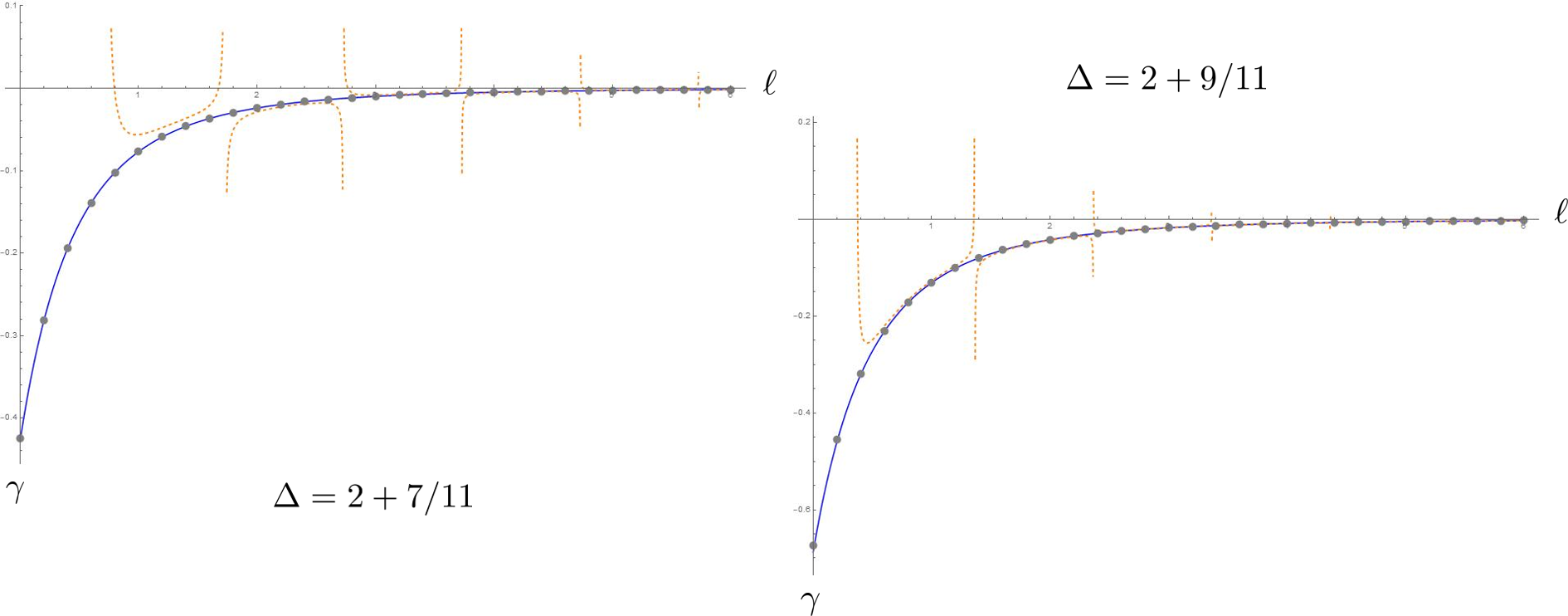}
    \caption{The plot of the expression \eqref{ScalarAnalytic} obtained by evaluating the spectral integral is in tick blue, the expression \eqref{ScalarExch} is in dashed orange and the first three terms of the large spin expansion in $1/\mathfrak{J}$ are the grey dots. In this plot we took $d=4$ and $\tau^\prime=4$, for $\Delta=2+7/11$ (LHS) and $\Delta=2+9/11$ (RHS). The expression \eqref{ScalarExch} displays some oscillating singular behaviour which diminishes as $\Delta$ approaches $\Delta=3$. For $\Delta\geq3$ no non-analytic behaviour in $1/\mathfrak{J}$ is observed. Remarkably the analytic result in spin matches the first three terms in the large spin expansion up to very low spin with a very small error that is indistinguishable in the graphs!}
    \label{fig:scalarLS}
\end{figure}

\begin{figure}[t]
    \centering
    \captionsetup{width=0.9\textwidth}
    \includegraphics[width=0.9\textwidth]{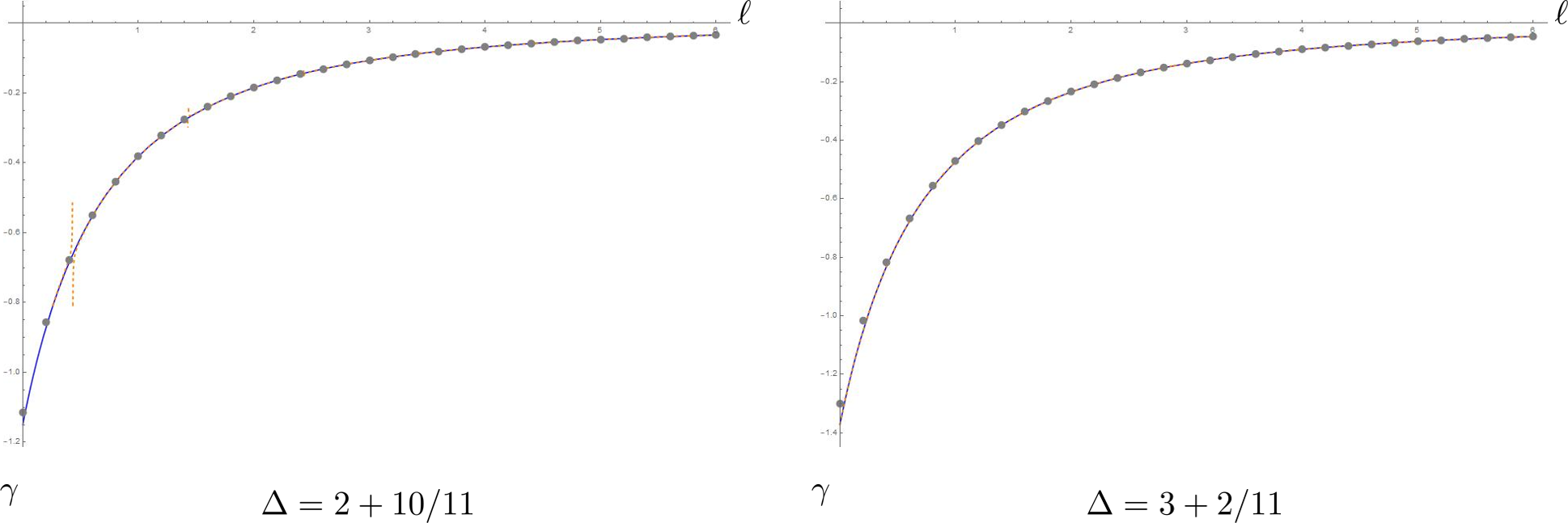}
    \caption{The plot of the expression \eqref{ScalarAnalytic} obtained by evaluating the spectral integral is in tick blue, the expression \eqref{ScalarExch} in dashed orange and the first three terms of the large spin expansion in $1/\mathfrak{J}$ in gray dots. We considered $d=4$ and $\tau^\prime=5/2$, with $\Delta=2+10/11$ (LHS) and $\Delta=3+2/11$ (RHS). In this case the expression \eqref{ScalarExch} precisely coincides with analytic in spin result \eqref{ScalarAnalytic} obtained by evaluating the spectral integral and with the first three terms of the asymptotic $1/\mathfrak{J}$ expansion. A small deviation can be observed for $\ell<1$.}
    \label{fig:scalarLS2}
\end{figure}

A few comments are in order:
\begin{itemize}
    \item We note that, for a limited range of values of $\Delta$ and $\tau^\prime$, the re-summation includes some non-analytic terms in $1/\mathfrak{J}$, which generate an oscillatory behaviour. Various plots of the general formula \eqref{ScalarExch} are presented in figures \ref{fig:scalarLS} and \ref{fig:scalarLS2}, for ranges of parameters with and without the oscillatory behaviour.\footnote{The reason for such oscillatory behaviour is related to the poor behaviour of a single conformal block, which is not a single valued function of the cross ratios as discussed e.g. in \cite{ElShowk:2011ag}. In particular, the projection of the shadow poles in a large-spin expansion is insensitive to exponentially suppressed terms. These terms would restore analyticity in spin at finite spin. The correct way of taking into account such exponentially suppressed terms is to perform the spectral integral with the appropriate measure -- as discussed in \S\ref{SpectralFull} and observed by Polyakov in \cite{Polyakov:1974gs}. The double-twist operators encoded in the poles of the weight-function \eqref{wa} ensure that the spectral integration is well defined and that the final result is single-valued, as opposed to the case of a single conformal block.} When there are no non-analytic terms this result (as is to be expected) behaviour matches the analytic result in spin \eqref{ScalarAnalytic} obtained in \S \ref{ExchScalar} by evaluating the spectral integral \eqref{anomdint}. In figure \ref{fig:scalarLS} it can be observed that, when there are non-analytic terms, the average of the oscillatory behaviour matches the analytic result in \eqref{ScalarAnalytic} which naturally sets to zero such contributions which are non-analytic at infinite spin. In all examples we plotted, this behaviour only arises for $\Delta<\tau^\prime$ for some (integer) values of $\tau^\prime$. Removing these non-analytic contributions for these values of $\Delta$ and $\tau^\prime$ requires a careful handling of the analytic continuation of the hypergeometric function.\footnote{We thank L. F. Alday for useful comments on this point.} This can be achieved by explicitly evaluating the spectral integral as in \S\ref{SpectralFull}. 
    \item The plots exhibit the standard convexity, monotonicity and negativity properties of double-twist operator anomalous dimensions \cite{NACHTMANN1973237,Komargodski:2012ek,Fitzpatrick:2012yx} down to finite spin.
    \item  Note that the anomalous dimension \eqref{ScalarAnalytic} and \eqref{ScalarExch} vanish identically when $\tau^\prime=2\Delta+2n$, owing to the following Gamma function factor in the denominator
    \begin{equation}
        \frac{1}{\Gamma\left(\Delta-\frac{\tau^\prime}{2}\right)}\overset{\tau^\prime=2\Delta+2n}{=} \frac{1}{\Gamma\left(-n\right)}=0,\qquad n \in \mathbb{N}_{0}.
    \end{equation}
We furthermore observe this explicitly for all other cases considered in this note. This implies double-trace operators don't contribute to the part of anomalous dimensions analytic in spin. This is in perfect agreement with the Lorentzian inversion formula \cite{Caron-Huot:2017vep}, which prescribes that the part of the anomalous dimensions analytic in spin is entirely fixed by single-trace operators while double-trace operators drop out. Non-analytic contributions in spin are thus consistently relegated to local contact interactions in the bulk involving a finite number of derivatives, which by definition must have finite support in spin (see e.g. \cite{Heemskerk:2009pn}) and for this reason are intrinsically non-analytic.\footnote{Incidentally this observation does not leave any room for pseudo-local contact interactions (i.e. interactions with an unbounded number of derivatives that generate only double-trace contributions), in accordance with the no-go result of \cite{Sleight:2017pcz} for field theories in AdS with higher-spin symmetry. Contact terms can only contribute with strictly a finite support in spin.} According to the inversion formula these contributions are further constrained by the leading Regge behaviour of the full CFT correlator. For scalar exchanges this leaves room only for $\phi^4$-type bulk contact interactions, where $\phi$ is the scalar field in AdS dual to the operator ${\cal O}$, while in general unitary CFTs contact terms may be allowed up to spin $\ell \leq 2$. In the case of correlators which are not bounded in the Regge-limit, contact terms are allowed for $0\leq\ell \leq\ell^\prime$ where $\ell^\prime$ is the highest spin single-trace operator dominating the Regge limit. This is for instance the case for the contact part of spin-$\ell^\prime$ exchange amplitudes \cite{Costa:2014kfa}.    
\end{itemize}

In the following, before presenting the large spin asymptotic expansion of the general formula \eqref{ScalarExch}, we highlight some potentially useful simplifications in its form for particular dimensions and twists. Further notable examples are also given in the applications section \S \ref{slightlyB}. 

\subsubsection*{Even $d$ and $\tau^\prime=d-2$}

It is interesting to note that in even dimensions and for integer $\tau^\prime$ the result \eqref{ScalarExch} drastically simplifies. For example, for $\tau^\prime=d-2$ we have
\begin{subequations}
\begin{align}
    d&=4\,,&\gamma_{0,\ell}&=-a_{\tau^\prime,0}\frac{2(\Delta -1)^2}{\mathfrak{J}^2-(\Delta -2) (\Delta -1)}\,,\\
    d&=6\,,&\gamma_{0,\ell}&=-a_{\tau^\prime,0}\frac{12 (\Delta -2)^2 (\Delta -1)^2}{\left(\mathfrak{J}^2-(\Delta -3) (\Delta -2)\right) \left(\mathfrak{J}^2-(\Delta -2) (\Delta -1)\right)}\,,
\end{align}
\end{subequations}
which are particular cases of the general even-dimensional result:
\begin{align}\label{genevend}
    \gamma_{0,\ell}=-a_{\tau^\prime,0}\,\frac{2\Gamma (d)}{\Gamma \left(\frac{d}{2}\right)^2}\prod_{i=0}^{\tfrac{d}2-1}\frac{(\Delta -i-1)^2}{\mathfrak{J}^2-(\Delta -i-1) (\Delta -i-2)}
\end{align}
where we expressed the result in terms of the conformal spin, which for leading double-twist operators is $\mathfrak{J}^2=\left(\ell+\Delta\right)\left(\ell+\Delta-1\right)$. 

Note that the expression \eqref{genevend} is analytic in spin and thus agrees with the expression \eqref{ScalarAnalytic} obtained by evaluating the spectral integral.

\subsubsection*{Large spin expansion}

As outlined in \S \ref{subsec::approach}, via the Mellin representation \eqref{mellinhyper} of the hypergeometric function ${}_4F_3$ we can straightforwardly determine the large spin expansion of the double-twist operator anomalous dimensions \eqref{ScalarExch} from the asymptotic expansion \eqref{prodgammaexp} of simple ratios of gamma functions. This gives
\begin{multline}\label{lsexpexchsc}
    \gamma_{0,\ell}=-\frac{a_{\tau^\prime,0}}{\mathfrak{J}^{\tau^\prime}}\frac{2\Gamma \left(\frac{d}{2}\right) \Gamma (\Delta )^2}{\Gamma \left(\frac{\tau^\prime}{2}\right) \Gamma \left(\frac{d-\tau^\prime}{2}\right)}\sum_{q=0}^\infty\sum_{k=0}^\infty d^k_{\Delta-1,\frac{\tau^\prime}{2}-\Delta +q+1}\\ \times \frac{(-1)^{q} \Gamma \left(q+\frac{\tau^\prime}{2}\right) \Gamma \left(\frac{d-\tau^\prime}{2}-q\right)}{q! \Gamma \left(\frac{d-\tau^\prime-2 q}{2}\right) \Gamma \left(\Delta -\frac{\tau^\prime}{2}-q\right)^2}\,\mathfrak{J}^{-2(q+k)}\,,
\end{multline}
which matches the result obtained by starting with the full crossing kernel \eqref{cpwanom0ell} and closing the integration contour only on the non-shadow poles associated with the $\Gamma$-function $ \Gamma \left(\frac{\tau^\prime-2 s}{2}\right)$ in the Mellin representation of the hypergeometric function ${}_4F_3$:\footnote{In this case picking the poles of the other $\Gamma$-function would give the shadow contributions.}
\begin{equation}
    \gamma_{0,\ell}=a_{\tau^\prime,0}\frac{2\Gamma \left(\frac{d}{2}\right) \Gamma (\Delta )^2}{\Gamma \left(\frac{\tau^\prime}{2}\right) \Gamma \left(\frac{d-\tau^\prime}{2}\right)}\int_{-i\infty}^{i\infty}\frac{ds}{2\pi i}\,\sum_{k=0}^\infty d^k_{\Delta-1,s+1-\Delta}\,\mathfrak{J}^{-2(s+k)}\frac{\Gamma (s) \Gamma \left(\tfrac{\tau^\prime-2 s}{2}\right) \Gamma \left(\tfrac{d-2 s-\tau^\prime}{2}\right)}{\Gamma \left(\tfrac{d}{2}-s\right) \Gamma (\Delta -s)^2}\,.
\end{equation}

The asymptotic expansion \eqref{lsexpexchsc} can be checked to be in complete agreement with previously known results \cite{Alday:2015ewa,Dey:2017fab} for the $1/\mathfrak{J}^2$ expansion when a direct comparison was feasible. 

\subsection{Exchanged spinning operators}
\label{subsec::exchspinop}

The generalisation of the large spin expansion discussed in the previous section to exchanges of spinning operators in the ${\sf t}$-channel (i.e. general $\ell^\prime$) follows in the same way as the $\ell^\prime=0$ case considered in \S \ref{subsec::exchscalarop}. 

The corresponding crossing kernel for a ${\sf t}$-channel CPW with exchanged twist $\tau$ and spin $\ell^\prime$ was determined in \cite{Sleight:2018epi} as a sum of ${}_4F_3$ hypergeometric functions. In latter expression, which we review in \eqref{KernelEqualTau}, not all hypergeometric functions in the sum are balanced for $\ell^\prime>0$. This is not an issue at this point since each hypergeometric function to a polynomial in this case. However, after projecting away the shadow contributions as described in \S \ref{subsec::approach}, the hypergeometric functions in the resulting expression do not truncate to polynomials and therefore not all of them converge manifestly at $z=1$. In this case it is therefore important to find a representation of the result which is manifestly non-singular at argument $z=1$. Fortunately, one may verify that the singularities cancel upon performing the finite sum over the hypergeometric functions in the shadow-projected expression. This suggests that the expression can be re-written in terms of hypergeometric functions which are $n$-balanced with $n > 0$ for all parameters. Indeed, one can re-visit the original expression for the crossing kernel of the CPW and express it as:
\begin{subequations}
\begin{align}\label{SpinningBalanced}
    \gamma^{\text{CPW}}_{0,\ell}&=a_{\tau^\prime,\ell^\prime}\,\widetilde{\mathcal{Z}}_{{\ell^\prime}}\,\sum_{k=0}^{
{\ell^\prime}}\sum_{p=0}^{{\ell^\prime}-k}{A}^{{\ell^\prime}}_{p,k}\, _4F_3\left(\begin{matrix}-\ell,2 \Delta +\ell-1,\frac{d}{2}-\ell^\prime+k-\frac{\tau^\prime}{2},k+\frac{\tau^\prime}{2}\\\frac{d}{2}-\ell^\prime+2 k+p,\Delta ,\Delta \end{matrix};1\right),\\
\widetilde{\mathcal{Z}}_{{\ell^\prime}}&=\tfrac{(-1)^{\ell^\prime+\ell} \ell^\prime!\, 2^{\ell^\prime+1} \Gamma \left(\frac{d}{2}\right) \Gamma \left(\frac{\tau^\prime}{2}\right)\Gamma \left(\frac{d}{2}-\tau^\prime\right) \Gamma \left({\ell^\prime}+\frac{\tau^\prime}{2}+\frac{1}{2}\right) \Gamma (d-{\ell^\prime}-\tau^\prime-1)}{\pi ^2 \Gamma \left(\frac{\tau^\prime+1}{2}\right) \Gamma \left(\frac{d}{2}+{\ell^\prime}-1\right) \Gamma (d-\tau^\prime-1) \Gamma \left({\ell^\prime}+\frac{\tau^\prime}{2}\right) \Gamma \left(\frac{d-2 ({\ell^\prime}+\tau^\prime)}{2}\right)}\,\sin ^2\left(\tfrac{\pi(d-\tau^\prime)}{2} \right), \\
{A}^{{\ell^\prime}}_{p,k}&=\tfrac{(-1)^{k-p} \Gamma \left(k+\frac{\tau^\prime}{2}\right) \Gamma \left(\frac{d}{2}+{\ell^\prime}-k-1\right) \Gamma \left(k+p+\frac{\tau^\prime}{2}\right) \Gamma \left(\frac{d}{2}-{\ell^\prime}+k-\frac{\tau^\prime}{2}\right) \Gamma \left(\frac{d}{2}-{\ell^\prime}+k+p-\frac{\tau^\prime}{2}\right)}{\Gamma (k+1) \Gamma ({\ell^\prime}-2 k-p+1) \Gamma \left(\frac{d}{2}-{\ell^\prime}+2 k+p\right)}\\\nonumber&\qquad\times\sum_{r=0}^{\ell^\prime}\tfrac{\Gamma \left(-\frac{d}{2}+k+r+\frac{\tau^\prime}{2}+1\right) \Gamma \left(\frac{\tau^\prime-d}{2}-k-r+{\ell^\prime}+1\right)}{\Gamma (p-r+1) \Gamma \left(k+r+\frac{\tau^\prime}{2}\right) \Gamma (-{\ell^\prime}+2 k+p+r+1) \Gamma \left(\frac{\tau^\prime}{2}+{\ell^\prime}-k-r)\right)}
\end{align}
\end{subequations}
which is a linear combination of $(1+p)$-balanced hypergeometric functions. This is the extension to general exchanged spin $\ell^\prime$ of equation \eqref{cpwanom0ell} for double-twist operator anomalous dimensions induced by crossed channel CPWs.

As before, to obtain the double-twist operator anomalous dimensions $\gamma_{0,\ell}$ induced by the physical conformal block we project away the contributions from the shadow conformal multiplet as prescribed in \S \ref{subsec::approach}. Starting from \eqref{SpinningBalanced}, this gives
\begin{multline}\label{KernelEqualTaucb}
    \gamma_{0,\ell}=a_{\tau^\prime,\ell^\prime}\widetilde{Z}_{\ell^\prime}\,\sum_{k=0}^{
{\ell^\prime}}\sum_{p=0}^{{\ell^\prime}-k}B^{{\ell^\prime}}_{p,k}\\\times\, _4F_3\left(\begin{matrix}k+\frac{\tau^\prime}{2},-\frac{d}{2}+J-k-p+\frac{\tau^\prime}{2}+1,-\Delta +k+\frac{\tau^\prime}{2}+1,-\Delta +k+\frac{\tau^\prime}{2}+1\\k+\ell+\frac{\tau^\prime}{2}+1,-2 \Delta +k-\ell+\frac{\tau^\prime}{2}+2,-\frac{d}{2}+J+\tau^\prime+1\end{matrix};1\right)
\end{multline}
where the coefficient $B^{{\ell^\prime}}_{p,k}$ is defined as 
\begin{equation}
   B^{{\ell^\prime}}_{p,k}=-A^{{\ell^\prime}}_{p,k}\tfrac{\Gamma (\Delta )^2 \ell!\, \Gamma \left(\frac{d}{2}-{\ell^\prime}-\tau^\prime\right) \Gamma \left(\frac{d}{2}-{\ell^\prime}+2 k+p\right) \Gamma \left(-k+\ell+2 \Delta -\frac{\tau^\prime}{2}-1\right)}{\Gamma (\ell+2 \Delta -1) \Gamma \left(-k+\Delta -\frac{\tau^\prime}{2}\right)^2 \Gamma \left(k+\ell+\frac{\tau^\prime}{2}+1\right) \Gamma \left(\frac{d}{2}-{\ell^\prime}+k-\frac{\tau^\prime}{2}\right) \Gamma \left(\frac{d}{2}-{\ell^\prime}+k+p-\frac{\tau^\prime}{2}\right)},
\end{equation}
This expression is the extension of the double-twist operator anomalous dimensions \eqref{ScalarExch} for $\ell^\prime=0$ to general spin $\ell^\prime$ exchanged in the ${\sf t}$-channel. 

As noted in \S \ref{subsec::exchscalarop} for the $\ell^\prime=0$ case, the above result vanishes identically for double-twist operators $\tau^\prime=2\Delta+2n$, as consistent with known inversion formulas. 

\subsubsection*{Large spin expansion}

In the same way as for the scalar exchange ($\ell^\prime=0$) we can determine the $1/\mathfrak{J}^2$ expansion of the anomalous dimensions \eqref{KernelEqualTaucb} from the large spin expansion of each hypergeometric function in the expression \eqref{KernelEqualTaucb}. Combining the expansions of each hypergeometric function in the finite sum, one then obtains the corresponding large spin expansion for the anomalous dimension as:
\begin{multline}
    \gamma_{0,\ell}=-\frac{a_{\tau^\prime,\ell^\prime}}{\mathfrak{J}^{\tau^\prime}}\frac{\Gamma (\Delta )^2\Gamma (\tau^\prime) \Gamma \left(\frac{d-\tau^\prime}{2}\right)^2}{\Gamma \left(\frac{d}{2}\right)\Gamma \left(\frac{\tau^\prime}{2}\right)^2 \Gamma \left(\frac{d}{2}-\tau^\prime \right)}\widetilde{Z}_{\ell^\prime}\sum_{k=0}^{\ell^\prime}\sum_{p=0}^{\ell^\prime-k}A_{p,k}^{\ell^\prime}\sum_{i,n=0}^\infty d_{\Delta -1,\frac{\tau^\prime}{2}-\Delta +k+n+1}^i\\\times\,\tfrac{(-1)^n \Gamma \left(k+n+\frac{\tau^\prime}{2}\right) \Gamma \left(\frac{d}{2}-\ell^\prime+2 k+p\right) \Gamma \left(\frac{d}{2}-\tau^\prime-\ell^\prime-n \right)}{n!\, \Gamma \left(k+\frac{\tau^\prime}{2}\right) \Gamma \left(\frac{d-\tau^\prime}{2}-\ell^\prime+k\right) \Gamma \left(\Delta -\frac{\tau^\prime}{2}-k-n\right)^2 \Gamma \left(\frac{d-\tau^\prime}{2}-\ell^\prime+k-n+p\right)}\mathfrak{J}^{-2 \left(i+k+n\right)}\,.\label{LargeSpinJ}
\end{multline}
We tested this expression with the examples given in \cite{Alday:2015ewa}, finding perfect agreement.

\subsubsection*{Exchanged conserved currents}

A simple application of the above result is when the exchanged operator is a conserved current, i.e. $\tau^\prime=d-2$. For $\ell^\prime=2$ this is the stress tensor. Plugging $\tau^\prime=d-2$ into \eqref{KernelEqualTaucb} we obtain directly:
\begin{equation}\label{conscurr}
    \gamma_{0,\ell}=-a_{\tau^\prime,\ell^\prime}\,c^{(0)}_0\frac{\Gamma (\ell +1) \Gamma \left(2\Delta+\ell-\frac{d}{2} \right)}{ \Gamma \left(\ell+\frac{d}{2}\right)\Gamma (2\Delta+\ell -1)},
\end{equation}
where $c^{(0)}_0$ takes the canonical form (identified in \cite{Fitzpatrick:2012yx,Komargodski:2012ek}):
\begin{equation}
    c^{(0)}_0=\frac{2\Gamma (\Delta )^2 \Gamma \left(\tau^\prime+2\ell^\prime\right)}{2^{\ell^\prime}\Gamma \left(\Delta -\frac{\tau^\prime}{2}\right)^2 \Gamma \left(\ell^\prime+\frac{\tau^\prime}{2}\right)}.
\end{equation}

Note that this re-summation of the large-spin expansion does not suffer from the divergence problems encountered when evaluating the spectral integral to obtain the analytic result in spin (see e.g. the discussion after eq.~\eqref{spurious}), for which we must consider also the contribution of the colliding spurious pole to get a finite result.

We also note that the dependence on the exchanged spin $\ell^\prime$ is completely factorised into $c^{(0)}_0$.  This confirms the observation in \cite{Alday:2015ewa} that the coefficients $c^{(k)}_0$ in the asymptotic expansion \eqref{intlsexpgamma} appear to be independent of $\ell^\prime$. This observation allowed the authors of \cite{Alday:2015ewa} to obtain \eqref{conscurr} by using the result for $\ell^\prime=0$ to re-sum the large spin expansion.\footnote{This factorisation of the $\ell^\prime$-dependence can be regarded as a consequence of higher-spin symmetry, and allows to reabsorb the exchange of higher-spin currents into the exchange of a scalar operator of twist $\tau^\prime=d-2$. We will draw on this general property in \S \ref{slightlyB} for applications to theories with slightly broken higher-spin symmetry.} The expansion of \eqref{conscurr} in powers of $1/\mathfrak{J^2}$ is given by equation \eqref{prodgammaexp}
\begin{equation}
    \gamma_{0,\ell}= -a_{\tau^\prime,\ell^\prime}\frac{\,c^{(0)}_0}{\mathfrak{J}^{d-2}}\sum_{k=0}^\infty d_{\Delta-1,\tfrac{d}2-\Delta}^k\mathfrak{J}^{-2k},
\end{equation}
which gives directly a formula for $c^{(k)}_0$ in terms of generalised Bernoulli polynomials (see equation \eqref{dbern})
\begin{equation}
    c^{(k)}_0=d_{\Delta-1,\tfrac{d}2-\Delta}^k.
\end{equation}

\subsection{Subleading double-twist operators}
\label{subsec::subleading}

In this section we consider the anomalous dimensions of subleading twist double-trace operators induced by the exchange of a scalar of twist $\tau^\prime$ in the crossed channel. Before projecting the shadow contributions from the crossing kernel \eqref{gamma0000nl}, we have that
\begin{equation}
    a^{(0)}_{n,\ell}\, \frac{\gamma^{\text{CPW}}_{n,\ell}}{2}=a_{\tau^\prime,0}\,{}^{({\sf t})}\hat{\mathfrak{I}}^{\text{projected}}_{\tau^\prime,0|\ell}\left(t=2\Delta+2n\right).
\end{equation}

As in the previous sections, we employ the Mellin representation of hypergeometric functions in the expression \eqref{gamma0000nl} for the crossing kernel to project away the contributions from the shadow conformal multiplet in the large spin limit. This boils down to the following replacement in equation \eqref{gamma0000nl}:
\begin{equation}
   T_{ij}^n \rightarrow \bar{T}_{ij}^n,
\end{equation}
with
\begin{multline}
    \bar{T}_{ij}^n=-t_{ij}^n\frac{\ell!\, \Gamma (n+\Delta )^2 \Gamma \left(\tfrac{d+2 i+2 j}{2}\right) \Gamma \left(\tfrac{d+2 i-2 j-2 \tau^\prime}{2}\right) \Gamma \left(\tfrac{-2 j+2 \ell+4 n+4 \Delta -\tau^\prime -2}{2}\right)}{\Gamma \left(\frac{d+2 i-\tau^\prime}{2}\right)^2 \Gamma \left(\tfrac{2 j+2\ell+\tau^\prime+2}{2}\right) \Gamma (\ell+2 n+2 \Delta -1) \Gamma \left(\tfrac{-2 j+2 n+2 \Delta -\tau^\prime}{2}\right)^2}\\\times\, _4F_3\left(\begin{matrix}-\frac{d}{2}-i+\frac{\tau^\prime}{2}+1,j+\frac{\tau^\prime}{2},-\Delta +j-n+\frac{\tau^\prime}{2}+1,-\Delta +j-n+\frac{\tau^\prime}{2}+1\\j+\ell+\frac{\tau^\prime}{2}+1,-2 \Delta +j-\ell-2 n+\frac{\tau^\prime}{2}+2,-\frac{d}{2}-i+j+\tau^\prime+1\end{matrix};1\right).
\end{multline}
In particular, the resulting anomalous dimensions are given by the expression
   \begin{align}\label{gamma0000nlCB}
    \frac12\,\alpha_n( 2\Delta+2n)a^{(0)}_{n,\ell}\,\gamma_{n,\ell}=a_{\tau^\prime,0}\frac{2\, \Gamma (\tau^\prime)}{\Gamma \left(\frac{\tau^\prime}{2}\right)^4 \Gamma \left(\frac{d}{2}-\tau^\prime\right)}\sum_{j=0}^n\,D_{j}\bar{T}^n_{n-j,j}\,.
\end{align}

As before, the large spin expansion \eqref{intlsexpgamma} of the anomalous dimensions can be worked out systematically from the Mellin representation of the ${}_4F_3$ following the steps outlined towards the end of \S \ref{subsec::approach}. A consistency check of \eqref{gamma0000nl} is that all pre-factors nicely combine into an expansion in the inverse conformal spin squared $1/\mathfrak{J}^2$. Finally, one can further check that the exchange of double-twist operators in the ${\sf t}$-channel give vanishing contributions as before.

\subsubsection*{Examples: $\tau^\prime=d-2$}

There are simplifications for particular integer values of $\tau^\prime$ and dimension $d$. For instance, taking $\tau^\prime=d-2$ the simplest expression is obtained in $d=4$, where we have 
\begin{equation}\label{taudm2d4}
    \gamma_{n,\ell}=a_{\tau^\prime,0}\frac{2(\Delta-1)^2}{\Delta ^2-3 \Delta +n^2+2 \Delta  n-3 n-\mathfrak{J}^2+2}\,,
\end{equation}
which matches the result of \cite{Alday:2017gde} obtained by considering an explicit re-summation of the large spin expansion. 

The explicit form of the result for general $d$ is more involved due to the complicated form of the coefficients $D_j$. For $n=1$ it reads
\begin{multline}
    \gamma^{(\tau^\prime=d-2)}_{1,\ell}=a_{d-2,0}\frac{2^{d-2} \Gamma \left(\frac{d-1}{2}\right) \Gamma (\Delta )^2 \Gamma (\ell+1)\Gamma \left(-\frac{d}{2}+\ell+2 \Delta +1\right)}{\sqrt{\pi } (d-2 (\Delta +1)) \Gamma \left(\frac{d}{2}-1\right) \Gamma \left(-\frac{d}{2}+\Delta +1\right)^2 \Gamma \left(\frac{d}{2}+\ell+1\right) \Gamma (\ell+2 \Delta +1)}\\
    \times \Big(d^2 \left(\Delta^2+\ell^2+2 \Delta  \ell+\ell+1\right)-d \left(4 \Delta ^2+6 \Delta +14 \Delta  \ell+7 \ell (\ell+1)+4\right)\nonumber\\
    +2 \left(6 \Delta +2 \Delta ^2 (\ell+2)+\Delta  \ell(\ell+11)+5\ell(\ell+1)+2\right)\Big)\,.
\end{multline}
We are able to obtain closed form expressions for any fixed value of $n$, though they are rather cumbersome. For instance, with $n=2$ and $d=3$ we have:
\begin{align}
    \gamma^{(\tau^\prime=1)}_{2,\ell}&=-a_{1,0}\frac{2(\Delta -1) (2 \Delta +1) \Gamma (\Delta )^2 \Gamma (\ell+1) \Gamma \left(\ell+2 \Delta +\frac{1}{2}\right)}{16 \pi  \Gamma \left(\Delta +\frac{3}{2}\right)^2 \Gamma \left(\ell+\frac{7}{2}\right) \Gamma (\ell+2 \Delta +3)}\nonumber\\&\times\,\Big(\Delta ^3 \left(4 \ell \left(8 \ell^2+34 \ell+41\right)+51\right)+2 \Delta ^2 (\ell+1) (2 \ell+5) \left(2 \ell^2+\ell-4\right)\nonumber\\&+\Delta ^4 (16 \ell (2 \ell+7)+89)-4 \Delta  (\ell+1)^2 (\ell+2) (\ell+3)-(\ell+1)^2 (\ell+2)^2\Big).
\end{align}
For $d=5$:
\begin{align}
    \gamma^{(\tau^\prime=3)}_{2,\ell}&=-a_{3,0}\frac{4 \Gamma (\Delta )^2 \Gamma (\ell+1) \Gamma \left(\ell+2 \Delta -\frac{1}{2}\right)}{\pi  (2 \Delta -1) \Gamma \left(\Delta -\frac{1}{2}\right)^2 \Gamma \left(\ell+\frac{9}{2}\right) \Gamma (\ell+2 \Delta +3)}\nonumber\\
    &\times \Big(\Delta ^4 \left(4 \ell \left(8 \ell^2+50 \ell+13\right)-791\right)+\Delta ^5 (16 \ell(2 \ell+15)+433)\nonumber\\
    &+\Delta ^3 (2 \ell (\ell (4 \ell (\ell+4)-107)-319)+275)-\Delta ^2 (4 \ell (\ell (\ell (\ell+27)+76)+9)+61)\nonumber\\
    &-3 \Delta  (\ell+1) (\ell (\ell (7 \ell+11)-58)-68)+18 (\ell+1) (\ell+2) (\ell (\ell+3)-1)\Big),
\end{align}
and for $d=6$:
\begin{align}
    \gamma^{(\tau^\prime=4)}_{2,\ell}&=-a_{4,0}\tfrac{12 (\Delta -1)^2 \left(2 \Delta ^3 (\ell+7)+\Delta ^2 (\ell (\ell+11)-8)+4 \Delta  (\ell-2) (\ell+1)-8 (\ell+1) (\ell+2)\right)}{(\ell+1) (\ell+2) (\ell+3) (2\Delta+\ell) (2 \Delta +\ell+1) (2 \Delta +\ell+2)}.
\end{align}

Note that all of the expressions above are analytic in spin.

\subsection{External spinning operators}
\label{subsec::extspinops}

In this section, we apply the present approach to the case of spinning external operators also considered in \S \ref{spinninWilson}. We shall take $\tau_i=\Delta$ in the correlator \eqref{0O1OOO2} (we give the result for generic $\tau_i$ in \S\ref{subsec::shadowpro}) and as before we consider a scalar of twist $\tau^\prime$ exchanged in the crossed channel and its contribution to the (averaged, due to operator mixing) anomalous dimensions of leading double-twist operators in the ${\sf s}$-channel. 

The relevant ${\sf t}$-channel crossing kernel is \eqref{CKj1j2t}, and after projecting away the shadow contributions as prescribed at the end of \S \ref{subsec::approach}, it reads 
 \begin{multline}\label{EqDelta}
    {}^{({\sf t})}\bar{\mathfrak J}_{\tau^\prime,0|\ell}
    =\frac{(-1)^{\ell+1}2^{-2\Delta -J_1-J_2-\ell+\tau^\prime+1} \Gamma \left(\frac{\tau^\prime+1}{2}\right) \Gamma (J_1-J_2+\ell+\Delta ) \Gamma \left(J_2+\ell+2 \Delta -\frac{\tau^\prime}{2}-1\right)}{\Gamma \left(\frac{\tau^\prime}{2}\right) \Gamma \left(\ell+\Delta -\frac{1}{2}\right) \Gamma \left(J_1+\Delta -\frac{\tau^\prime}{2}\right) \Gamma \left(J_2+\Delta-\frac{\tau^\prime}{2}\right) \Gamma \left(-J_2+\ell+\frac{\tau^\prime}{2}+1\right)}\\\times\, _4F_3\left(\begin{matrix}J_1+\frac{\tau^\prime}{2},-\frac{d}{2}-J_2+\frac{\tau^\prime}{2}+1,-\Delta -J_1+\frac{\tau^\prime}{2}+1,-\Delta -J_2+\frac{\tau^\prime}{2}+1\\-J_2+\ell+\frac{\tau^\prime}{2}+1,-2 \Delta -J_2-\ell+\frac{\tau^\prime}{2}+2,-\frac{d}{2}+\tau^\prime+1\end{matrix};1\right)\,,
\end{multline}
which is a $1$-balanced hypergeometric function. As in \S \ref{spinninWilson}, for brevity here we only present explicitly the result for the $\sf{t}$-channel crossing kernel. The shadow-projected $\sf{u}$-channel crossing kernels are given in \S \ref{subsec::shadowpro}.  Since we are just considering the exchange of a scalar operator in the ${\sf t}$- and ${\sf u}$-channels, the corresponding ${\sf t}$- and ${\sf u}$-channel CPWs are unique. One can furthermore check that double-twist operators (i.e. $\tau^\prime=2\Delta+2n$) give vanishing contributions, consistent in this case with spinning inversion formulas.

The corresponding expression for the averaged anomalous dimensions \eqref{0avgamma} is
\begin{equation}\label{3DTANOMspin}
   \frac{1}{2}  c^{(0)}_{\mathcal{O}_{J_1}\mathcal{O}_{J_2}[\mathcal{O}_{J_1}\mathcal{O}_{J_2}]_{(\ell)}}c^{(0)}_{\mathcal{O}\mathcal{O}[\mathcal{O}\mathcal{O}]_{(\ell)}}\gamma_{0,\ell}= a_{\tau^\prime,0} {}^{({\sf t})}\bar{\mathfrak J}_{\tau^\prime,0|\ell}.
\end{equation}
As before, the mean field theory OPE coefficients $ c^{(0)}_{\mathcal{O}\mathcal{O}[\mathcal{O}\mathcal{O}]_{(\ell)}}$ are given by setting $n=0$ in equation \eqref{OPEnl}. On the other hand, the coefficients $c^{(0)}_{\mathcal{O}_{J_1}\mathcal{O}_{J_2}[\mathcal{O}_{J_1}\mathcal{O}_{J_2}]_{(\ell)}}$ are only known explicitly so far for $J_2=0$ or $J_1=0$ \cite{Sleight:2018epi}. Setting $J_1=J$ and $J_2=0$ the coefficients are given by equation \eqref{0spinningMean}. The corresponding expression, for $\gamma_{0,\ell}$ is then given by 
\begin{multline}\label{spinninggenanom}
    \gamma_{0,\ell}=-a_{\tau^\prime,0}\sqrt{\frac{\Gamma (\Delta ) \Gamma (\ell+1) \Gamma (2 J+\Delta ) \Gamma (-J+\ell+1) \Gamma (J+\ell+\Delta )}{\Gamma (\ell+2 \Delta -1) \Gamma (-J+\ell+\Delta ) \Gamma (J+\ell+2 \Delta -1)}}\\ \times \frac{2^{1-\frac{J}{2}} \Gamma \left(\frac{d}{2}\right) \Gamma (\Delta ) \Gamma \left(\frac{d}{2}-\tau^\prime\right) \Gamma \left(\ell+2 \Delta -\frac{\tau^\prime}{2}-1\right)}{\Gamma \left(\frac{d-\tau^\prime}{2}\right)^2 \Gamma \left(\Delta -\frac{\tau^\prime}{2}\right) \Gamma \left(\ell+\frac{\tau^\prime}{2}+1\right) \Gamma \left(J+\Delta -\frac{\tau^\prime}{2}\right)}\\ \times\, _4F_3\left(\begin{matrix}\frac{\tau^\prime-d }{2}+1,J+\frac{\tau^\prime}{2},\frac{\tau^\prime}{2}-\Delta+1,\frac{\tau^\prime}{2}-\Delta -J+1\\\ell+\frac{\tau^\prime}{2}+1,\frac{\tau^\prime}{2}-2 \Delta -\ell+2,\tau^\prime-\frac{d}{2} +1\end{matrix};1\right)\,.
\end{multline}

\subsection*{Large spin expansion}

As before, we can systematically derive the expansion of the result \eqref{spinninggenanom} in $1/\mathfrak{J^2}$. To this end, it is useful to note that the crossing kernel admits the following expansion:
\begin{align}
    \gamma_{0,\ell}&=a_{\tau^\prime,0}\frac{(-1)^{\ell}2^{\tau^\prime-\frac{J}{2}} \Gamma \left(\frac{\tau^\prime+1}{2}\right)\sqrt{\Gamma (\Delta ) \Gamma (2 J+\Delta )}\Gamma (\Delta ) \Gamma \left(\frac{d-\tau^\prime}{2}\right)}{\sqrt{\pi }\Gamma \left(\frac{\tau^\prime}{2}\right) \Gamma \left(\frac{d}{2}-\tau^\prime\right) \Gamma \left(J+\frac{\tau^\prime}{2}\right)}\\\nonumber
    &\times\sum_{n=0}^\infty\frac{(-1)^{n}\Gamma \left(\frac{d-2 (n+\tau^\prime)}{2}\right) \Gamma \left(J+n+\frac{\tau^\prime}{2}\right)}{ n!  \Gamma \left(\frac{d-2 n-\tau^\prime}{2}\right) \Gamma \left(-n+\Delta -\frac{\tau^\prime}{2}\right) \Gamma \left(J-n+\Delta -\frac{\tau^\prime}{2}\right)}\\\nonumber
    &\times\,\frac{\sqrt{(\lambda -J)_{2 J}}}{\sqrt{(\lambda-J-\Delta +1)_{J} (\lambda+\Delta -1)_{J}}}\,\frac{\Gamma (-\Delta +\lambda +1) \Gamma \left(-n+\Delta +\lambda -\frac{\tau^\prime}{2}-1\right)}{\Gamma (\Delta +\lambda -1) \Gamma \left(n-\Delta +\lambda +\frac{\tau^\prime}{2}+1\right)}\,,
\end{align}
where we have conveniently introduced $\lambda=\Delta+\ell$ and we gave the terms which depend on the spin $\ell$ on the third line. Of these, the ratio of Gamma functions independent of the external spin $J$ is also present in the case of external scalar operators, and we thus already know its large spin expansion from the result \eqref{lsexpexchsc}. The only difference in the case of external spins is the factor
\begin{equation}
    \frac{\sqrt{(\lambda -J)_{2 J}}}{\sqrt{(\lambda-J-\Delta +1)_{J} (\lambda+\Delta -1)_{J}}}=\sum_{k=0}^{\infty}\frac{p_k}{\mathfrak{J}^{2k}}\,,\qquad p_0=1,
\end{equation}
which admits an expansion in $1/\mathfrak{J}^2$ with coefficients $p_k$. The corresponding expansion of the anomalous dimensions then reads:
\begin{equation}
\gamma_{0,\ell}=\frac{2^{-\frac{J}{2}+\tau^\prime} \Gamma (\Delta ) \Gamma \left(\frac{\tau^\prime+1}{2}\right) \Gamma \left(\frac{d-\tau^\prime}{2}\right) \sqrt{\Gamma (\Delta ) \Gamma (2 J+\Delta )}}{\sqrt{\pi } \Gamma \left(\frac{\tau^\prime}{2}\right) \Gamma \left(\frac{d}{2}-\tau^\prime\right) \Gamma \left(J+\frac{\tau^\prime}{2}\right)}\frac{a_{\tau^\prime,0}}{\mathfrak{J}^{\tau^\prime}}\sum_{n=0}^\infty \frac{f_{J,n}}{\mathfrak{J}^{2n}}\,,
\end{equation}
with
{\footnotesize\begin{align}
f_{J,n}=\sum_{k_1,k_2=0}^n\frac{p_{k_2}\,d^{k_1}_{\Delta -1,\frac{\tau^\prime}{2}-\Delta -k_1-k_2+n+1}\Gamma \left(\frac{d}{2}+k_1+k_2-n-\tau^\prime\right) \Gamma \left(J-k_1-k_2+n+\frac{\tau^\prime}{2}\right)}{(n-k_1-k_2)! \Gamma \left(\frac{d}{2}+k_1+k_2-n-\frac{\tau^\prime}{2}\right) \Gamma \left(k_1+k_2-n+\Delta -\frac{\tau^\prime}{2}\right) \Gamma \left(J+k_1+k_2-n+\Delta -\frac{\tau^\prime}{2}\right)}\,.
\end{align}}
The leading term $c^{(0)}_0$ in the large spin expansion \eqref{intlsexpgamma} in this case reads
\begin{equation}
  \gamma_{0,\ell}\approx-\frac{2^{-\frac{J}{2}+\tau^\prime}\sqrt{\Gamma (\Delta ) \Gamma (2 J+\Delta )} \Gamma (\Delta )\Gamma \left(\frac{\tau^\prime+1}{2}\right) \Gamma \left(\frac{d-2\tau^\prime}{2}\right)}{\sqrt{\pi } \Gamma \left(\frac{\tau^\prime}{2}\right) \Gamma \left(\frac{d}{2}-\tau^\prime\right) \Gamma \left(\Delta -\frac{\tau^\prime}{2}\right) \Gamma \left(J+\Delta -\frac{\tau^\prime}{2}\right)}\frac{a_{\tau^\prime,0}}{\mathfrak{J}^{\tau^\prime}}\,
\end{equation}
which reduces to the case of identical external scalar operators for $J=0$.

\section{Applications: CFTs with slightly broken higher spin symmetry}
\label{slightlyB}

A possible application of our results is to CFTs with slightly broken higher-spin symmetry in the large $N$ limit, where the concept of double-twist operator acquires the natural interpretation of a double-trace operator. Such theories have a tower of single-trace operators of spins $\ell^\prime=2,4,6,...$\footnote{It is also possible to have higher-spin currents of each integer spin $\ell^\prime=1,2,3,...$ but here we will only consider the minimal spectrum consistent with higher-spin symmetry \cite{Fradkin:1986ka,Mikhailov:2002bp,Eastwood:2002su,Vasiliev:2003ev,Joung:2014qya}.} and twist $\tau^\prime=d-2+{\cal O}\left(1/N\right)$, which thus become conserved currents in the limit of large $N$. For previous work on double-trace anomalous dimensions in such theories, see \cite{Lang:1992zw,Leonhardt:2003du,Alday:2015ota,Giombi:2016zwa,Alday:2016njk,Giombi:2017rhm,Turiaci:2018nua,Aharony:2018npf}. 

These theories have the attractive feature - which we shall sometimes exploit in the following sections - that the higher-spin symmetry allows to express the contributions from the tower of higher spin currents to the ${\cal O}\left(1/N\right)$ anomalous dimensions of double-trace operators in terms of an effective contribution from a scalar operator of twist $\tau^\prime=d-2$ in the higher-spin multiplet:
\begin{equation}
    \sum_{\ell^\prime=0,2,\ldots}\gamma_{n,\ell|\ell^\prime}={\cal O}\left(\tfrac{1}{N^2}\right)\qquad\rightarrow\qquad \sum_{\ell^\prime=2,4,\ldots}\gamma_{n,\ell|\ell^\prime}=-\gamma_{n,\ell|0}+{\cal O}\left(\tfrac{1}{N^2}\right)\,.\label{HSeff}
\end{equation}
Here we introduced the notation $\gamma_{n,\ell|\ell^\prime}$ which labels the spin $\ell^\prime$ of the twist $\tau^\prime=d-2$ operator in the ${\sf t}$-channel which generates this contribution to the anomalous dimension of the double-trace operator $\left[{\cal O}{\cal O}\right]_{n,\ell}$. 

In this way, at leading order in $1/N$, it is often possible to re-package an infinite sum over higher-spin currents as the exchange of an effective scalar of twist $\tau^\prime=d-2$, simplifying the analysis. From a bootstrap perspective this means that, to leading order in $1/N$, \emph{the anomalous dimensions will be often encoded in the crossing kernels of a scalar exchange in the crossed-channel.}

In general $d$, the most well-known are the critical Boson and critical Fermion theories which involve $N$ scalar or Fermion fields in the fundamental representation where, in addition to the tower of higher-spin currents, there is a scalar operator of dimension $\Delta=2$ in the Bosonic theory and $\Delta=1$ in the Fermionic theory. Such scalar operators are usually denoted by $\sigma$ in the literature but we shall call them ${\cal O}$. In the following sections we shall also consider the special case $d=3$, where there are two main classes of theories with slightly broken higher-spin symmetry \cite{Maldacena:2012sf}: The quasi-Boson ($\Delta=1$) and the quasi-Fermion ($\Delta=2$) theories in which the fundamental scalars/Fermions are coupled to $O\left(N\right)_k$ Chern-Simons gauge fields. The quasi-Boson/Fermion theory is believed to be equivalent to the critical-Fermion/Boson theory in $d=3$ with Chern-Simons coupling \cite{Giombi:2011kc,Aharony:2011jz,Aharony:2012nh}. 
It will be convenient to encode (some of\footnote{In the following we shall not consider explicitly the deformation of spinning operator  OPE coefficients. In $d=3$, for instance, when the Chern-Simons level is non-vanishing such contributions appear \cite{Maldacena:2012sf} due to the deformation of the corresponding $J_1$-$J_2$-$J_3$ structures away from the free-theory point. They are usually referred to as ``odd'' structures, which start contributing when at least two of the operators are spinning. Taking explicitly into account these additional structure will require the extension to the parity odd case of the techniques of \cite{Sleight:2018epi}.}) the information about these theories at leading order in $1/N$ into three effective parameters which encode the deviation from the higher-spin symmetric point (the free theory):
\begin{subequations}\label{alphacoeffs}
\begin{align}
    \alpha_1&:=(c_{\mathcal{O}\mathcal{O}\mathcal{O}})^2-({c}^{\text{free}}_{\mathcal{O}\mathcal{O}\mathcal{O}})^2,\\
    \alpha_2&;=c_{\mathcal{O}\mathcal{O}\mathcal{O}}-{c}^{\text{free}}_{\mathcal{O}\mathcal{O}\mathcal{O}},\\
    \alpha_3&:= c^{}_{{\cal O}{\cal O}{\cal O}_{d-2}},
\end{align}
\end{subequations}
where by ${c}_{\mathcal{O}\mathcal{O}\mathcal{O}}$ we mean the OPE coefficient of the scalar ${\cal O}$ with itself in the deformed theory. The third parameter $\alpha_3$ is an effective parameter which encodes, in the spirit of equation \eqref{HSeff}, the re-summation of the tower of higher-spin currents into an effective OPE coefficient of a scalar ${\cal O}_{d-2}$ of twist $\tau^\prime=d-2$. 

\subsection{Scalar correlators}
\label{subsubsec::scalarcrr}

Let us first focus on the exchange of a scalar operator of twist $\tau^\prime=d-2$ in the ${\sf t}$-channel between identical external scalar operators ${\cal O}$ of dimension $\Delta$, which has various applications.

The double-trace anomalous dimensions $\gamma_{0,\ell}$ in this case are given by simply plugging into the general formula \eqref{ScalarAnalytic}, which simplifies to: 
\begin{equation}\label{gamma0}
    \gamma_{0,\ell}=-\alpha^2_3\frac{2^{d-2} \Gamma \left(\frac{d-1}{2}\right) \Gamma (\Delta )^2 \Gamma (\ell+1) \Gamma \left(-\frac{d}{2}+\ell+2 \Delta \right)}{\sqrt{\pi } \Gamma \left(\frac{d}{2}-1\right) \Gamma \left(-\frac{d}{2}+\Delta +1\right)^2 \Gamma \left(\frac{d}{2}+\ell\right) \Gamma (\ell+2 \Delta -1)}\,.
\end{equation}
As a function of $\Delta$, the result for $n>0$ becomes increasingly cumbersome for increasing $n$ and the explicit form is not very instructive.
In $d=3$, the $n>0$ the results simplify a bit and a few examples read:
\begin{subequations}
\begin{align}
    \label{gamma1}\gamma_{1,\ell}&=-\alpha^2_3(\Delta -1) \frac{2\Gamma (\Delta )^2 (5 \Delta +2 \ell (2 \Delta +\ell+1)-1) \Gamma (\ell+1) \Gamma \left(\ell+2 \Delta -\frac{1}{2}\right)}{\pi  (2 \Delta -1) \Gamma \left(\Delta -\frac{1}{2}\right)^2 \Gamma \left(\ell+\frac{5}{2}\right) \Gamma (\ell+2 \Delta +1)},\\
    \label{gamma2}\gamma_{2,\ell}&=-\alpha^2_3(\Delta -1) \frac{(2 \Delta +1)\Gamma (\Delta )^2 \Gamma (\ell+1) \Gamma \left(\ell+2 \Delta +\frac{1}{2}\right)}{8 \pi  \Gamma \left(\Delta +\frac{3}{2}\right)^2 \Gamma \left(\ell+\frac{7}{2}\right) \Gamma (\ell+2 \Delta +3)}\,\nonumber \\&\times\Big(\Delta ^3 \left(4 \ell \left(8 \ell^2+34 \ell+41\right)+51\right)+2 \Delta ^2 (\ell+1) (2 \ell+5) \left(2 \ell^2+\ell-4\right)\nonumber\\
    &+\Delta ^4 (16 \ell (2 \ell+7)+89)-4 \Delta  (\ell+1)^2 (\ell+2) (\ell+3)-(\ell+1)^2 (\ell+2)^2\Big)\,, \\ \nonumber
    &\vdots
\end{align}
\end{subequations}
with expressions of increasing complexity for $n=3,4,5,...\,$, though for each $n>0$ they are proportional to $\Delta-1$.

For $\Delta=1$ and $d=3$, relevant for the quasi-Boson theory, we find in precise agreement with the results of \cite{Giombi:2017rhm}
\begin{equation}\label{scalar}
    \gamma_{0,\ell}=-\frac{4}{\pi ^2}\frac{\alpha_1}{2 \ell+1}\,,
\end{equation}
which in this case should be multiplied by the corresponding $\alpha_1$ parameter introduced in \S\ref{slightlyB}. Moreover, for $n>0$, since each $\gamma_{n>0,\ell}$ is proportional to $\Delta-1$ we have
\begin{equation}\label{dtanomvand3}
   \gamma_{n>0,\ell}=0,  
\end{equation}
in agreement with the recent result of \cite{Aharony:2018npf}. 

Considering again general $d$ while keeping $\Delta=1$, the contributions to the $n>0$ anomalous dimensions simplifies drastically and we obtain the following closed formula:
\begin{align}\label{anomDelta1gend}
    \gamma_{n,\ell}&=-\alpha^2_3\frac{2(-1)^n}{\pi ^2 n!} \sin ^2\left(\frac{\pi  d}{2}\right) \Gamma (d-2) (d-n-2)_n\frac{\Gamma \left(\ell+n+2-\frac{d}{2}\right)}{\Gamma \left(\frac{d}{2}+\ell+n\right)}\,.
\end{align}
As far as we are aware, the result for general $n$ is new. We note that this gives an appealing extension to general $d$ of the result \eqref{dtanomvand3} for the vanishing of the sub-leading double-trace anomalous dimensions in $d=3$:\footnote{From a bootstrap perspective, that the critical Fermion is equivalent to a quasi-Boson theory follows from higher-spin symmetry - which allows to rewrite an infinite sum of conserved currents in terms of an effective scalar contribution \eqref{HSeff}.}
\begin{equation}
     \gamma_{n>d-3,\ell}=0.
\end{equation}

For the critical Fermion theory, \eqref{anomDelta1gend} is the full contribution to the anomalous dimension since the OPE coefficient of the scalar ${\cal O}$ in the OPE of ${\cal O}$ with itself must vanish due to parity. In particular, this means that the leading contribution to $\gamma_{n,\ell}$ in $1/N$ comes from the tower of higher-spin currents. As explained at the beginning of \S \ref{slightlyB}, the contribution from the higher-spin tower is given by the exchange of a scalar operator ${\cal O}_{d-2}$ of twist $\tau^\prime=d-2$. The effective OPE coefficient $\alpha_3^2$ multiplying the $\gamma_{n,\ell}$ in \eqref{anomDelta1gend} is given in accordance with \cite{Giombi:2017rhm} as:
\begin{equation}
    \alpha_3^2=\frac{8}{N\left(d-4\right)}.
\end{equation}
This recovers the $n=0$ anomalous dimensions computed in \cite{Giombi:2017rhm} and moreover allows us to extend them to arbitrary $n$. The special case $d=4$ requires to consider an $\epsilon$-expansion:
\begin{equation}
    \gamma_{n,\ell}\Big|_{d=4-\epsilon}=\frac{4 (-1)^n}{n! \Gamma (2-n) (\ell+n)(\ell+n+1)}\,\epsilon+O\left(\epsilon ^2\right)\,,
\end{equation}
which anyway vanishes for $n>1$.
 
Another interesting case is $\Delta=2$ and $\tau^\prime=d-2$. Like the $\Delta=1$ case considered above, the corresponding results for $\gamma_{n,\ell}$ can be used to compute the contribution from the higher-spin tower to double-trace operator anomalous dimensions in the critical Boson theory. In this case we have to multiply all results by \cite{Giombi:2017rhm}\footnote{The following results also apply to the critical Boson with Chern-Simons coupling in $d=3$, just one has to multiply by the corresponding $\alpha^2_3$ which can be extracted from \cite{Aharony:2012nh}.}
\begin{equation}
    \alpha^2_3=\frac{16 (d-3)}{N(d-6)}.
\end{equation}
In particular, from \eqref{gamma0} we immediately have 
\begin{equation}
    \gamma_{0,\ell}=-\alpha_3^2\frac{8 \sin ^2\left(\frac{\pi  d}{2}\right) \Gamma (d-2) \Gamma \left(-\frac{d}{2}+\ell+4\right)}{\pi ^2 (d-4)^2 (1 + \ell) (2 + \ell) \Gamma \left(\frac{d}{2}+\ell\right)},
\end{equation}
recovering the result of \cite{Giombi:2017rhm}, and for $n=1$ we have
\begin{multline}
    \gamma_{1,\ell}=-\alpha_3^2\frac{8\sin ^2\left(\frac{\pi  d}{2}\right) \Gamma (d-2) \Gamma \left(\ell+5-\frac{d}{2}\right)}{\pi ^2 (\ell+1) (\ell+2) (\ell+3) (\ell+4)\,\Gamma \left(\frac{d}{2}+\ell+1\right)}\Big[\frac{2-d}{(d-4)^3}\\-\frac{((d-7) d+14) (d \ell(\ell+5)+5 d-4 \ell (\ell+5)-18)}{(d-6) (d-4)^3}\Big]\,,
\end{multline}
with increasingly complicated expressions for higher $n$ for general $d$.\footnote{In $d=4$ there is a simple expression for general $n$, given by \eqref{taudm2d4} with $\Delta=2$.} In the critical Boson theory there is also a contribution from the scalar of twist $\tau^\prime=2$, which gives\footnote{The expression \eqref{gamma000} for $\gamma_{0,\ell}$ agrees with the result in \cite{Aharony:2018npf} obtained in $d=3$ for the critical Boson theory with Chern-Simons coupling. The subleading twist results are new.}
\begin{subequations}
\footnotesize{\begin{align}
    \gamma_{0,\ell}&=-\alpha_1\frac{2}{(\ell+1) (\ell+2)}\,,\label{gamma000}\\
    \gamma_{1,\ell}&=-\alpha_1\frac{2(d+\ell (\ell+5)+2)}{(\ell+1) (\ell+2) (\ell+3) (\ell+4)}\,,\\
    \gamma_{2,\ell}&=-\alpha_1\frac{2(2 d^3+2 d^2 (\ell (\ell+7)-3)+d (\ell (\ell+7) (\ell (\ell+7)-2)-20)-2 (\ell+1) (\ell+6) (3 \ell (\ell+7)+16))}{(d-6) (\ell+1) (\ell+2) (\ell+3) (\ell+4) (\ell+5) (\ell+6)}\,,\\
    \vdots\nonumber
\end{align}}
\end{subequations}
in which case we have to multiply by (see definitions \eqref{alphacoeffs})\footnote{As before, this result can also be applied to the critical Boson with Chern-Simons coupling in $d=3$, using instead the corresponding $\alpha_1$ which can be extracted from \cite{Aharony:2012nh}.}
\begin{equation}
    \alpha_1=\frac{1}{N}\frac{2^d (d-3)^3 \sin \left(\frac{\pi  d}{2}\right) \Gamma \left(\frac{d-3}{2}\right)}{\pi ^{3/2} (d-4) \Gamma \left(\frac{d}{2}-1\right)}.
\end{equation}
The result \eqref{gamma000} was for $\Delta=2$, but we can also obtain a closed expressions for $\Delta$ arbitrary:
\begin{multline}
    \gamma_{0,\ell}=-\frac{2(\Delta -1)^2 (-d+4 \Delta +2 \ell+2)}{(\Delta +\ell)^2 (-d+4 \Delta +2 \ell)}\\\times\, _6F_5\left(\begin{matrix}1,1,-\frac{d}{2}+\Delta +1,-\frac{d}{2}+\Delta +1,-\frac{d}{4}+\frac{\ell}{2}+\Delta +\frac{3}{2},\ell+2 \Delta -1\\3-\frac{d}{2},-\frac{d}{4}+\frac{\ell}{2}+\Delta +\frac{1}{2},\ell+\Delta +1,\ell+\Delta +1,-\frac{d}{2}+\ell+2 \Delta +1\end{matrix};1\right)\,.
\end{multline}
For $\tau=1$ and $\Delta=1$ with $d$ arbitrary we get instead:
\begin{subequations}
\begin{align}
    \gamma_{0,\ell}=&-\frac{2\,\alpha_1}{\pi ^2}\frac{(\ell!)^2 \Gamma \left(-\frac{d}{2}+\ell+2\right) \Gamma \left(-\frac{d}{2}+\ell+3\right)}{\Gamma \left(\ell+\frac{3}{2}\right)^2 \Gamma \left(-\frac{d}{2}+\ell+\frac{5}{2}\right)^2}\\\nonumber
    &\times\, _7F_6\left(\begin{matrix}\frac{1}{2},\frac{1}{2},\frac{3}{2}-\frac{d}{2},\frac{3}{2}-\frac{d}{2},-\frac{d}{4}+\frac{\ell}{2}+2,\ell+1,-\frac{d}{2}+\ell+2\\2-\frac{d}{2},-\frac{d}{4}+\frac{\ell}{2}+1,\ell+\frac{3}{2},\ell+\frac{3}{2},-\frac{d}{2}+\ell+\frac{5}{2},-\frac{d}{2}+\ell+\frac{5}{2}\end{matrix};1\right)\,\\
    &\vdots\,.\nonumber
\end{align}
\end{subequations}
and so on for higher $n$ according to \eqref{N>0}.

\subsection{Spinning correlators}\label{sec::spinningcorrelators}

Here we consider external operators with spin, focusing primarily on four-point correlators of the type \eqref{0O1OOO2} in the context of theories with slightly broken higher-spin symmetry. In this case there is indeed a possibility for double-trace operator mixing \eqref{0mix} at ${\cal O}\left(1/N\right)$ owing to the twist degeneracy of the tower of higher-spin currents (which in this discussion we denote by ${\cal O}_J$) at ${\cal O}\left(1\right)$, which have twist $\tau^\prime=d-2+{\cal O}\left(1/N\right)$ and $J=2,4,6,...\,$. 

In the following we work in $d=3$ with $\Delta=1$ and $\tau^\prime=1$, as relevant for the quasi Boson theory. We shall not consider the contribution of exchanges of spinning operators in the crossed channels, which are proportional to conformal structures different from those of the free boson at ${\cal O}\left(1/N\right)$.\footnote{I.e. we are not considering the deformation of the OPE structure $\left\langle\mathcal{O}_{J_1}\mathcal{O}_{J_2}\mathcal{O}\right\rangle$ at this order. In particular, in the presence of a Chern-Simons term, the $J_1$-$J_2$-$0$ conformal structure is deformed by some parity odd structure \cite{Maldacena:2012sf} and such additional 3d conformal structures will induce a term in the 4pt correlator proportional to the $\epsilon_3$ tensor on top of the parity preserving term studied here. It has to be verified how this term modifies to the anomalous dimensions in general.} For the correlator \eqref{0O1OOO2}, when $J_1=0$ or $J_2=0$ the crossed-channel scalar crossing kernels contribute to double-trace anomalous dimensions at this order, which are proportional to the deformation parameter $\alpha_2$ defined in \eqref{alphacoeffs}. From equation \eqref{DTANOMspin} we have
\begin{multline}
    c^{(0)}_{\mathcal{O}_{J_1}\mathcal{O}_{J_2}[\mathcal{O}_{J_1}\mathcal{O}_{J_2}]}c^{(0)}_{\mathcal{O}\mathcal{O}[\mathcal{O}\mathcal{O}]}\gamma_{0,\ell}=c_{\mathcal{O}_{J_1}\mathcal{O}\mathcal{O}}c_{\mathcal{O}_{J_2}\mathcal{O}\mathcal{O}}\frac{\ell!\, 2^{\ell-J_1-J_2+1} \Gamma \left(J_2+\ell+\frac{1}{2}\right) \Gamma (J_1-J_2+\ell+1)}{\pi\,(2\ell)!\,  \Gamma \left(J_1+\frac{1}{2}\right) \Gamma \left(J_2+\frac{1}{2}\right) \Gamma \left(\frac32-J_2+\ell\right)}\\\times\,_4F_3\left(\begin{matrix}\frac{1}{2}-J_1,J_1+\frac{1}{2},\frac{1}{2}-J_2,-J_2\\\frac{1}{2},\frac{1}{2}-J_2-\ell,\frac{3}{2}-J_2+\ell\end{matrix};1\right)\,,
\end{multline}
where $c_{\mathcal{O}_{J_1}\mathcal{O}\mathcal{O}}$ and $c_{\mathcal{O}_{J_2}\mathcal{O}\mathcal{O}}$ are the OPE coefficients of the crossed channel conformal blocks. Setting $J_2=0$ we can also use \eqref{0spinningMean} to normalise the result, obtaining:
\begin{equation}
\gamma_{0,\ell}=-\alpha_2\,c_{\mathcal{O}_{J_1}\mathcal{O}\mathcal{O}}\frac{2^{2-\frac{J_1}{2}} \sqrt{\Gamma (2 J_1+1)}}{\pi ^{3/2} \Gamma \left(J_1+\frac{1}{2}\right)}\frac{1}{2\ell+1}\,,
\end{equation}
Remarkably, this result displays the same $\ell$ dependence regardless of $J_1$ and reduces to \eqref{scalar} for $J_1=0$. Including the $J_1$ dependence in the free Boson OPE coefficients in the crossed channels \cite{Diaz:2006nm}:
\begin{equation}
c_{\mathcal{O}_{J_1}\mathcal{O}\mathcal{O}}=\frac{1}{\sqrt{N}}\frac{2^{2-\tfrac{J_1}2}}{\pi^{1/4}} \sqrt{\frac{\Gamma \left(J_1+\frac{1}{2}\right)}{\Gamma (J_1+1)}}\,,
\end{equation}
the above result further simplifies to:\footnote{Note that the spin $\ell$ of the double-trace operator $\left[{\cal O}_{J_1}{\cal O}\right]_{\ell}$ satisfies by definition $\ell \geq J_1$.}
\begin{equation}\label{gammaL}
\gamma_{0,\ell}=-\frac{16}{\pi ^2}\frac{\alpha_2}{ 2 \ell+1}\,,\qquad \ell \geq J_1,
\end{equation}
which no longer depends on $J_1$. Instead, for correlators 
\begin{equation}\label{1o2o}
    \langle {\cal O}_{J_1}\left(x_1\right){\cal O}\left(x_2\right){\cal O}_{J_2}\left(x_3\right){\cal O}\left(x_4\right) \rangle,
\end{equation}
the $\sf{u}$-channel crossing kernels can induce non-trivial double-trace operator anomalous dimensions at ${\cal O}\left(1/N\right)$ with both $J_1$ and $J_2$ non-zero.\footnote{Similarly, for correlators of the type
\begin{equation}\label{1oo2}
    \langle {\cal O}_{J_1}\left(x_1\right){\cal O}\left(x_2\right){\cal O}\left(x_3\right){\cal O}_{J_2}\left(x_4\right) \rangle,
\end{equation}
 such contributions to the double-trace anomalous dimensions come from ${\sf t}$-channel crossing kernels.} In the following we take $J_1=J_2=J$. Complete expressions for such crossing kernels were obtained in \cite{Sleight:2018epi} for arbitrary $\ell$ up to $J=2$, and we quote the result for $J=1$ in \eqref{Jeq1}. The shadow projection discussed in this note can be applied with little effort, but the corresponding expressions look rather lengthy in general. They however simplify drastically in some notable examples, such as the $d=3$ case under consideration with $\tau^\prime=1$ and $\Delta=1$. Considering the contribution of a scalar exchange in the ${\sf u}$-channel, we obtain: 
\begin{equation}
    \gamma_{0,\ell}=-\frac{16}{\pi ^2}\,\frac{\alpha_2}{2 \ell+1}c_{\mathcal{O}_J\mathcal{O}_J\mathcal{O}},
\end{equation}
for $J=1$, and
\begin{equation}
   \gamma_{0,\ell}=-\frac{64}{3 \pi ^2}\,\frac{\alpha_2}{2 \ell+1}c_{\mathcal{O}_J\mathcal{O}_J\mathcal{O}},
\end{equation}
for $J=2$. Plugging in the expression for the corresponding free scalar OPE coefficients \cite{Sleight:2016dba}
\begin{equation}
c_{\mathcal{O}_J\mathcal{O}_J\mathcal{O}}=\frac{1}{\sqrt{N}}\frac{2^{\frac{5}{2}-J} \Gamma \left(J+\frac{1}{2}\right)}{\sqrt{\pi } \Gamma (J+1)},
\end{equation}
we then obtain
\begin{align}
J&=1\,,& \gamma_{0,\ell}&=-\frac{8\sqrt{2}}{\pi^2}\frac{\alpha_2}{2\ell+1}\,,& \ell&\geq1 \,,\\
J&=2\,,& \gamma_{0,\ell}&=-\frac{8\sqrt{2}}{\pi^2}\frac{\alpha_2}{2\ell+1}\,,& \ell&\geq2\,.
\end{align}
The above result is remarkably simple and suggests that, as in the case of \eqref{gammaL}, the dependence on $J$ disappears from the averages \eqref{0avgamma} while the $\ell$-dependence is overall in the mixing matrix. This result, if confirmed by explicit computation and extended (plausibly) to more general correlators $\left\langle \mathcal{O}_{J_1}\mathcal{O}\mathcal{O}_{J_2}\mathcal{O}\right\rangle$ using the methods of \cite{Sleight:2018epi}, would produce a very simple mixing matrix whose entries are independent on $\ell$ for the $[\mathcal{O}_J\mathcal{O}]_\ell$ double-trace operators. It would be very interesting to explicitly verify whether or not this statement holds when also taking into account the odd conformal structure.\footnote{It is tempting to argue the odd structure will only affect the anomalous dimensions by an overall $\lambda$-dependent factor at this order due to higher-spin symmetry.}

\section*{Acknowledgments}

We are grateful to Ofer Aharony and Fernando Alday for useful correspondence, as well as Shai Chester, Alessandro Vichi and Yifan Wang for useful discussions. We thank Xinan Zhou for bringing some typos to our attention. M.T. thanks Caltech and the Simons Bootstrap Collaboration for hospitality and support during the Bootstrap 2018 workshop. C.S. holds a Marina Solvay Fellowship and was partially partially supported by the European Union's Horizon 2020 research and innovation programme under the Marie Sklodowska-Curie grant agreement No 793661. The
research of M.T. is partially supported by the European Union's Horizon 2020 research and innovation programme under the Marie Sklodowska-Curie grant agreement No 747228 and by the Russian Science Foundation grant 14-42-00047 in association with Lebedev Physical Institute.

\begin{appendix}

\section{OPE data from $6j$ symbols}\label{6jsymbols}

In this appendix we discuss the Mellin Barnes integral representation for $6j$ symbols of the conformal group and their relation to OPE data. The residues of $6j$ symbols encode the coefficients of the conformal block expansion of a given CPW into another channel. These singularities were extracted in \cite{Sleight:2018epi} in general $d$ for both spinning external operators and spinning exchanged operators, which was done by directly employing the orthogonality properties \cite{Costa:2012cb,Gopakumar:2016cpb} of conformal partial waves in Mellin space. That approach is partially reviewed in \S \ref{subsec::approach} (see e.g. equation \eqref{cpwccint}) and some of the results for the crossing kernels are reviewed in appendix \ref{sec::crossingkernels}. Here we discuss the extraction of the conformal block expansion coefficients directly from the $6j$ symbols and how this is related to the procedure of \cite{Sleight:2018epi}.

Writing an expression for $6j$ symbols is straightforward using Mellin space. A bit more work has to be done to extract the OPE data from a given expression for a $6j$ symbol, which entails evaluating a spectral integral. To this end it is convenient to use their representation as Mellin Barnes integrals, as we discuss below. As far as we are aware, this representation of $6j$ symbols was first adopted for scalar principal series representations by Krasnov and Louko in \cite{Krasnov:2005fu}.

$6j$ symbols can be defined quite generally for arbitrary totally symmetric operator representations as:
\begin{align}\label{6jdef}
    \left\{\begin{matrix}
    \mathcal{O}_{\tau_1,\ell_1}& \mathcal{O}_{\tau_2,\ell_2}& \mathcal{O}_{\tau,\ell^\prime}\\
    \mathcal{O}_{\tau_3,\ell_3}&\mathcal{O}_{\tau_4,\ell_4}&\mathcal{O}_{\tilde{\tau},\ell}    \end{matrix}\right\}^{{\bf n},{\bf \bar n};{\bf m},{\bf \bar m}}&\\&\hspace{-50pt}=\int d^dx_1d^dx_2d^dx_3d^dx_4\,\text{Tr}\left[{}^{(\sf{s})}\bar{\mathcal{F}}^{{\bf n},{\bf \bar n}}_{d-\bar{\tau}-2\ell,\ell}(x_i;z_i)\ {}^{(\sf{t})}{\mathcal{F}}^{{\bf m},{\bf \bar m}}_{\tau,\ell^\prime}(x_i;z_i)\right]\nonumber\\
    &\hspace{-50pt}\equiv\left\langle {}^{(\sf{s})}\bar{\mathcal{F}}^{{\bf n},{\bf \bar n}}_{d-\bar{\tau}-2\ell,\ell}(x_i;z_i)\Big|\ {}^{(\sf{t})}{\mathcal{F}}^{{\bf m},{\bf \bar m}}_{\tau,\ell^\prime}(x_i;z_i)\right\rangle\,,
\end{align}
which is the projection of a ${\sf t}$-channel CPW onto a ${\sf s}$-channel CPW. Above we combined the conventions and notations of \cite{Krasnov:2005fu} for $6j$ symbols and \cite{Sleight:2018epi} for spinning CPWs. The $\bf{n}$, $\bf{\bar{n}}$, $\bf{m}$, $\bf{\bar{m}}$ label the spinning 3pt conformal structures entering each CPW, see e.g. subsection 2.2 of \cite{Sleight:2018epi}. The notation ${}^{(\sf{t})}{\mathcal{F}}^{{\bf m},{\bf \bar m}}_{\tau,\ell^\prime}(x_i;z_i)$ refers to a ${\sf t}$-channel CPW with external operators $\mathcal{O}_{\tau_1,\ell_1}$, $\mathcal{O}_{\tau_2,\ell_2}$, $\mathcal{O}_{\tau_3,\ell_3}$, $\mathcal{O}_{\tau_4,\ell_4}$ and internal operator $\mathcal{O}_{\tau,\ell^\prime}$, while ${}^{(\sf{s})}\bar{\cal F}^{{\bf n},{\bf \bar n}}_{d-\bar{\tau}-2\ell,\ell}(x_i;z_i)$ is the $\sf{s}$-channel CPW for dual external and internal operators of dimension $d-\Delta_i$.\footnote{Note that in the spinning case this will also require to determine the corresponding shadow 3pt conformal structures for each $\bf{n}$ and $\bf{\bar{n}}$. For the shadow transform of $0$-$0$-$\ell$ 3pt conformal structures, the shadow structures coincide with the original structures while in the general spinning case $\ell_1$-$\ell_2$-$\ell_3$ they where computed in \cite{Sleight:2018epi} (see Appendix A of that paper).} The shadow of the ${\sf s}$-channel CPW appears in the above expression since we are taking the adjoint with respect to the conformal invariant bilinear form. The $\left(z_i\right)_\mu$ are the standard auxiliary vectors which package the indices of each spinning CPW. The trace operation is the traceless contraction of all indices labelled by $z_i$, as implemented for instance in \cite{Giombi:2017hpr,Sleight:2018epi}.

A simple and straightforward strategy to evaluate the integrals in \eqref{6jdef} is to re-cast each CPW as a Mellin integral: 
\begin{subequations}\label{scmrep}
\begin{align}
     {}^{(i)}\mathcal{F}_{\tau,\ell}\left(y_i\right) &=\tfrac{1}{\left(y_{12}^2\right)^{\frac{1}{2}(\tau_1 + \tau_2)} \left(y_{34}^2\right)^{\frac{1}{2}(\tau_3 + \tau_4)}}\left(\tfrac{y_{24}^2}{y_{14}^2}\right)^{\tfrac{\tau_1-\tau_2}2}\left(\tfrac{y_{14}^2}{y_{13}^2}\right)^{\tfrac{\tau_3-\tau_4}2}  {}^{(i)}\mathcal{F}_{\tau,\ell}\left(u,v\right),
     \\ 
   {}^{(i)}\mathcal{F}_{\tau,\ell}\left(u,v\right) &= \int \frac{ds\, dt}{(4\pi i)^2}\,u^{t/2}v^{-(s+t)/2}\rho_{\left\{\tau_i\right\}}\left(s,t\right)\, {}^{( i)}\mathcal{F}_{\tau,\ell}\left(s,t\right),
   \end{align}
\end{subequations}
with Mellin measure
\begin{align}\label{rhoM}
\rho_{\left\{\tau_i\right\}}\left(s,t\right)&=\Gamma \left(\tfrac{-t+\tau_1+\tau_2}2\right) \Gamma \left(\tfrac{-t+\tau_3+\tau_4}2\right)\nonumber\\&\hspace{100pt}\times\Gamma \left(\tfrac{s+t}{2}\right) \Gamma \left(\tfrac{-s-\tau_1+\tau_2}2\right) \Gamma \left(\tfrac{-s+\tau_3-\tau_4}2\right) \Gamma \left(\tfrac{ s+t+\tau_1-\tau_2-\tau_3+\tau_4}2\right),
\end{align}
and we are considering ${i}={\sf s}$, ${\sf t}$, where 
\begin{equation}
    {}^{(\sf{t})}\mathcal{F}_{\tau_1,\tau_2,\tau_3,\tau_4|\tau,\ell}(s,t)={}^{(\sf{s})}\mathcal{F}_{\tau_1,\tau_4,\tau_3,\tau_2|\tau,\ell}(s-\tau_2+\tau_4,-s-t+\tau_2+\tau_3).
\end{equation}
In this way the conformal integrals in \eqref{6jdef} trivialise and give just:
\begin{align}\label{Ident}
    %\int\frac{ds\,dt\,d\bar{s}\,d\bar{t}}{(2\pi i)^4}
    \int d^dx_1 d^dx_2 d^dx_3 d^dx_4\,\frac{u^{(t+\bar{t})/2}v^{-(s+t+\bar{s}+\bar{t})/2}}{\left(y_{12}^2\right)^{d} \left(y_{34}^2\right)^{d}}=\frac{\pi ^{d/2} \Gamma \left(\frac{d+s+\bar{s}}{2}\right) \Gamma \left(\frac{t+\bar{t}-d}{2}\right) \Gamma \left(\frac{d-s-\bar{s}-t-\bar{t}}{2}\right)}{\Gamma \left(-\frac{s}{2}-\frac{\bar{s}}{2}\right) \Gamma \left(d-\frac{t}{2}-\frac{\bar{t}}{2}\right) \Gamma \left(\frac{s+\bar{s}+t+\bar{t}}{2}\right)}\,.
\end{align}
In the case of external spinning legs, after taking the trace, one is left with analogous scalar integrals which can be evaluated in Mellin space in exactly the same way.

One can then combine \eqref{scmrep} with \eqref{6jdef}, and after performing the conformal integral with \eqref{Ident} one ends up with the Mellin Barnes integral representation 
\begin{multline}\label{6jscalar}
\left\{\begin{matrix}
    \mathcal{O}_{\tau_1}& \mathcal{O}_{\tau_2}& \mathcal{O}_{\tau,\ell^\prime}\\
    \mathcal{O}_{\tau_3}&\mathcal{O}_{\tau_4}&\mathcal{O}_{\tilde{\tau},\ell}    \end{matrix}\right\}=\pi^{d/2}\int\frac{ds\,dt\,d\bar{s}\,d\bar{t}}{(2\pi i)^4}\frac{\Gamma \left(\frac{d+s+\bar{s}}{2}\right) \Gamma \left(\frac{t+\bar{t}-d}{2}\right) \Gamma \left(\frac{d-s-\bar{s}-t-\bar{t}}{2}\right)}{\Gamma \left(-\frac{s}{2}-\frac{\bar{s}}{2}\right) \Gamma \left(d-\frac{t}{2}-\frac{\bar{t}}{2}\right) \Gamma \left(\frac{s+\bar{s}+t+\bar{t}}{2}\right)}\\
    \rho_{\left\{d-\tau_i\right\}}(\bar{s},\bar{t}) \,{}^{(\sf{s})}\mathcal{F}_{d-\tau_1,d-\tau_2,d-\tau_3,d-\tau_4|d-\bar{\tau}-2\ell,\ell}(\bar{s},\bar{t})\,\underbrace{\rho_{\left\{\tau_i\right\}}(s,t)\,{}^{(\sf{t})}{\mathcal{F}}_{\tau_1,\tau_2,\tau_3,\tau_4|\tau,\ell^\prime}(s,t)}_{M^{(\sf{t})}(s,t)}\,.
\end{multline}
The Mellin representation of CPWs can be conveniently given in terms of Mack polynomials \cite{Mack:2009mi}
\begin{align}\label{mrepexsc}
    {}^{(\sf s)}\mathcal{F}_{\tau_1,\tau_2,\tau_3,\tau_4|\tau,\ell}\left(s,t\right)&= \mathcal{C}_{\ell,\tau}(\tau_i)\,\Omega_{\ell}(t)\, {}^{(\sf s)}P_{\ell,\tau}(s,t)\,,
\end{align}
where $\mathcal{C}_{\ell,\tau}(\tau_i)$ is a coefficient (given in our conventions by equation (3.7) in \cite{Sleight:2018epi}) and 
\begin{equation}
    \Omega_\ell(t)=\frac{\Gamma \left(\tfrac{\tau -t}{2}\right) \Gamma \left(\tfrac{d-2 \ell-t-\tau}{2}\right)}{\Gamma \left(\tfrac{-t+\tau_1+\tau_2}{2}\right) \Gamma \left(\tfrac{-t+\tau_3+\tau_4}{2}\right)}.
\end{equation}
The Mack polynomial ${}^{(\sf s)}P_{\ell,\tau}(s,t)$ is a degree $\ell$ polynomial in the Mellin variables $s$ and $t$. Explicit expressions for Mack polynomials are complicated in general, however there is a crucial simplification for $t=\tau$, when they are given by
orthogonal polynomials \cite{Costa:2012cb}
\begin{equation}\label{Qpolyapp}
    {\cal Q}_{\ell,\tau}\left(s\right) = \,\mathcal{C}_{\ell,\tau}(\tau_i)\,\Gamma \left(\tfrac{d}{2}-\ell-\tau \right)\, {}^{({\sf s})}P_{\ell,\tau}(s,t=\tau),
\end{equation}
which are related to continuous Hahn polynomials as given in equation \eqref{mackpoly}.

Let us make some comments:

\begin{itemize}
    \item From the expression \eqref{6jscalar}, $6j$ symbols appear to be generally quite complicated functions which may simplify in some cases.\footnote{\label{foo::2}Simplifications also arise when instead of a CPW one considers for $M^{({\sf t})}\left(s,t\right)$ in \eqref{6jscalar} simple examples of CFT correlators proportional to distributions in Mellin space. This is the case for Mean Field Theories or for instance the free boson and the free fermion theories. In this case the Mellin integrations are reduced from four to two. An example of such will be demonstrated in the sequel.} The residues of the $6j$ symbol in $\bar{\tau}$ encode the OPE data of the expansion of a ${\sf s}$-channel CPW into the ${\sf t}$-channel. The residues can be evaluated directly from the Mellin integral expression \eqref{6jscalar} by identifying the points in which the integration contours are pinched, thus collapsing some of the Mellin integrals to a single point. At this point the ${\sf s}$-channel CPW in \eqref{6jscalar} reduces to a continuous Hahn polynomial \eqref{Qpolyapp}, as we shall see more concretely below.\footnote{In particular, for the $\ell=0$ case we consider in \eqref{scalarrep6japp} we evaluate residues at ${\bar t}=d-\bar{\tau}$ which, via \eqref{Qpolyapp}, gives rise to a continuous Hahn polynomial.} One can then evaluate the remaining Mellin integrals via Barnes Lemmas which gives an expression for the residues as a sum of ${}_4F_3$ hypergeometric functions as in \cite{Sleight:2018epi}.
    
    \item The Mellin representation \eqref{6jscalar} is a transparent way to express $6j$ symbols which naturally encodes the residues for each conformal module in families of Gamma function poles. We shall demonstrate this below by extracting the residues of $6j$ symbols for scalar representations, matching the corresponding result obtained in \cite{Sleight:2018epi}. The representation \eqref{6jscalar} and the results we derive from it moreover hold for general $d$.

\end{itemize}

Before considering the extraction of residues of $6j$ symbols from their Mellin Barnes representation \eqref{6jscalar}, let us first consider a simpler example (mentioned in footnote \ref{foo::2}) to demonstrate how to extract residues in $\bar{\tau}$ from Mellin Barnes integrals of the type \eqref{6jscalar}.

\paragraph{Mean Field Theory OPE coefficients}

Here we consider the mean field theory correlation function 
\begin{subequations}
\begin{align}
    \langle {\cal O}_{\Delta}\left(x_1\right){\cal O}_{\Delta}\left(x_2\right){\cal O}_{\Delta}\left(x_3\right){\cal O}_{\Delta}\left(x_4\right) \rangle &= \frac{{\cal A}_{\text{MFT}}\left(u,v\right)}{\left(x^2_{12}x^{2}_{34}\right)^{\Delta}} \\
    {\cal A}_{\text{MFT}}\left(u,v\right) &= 1+u^{\Delta}+\left(\frac{u}{v}\right)^{\Delta}
\end{align}
\end{subequations}
for scalar operators ${\cal O}_{\Delta}$ of scaling dimension $\Delta$. The Mellin transform is given by a distribution \cite{Bekaert:2016ezc,Taronna:2016ats}
\begin{equation}
    M_{\text{MFT}}\left(s,t\right)=M^{({\sf s})}_{\text{MFT}}\left(s,t\right)+M^{({\sf t})}_{\text{MFT}}\left(s,t\right)+M^{({\sf u})}_{\text{MFT}}\left(s,t\right)
\end{equation}
with
\begin{subequations}
\begin{align}
    M^{({\sf s})}_{\text{MFT}}\left(s,t\right)&=\delta\left(s\right)\delta\left(t\right)\\
    M^{({\sf t})}_{\text{MFT}}\left(s,t\right)&=\delta\left(s+2\Delta\right)\delta\left(t-2\Delta\right)\\
    M^{({\sf u})}_{\text{MFT}}\left(s,t\right)&=\delta\left(s\right)\delta\left(t-2\Delta\right).
\end{align}
\end{subequations}
The expansion of $M^{({\sf t})}\left(s,t\right)$ and $M^{({\sf u})}\left(s,t\right)$ in the ${\sf s}$-channel consists of double-trace operator $\left[{\cal O}_{\Delta}{\cal O}_{\Delta}\right]_{n,\ell}$ contributions of twist $2\Delta+2n$. This means that \eqref{6jscalar} with $M^{({\sf t})}\left(s,t\right)=M^{({\sf t})}_{\text{MFT}}\left(s,t\right)$ or $M^{({\sf u})}_{\text{MFT}}\left(s,t\right)$ will have poles at ${\bar \tau}+\ell=(2\Delta+2n)-\ell$, which give the mean field theory OPE coefficients \eqref{OPEnl}.

The ${\sf s}$-channel contributions of double-trace operators $\left[{\cal O}_{\Delta}{\cal O}_{\Delta}\right]_{n,\ell}$ in $M^{({\sf t})}_{\text{MFT}}\left(s,t\right)$ and $M^{({\sf u})}_{\text{MFT}}\left(s,t\right)$ differ by a factor of $(-1)^\ell$, so from this point we shall focus on $M^{({\sf t})}_{\text{MFT}}\left(s,t\right)$. We shall also only consider the contributions of scalar double-trace operators ($\ell=0$) for simplicity.

Since the Mellin representation of the mean field theory correlator is a distribution in $s$ and $t$, the integration over these variables in \eqref{6jscalar} is trivial and we obtain
\begin{multline}\label{mftint}
    \left\langle {}^{(\sf{s})}\bar{\mathcal{F}}^{{\bf 0},{\bf \bar 0}}_{d-\bar{\tau},0}\Big|M^{(\sf t)}_{\text{MFT}}\right\rangle\\=\int\frac{d\bar{s}\,d\bar{t}}{(2\pi i)^2}\,\underbrace{\tfrac{\pi ^{d/2} \Gamma (d-\bar{\tau})\Gamma \left(-\frac{\bar{s}}{2}\right) \Gamma \left(\frac{d+\bar{s}}{2}\right) \Gamma \left(\frac{\bar{s}-\bar{t}}{2}\right)^2  \Gamma \left(-\frac{d}{2}+\Delta -\frac{\bar{t}}{2}\right) \Gamma \left(\frac{\bar{t}+d-\bar{\tau}}{2}\right) \Gamma \left(\frac{\bar{t}+\bar{\tau}}{2}\right) \Gamma \left(\frac{d-\bar{s}+\bar{t}-2 \Delta}{2}\right)}{4\Gamma \left(\frac{\bar{\tau}}{2}\right)^2 \Gamma \left(\bar{\tau}-\frac{d}{2}\right) \Gamma \left(\frac{d-\bar{\tau}}{2}\right)^2 \Gamma \left(d+\frac{\bar{t}}{2}-\Delta \right) \Gamma \left(\frac{\bar{s}-\bar{t}}{2}+\Delta \right)}}_{\mathcal{I}}\,.
\end{multline}
The above Mellin-Barnes form makes the evaluation of residues in $\bar{\tau}$ rather straightforward. Poles in ${\bar \tau}$ arise when the integration contour is pinched between poles in the Mellin variable $\bar{t}$ belonging to two different families of Gamma function poles: One with poles at $\bar{t}=\alpha+n$ and the other with poles at $\bar{t}=\beta-n$, where $n=0,1,2,...$ and $\alpha$ and $\beta$ are real numbers with at least one of them depending on $\bar{\tau}$.\footnote{Note that, in studying the pole pinching in $\bar{\tau}$, it is convenient to first perform the integral in $\bar{t}$. When performing this integral one can, without loss of generality, analytically continue $\bar{s}$ in a region such that no pole pinching can arise for all $\Gamma$ functions which depend on $\bar{s}$. Evaluating the integral in such regime is automatically consistent with analytic continuation in $\bar{s}$ when evaluating the $\bar{s}$-integral.}

Of the three Gamma functions that depend only on ${\bar t}$ in the numerator of integrand in \eqref{mftint}, there are two possible types of pole pinching. The first is between the Gamma function pair
\begin{equation}\label{nonshadopinch}
     \Gamma \left(-\frac{d}{2}+\Delta -\frac{\bar{t}}{2}\right) \Gamma \left(\frac{\bar{t}+d-\bar{\tau}}{2}\right),
\end{equation}
which exhibits pole pinching for $\bar \tau=2\Delta+2n$ and whose poles in ${\bar t}$ consequently give rise to single poles at $\bar \tau=2\Delta+2n$. The second is
\begin{equation}
     \Gamma \left(-\frac{d}{2}+\Delta -\frac{\bar{t}}{2}\right) \Gamma \left(\frac{\bar{t}+\bar{\tau}}{2}\right),
\end{equation}
giving single poles at $\bar \tau=d-2\Delta-2n$ corresponding to the shadow conformal multiplets. We will focus on the non-shadow poles \eqref{nonshadopinch}.

The way to evaluate the residue at $\bar{\tau}=2\Delta+2n$ just requires to shift the contour across the $\bar{t}$ poles that would pinch the contour when $\bar{\tau}=2\Delta+2n$. This requires to evaluate the residues of the finite  number of $\bar{t}$ poles that are crossed in shifting the contour, and in this way one removes the pinching when $\bar{\tau}\to 2\Delta+2n$. Evaluating the $\bar{t}$ residues:
\begin{equation}
    \mathcal{I}^{(m)}=\text{Res}_{\bar{t}=\bar{\tau}-d-2m}\left[\mathcal{I}\right],
\end{equation}
we obtain: 
\begin{subequations}
\begin{align}
   a_{0,n} &=\text{Res}_{\bar{\tau}=2\Delta+2n}\left[N_{0}(\bar{\tau})^{-1}\left\langle {}^{(\sf{s})}\bar{\mathcal{F}}^{{\bf 0},{\bf \bar 0}}_{d-\bar{\tau},0}\Big|M^{({\sf t})}_{\text{MFT}}\right\rangle\right]\\&=\int\frac{d\bar{s}}{2\pi i}\,\text{Res}_{\bar{\tau}=2\Delta+2n}\left[N_{0,n}(\bar{\tau})^{-1}\sum_{m=0}^n\mathcal{I}^{(m)}\right]\,,
\end{align}
\end{subequations}
where we have included the normalisation factor:
\begin{equation}\label{Normal}
    N_{0}(\bar{\tau})=\frac{\pi ^{d/2} \Gamma \left(\frac{d}{2}-\frac{\bar{\tau}}{2}\right)^2}{\Gamma \left(\frac{d}{2}\right) \Gamma \left(\frac{\bar{\tau}}{2}\right)^2}
\end{equation}
and where the leftover Mellin integral reduces to an integral of the type:
\begin{align}
    \int\frac{d\bar{s}}{2\pi i}\,\Gamma \left(-\tfrac{\bar{s}}{2}\right)^2 \Gamma \left(\tfrac{d+\bar{s}-2 \Delta -2n}{2}\right)^2\,p(\bar{s}),\\
    p(\bar{s})=\frac{\text{Res}_{\bar{\tau}=2\Delta+2n}\left[N_{0,n}(\bar{\tau})^{-1}\sum_{m=0}^n\mathcal{I}^{(m)}\right]}{\Gamma \left(-\tfrac{\bar{s}}{2}\right)^2 \Gamma \left(\tfrac{d+\bar{s}-2 \Delta -2n}{2}\right)^2}\,,
\end{align}
with $p(\bar{s})$ a polynomial in the variable $\bar{s}$. We can therefore evaluate the leftover integral by simple applications of the first Barnes lemma. Combining everything, performing the integrations and taking into account the normalisation to read off the actual OPE coefficient we finally recover the known Mean-field theory OPE \cite{Dolan:2000ut,Heemskerk:2009pn,Fitzpatrick:2011dm}:
\begin{equation}\label{mftopeapp}
    a_{0,n}=\frac{(-1)^n (\Delta )_n^2 \left(-\frac{d}{2}+\Delta +1\right)_n^2}{n! \left(\frac{d}{2}\right)_n \left(-\frac{d}{2}+n+2 \Delta \right)_n (d-2 (n+\Delta ))_n}\,.
\end{equation}
It is interesting to contrast the above computation with the approach taken in \cite{Sleight:2018epi}, which effectively re-sums the above residues via the action of the twist block operator (this is briefly reviewed in \S \ref{subsec::subleading}). This automatically subtracts away the contributions from descendants of subleading double-twist operators. In the latter equivalent approach the above anomalous dimensions are neatly re-summed into continuous Hahn polynomials evaluated at specific points.

\paragraph{Residues of scalar $6j$ symbols}

Another instructive example is given by the scalar $6j$ symbol, setting again for simplicity $\tau_i=\Delta$. In this case, from \eqref{6jscalar} we have the following Mellin integral:
\begin{multline}\label{scalarrep6japp}
\left\{\begin{matrix}
\mathcal{O}_{\tau_1}& \mathcal{O}_{\tau_2}& \mathcal{O}_{\tau}\\
\mathcal{O}_{\tau_3}&\mathcal{O}_{\tau_4}&\mathcal{O}_{\tilde{\tau}}    \end{matrix}\right\}=\frac{\pi ^{d/2}\Gamma (d-\bar{\tau})\Gamma (\tau )}{16\,\Gamma \left(\frac{\tau }{2}\right)^4 \Gamma \left(\frac{\bar{\tau}}{2}\right)^2 \Gamma \left(\frac{d}{2}-\tau \right) \Gamma \left(\frac{d-\bar{\tau}}{2}\right)^2 \Gamma \left(\bar{\tau}-\frac{d}{2}\right)}\\\times\,\int \frac{ds\,dt\,d\bar{s}\,d\bar{t}}{(2\pi i)^4}\,\frac{\Gamma \left(\frac{d+s+\bar{s}}{2}\right) \Gamma \left(\frac{-d+t+\bar{t}}{2}\right)\Gamma \left(\frac{d-s-\bar{s}-t-\bar{t}}{2}\right)}{\Gamma \left(-\frac{s+\bar{s}}{2}\right) \Gamma \left(d-\frac{t}{2}-\frac{\bar{t}}{2}\right) \Gamma \left(\frac{s+\bar{s}+t+\bar{t}}{2} \right)}\\\times\Gamma \left(-\tfrac{s}{2}\right)^2 \Gamma \left(-\tfrac{\bar{s}}{2}\right)^2 \Gamma \left(\tfrac{\bar{\tau}-\bar{t}}{2}\right)\Gamma \left(\tfrac{d-\bar{t}-\bar{\tau}}{2}\right) \Gamma \left(\tfrac{\bar{s}+\bar{t}}{2}\right)^2 \Gamma \left(\tfrac{2 \Delta -t}{2} \right)^2 \Gamma \left(\tfrac{s+t-2 \Delta +\tau}{2}\right)  \Gamma \left(\tfrac{d+s+t-2 \Delta -\tau}{2}\right)\,.
\end{multline}
To identify the poles in $\bar{\tau}$ from the above it is not much more involved that for the MFT case considered previously although there are two more Mellin integrals to handle. The poles at ${\bar \tau}=2\Delta+2n$ arise from the pinching of the following four $\Gamma$-functions:
\begin{equation}
\Gamma \left(\tfrac{-d+t+\bar{t}}{2}\right)\Gamma \left(\tfrac{d-\bar{t}-\bar{\tau}}{2}\right) \Gamma \left(\tfrac{2 \Delta -t}{2} \right)^2\,.
\end{equation}
Imposing that the $t$ and $\bar{t}$ contour separates the poles of the above $\Gamma$-functions requires:
\begin{align}
\text{Re}(\bar{t})&<d-\bar{\tau}\,,& \text{Re}(\bar{t})&>d-\text{Re}(t)\,,& \text{Re}(t)<2\Delta\,.
\end{align}
Therefore the $\bar{t}$ contour has to lie at
\begin{align}
\text{Re}(\bar{t})&<d-\bar{\tau}\,,& \text{Re}(\bar{t})&>d-2\Delta\,,
\end{align}
which implies contour pinching for $\bar{\tau}=2\Delta+2n$. 

Let us focus on leading double-twist operators for ease of presentation, which have $n=0$. As for the MFT example, when contour pinching arises one should simply evaluate the residues of the offending poles and shift the contour to remove the pinching. The $n>0$ case thus follows in the same way as for the $n=0$ case, just there are more residues to evaluate. For the case at hand one has to simply evaluate the residues at $t=d-\bar{t}$ and $\bar{t}=d-\bar{\tau}$, giving
\begin{multline}
\left\{\begin{matrix}
\mathcal{O}_{\tau_1}& \mathcal{O}_{\tau_2}& \mathcal{O}_{\tau}\\
\mathcal{O}_{\tau_3}&\mathcal{O}_{\tau_4}&\mathcal{O}_{\tilde{\tau}}    \end{matrix}\right\}=\frac{\pi ^{d/2} \Gamma (\tau ) \Gamma (d-\bar{\tau}) \Gamma \left(\Delta -\frac{\bar{\tau}}{2}\right)^2}{4 \Gamma \left(\frac{d}{2}\right) \Gamma \left(\frac{\tau }{2}\right)^4 \Gamma \left(\frac{\bar{\tau}}{2}\right)^2 \Gamma \left(\frac{d}{2}-\tau \right) \Gamma \left(\frac{d-\bar{\tau}}{2}\right)^2}\\\times\,\int \frac{ds\,d\bar{s}}{(2\pi i)^4}\Gamma \left(-\tfrac{s}{2}\right)^2 \Gamma \left(-\tfrac{\bar{s}}{2}\right)^2  \Gamma \left(\tfrac{d+\bar{s}-\bar{\tau}}{2}\right)^2 \Gamma \left(\tfrac{s-2 \Delta +\tau +\bar{\tau}}{2} \right) \Gamma \left(\tfrac{d+s-2 \Delta -\tau +\bar{\tau}}{2} \right)+\ldots\,,
\end{multline}
which now explicitly displays infinitely many double poles\footnote{We remind the reader that in order to extract the the full residue at $\bar{\tau}=2\Delta+2n$ we have to take into account more contour pinching for $n>0$, as we did in the MFT example. We do not present the explicit computation for brevity here and it follows along the same lines as for the MFT example.} at $\bar{\tau}=2\Delta+2n$. The remaining two Mellin integrals nicely reduce to the application of the first Barnes lemma twice. In the spinning case the above integrand would be dressed by a polynomial in $s$ and $\bar{s}$, whose integral reduces again to the application of the first Barnes lemma. After performing the remaining integrals we arrive to:
\begin{multline}
\left\{\begin{matrix}
\mathcal{O}_{\tau_1}& \mathcal{O}_{\tau_2}& \mathcal{O}_{\tau}\\
\mathcal{O}_{\tau_3}&\mathcal{O}_{\tau_4}&\mathcal{O}_{\tilde{\tau}}    \end{matrix}\right\}=\frac{\pi ^{d/2} \Gamma (\tau ) \Gamma \left(\frac{d-\bar{\tau}}{2}\right)^2 \Gamma \left(\Delta -\frac{\bar{\tau}}{2}\right)^2 \Gamma \left(\frac{-2 \Delta +\tau +\bar{\tau}}{2}\right)^2 \Gamma \left(\frac{d-2 \Delta -\tau +\bar{\tau}}{2}\right)^2}{4 \Gamma \left(\frac{d}{2}\right) \Gamma \left(\frac{\tau }{2}\right)^4 \Gamma \left(\frac{\bar{\tau}}{2}\right)^2 \Gamma \left(\frac{d}{2}-\tau \right) \Gamma \left(\frac{d}{2}-2 \Delta +\bar{\tau}\right)}+\ldots\,.
\end{multline}
Taking into account the normalisation $N_{0}(\bar{\tau})$ in \eqref{Normal} we finally obtain:
\begin{align}
\frac1{N_{\bar{\tau},0}}\left\{\begin{matrix}
\mathcal{O}_{\tau_1}& \mathcal{O}_{\tau_2}& \mathcal{O}_{\tau}\\
\mathcal{O}_{\tau_3}&\mathcal{O}_{\tau_4}&\mathcal{O}_{\tilde{\tau}}    \end{matrix}\right\}=\frac{\Gamma (\tau ) \Gamma \left(\Delta -\frac{\bar{\tau}}{2}\right)^2 \Gamma \left(\frac{-2 \Delta +\tau +\bar{\tau}}{2}\right)^2 \Gamma \left(\frac{d-2 \Delta -\tau +\bar{\tau}}{2}\right)^2}{4 \Gamma \left(\frac{\tau }{2}\right)^4 \Gamma \left(\frac{d}{2}-\tau \right) \Gamma \left(\frac{d}{2}-2 \Delta +\bar{\tau}\right)}+\ldots\,.
\end{align}
The above precisely matches the result for the crossing kernel obtained in \cite{Sleight:2018epi}:
\begin{equation}\label{ccscalar}
\mathfrak{J}^{(\sf{t})}_{\tau,0|0}\left(t\right)=\frac{\Gamma (\tau ) \Gamma \left(\frac{t-2 \Delta +\tau }{2}\right)^2 \Gamma \left(\frac{d+t-2 \Delta -\tau}{2}\right)^2}{\Gamma \left(\frac{\tau }{2}\right)^4 \Gamma \left(\frac{d}{2}-\tau \right) \Gamma \left(\frac{d}{2}+t-2 \Delta \right)},
\end{equation}
where we have re-expressed in terms of the conventions of \cite{Sleight:2018epi}, which replaces $\bar{\tau}$ in \eqref{ccscalar} with $t$ and divides by the $\Gamma$-function factor $\frac{1}{4} \Gamma \left(\Delta -\frac{\bar{\tau}}{2}\right)^2$ which in the conventions of \cite{Sleight:2018epi} appears within the Mellin measure $\rho$. 

Finally, taking the coefficient of the double pole in \eqref{ccscalar} at $\bar{\tau}=2\Delta$ reproduces the anomalous dimension \eqref{cpwanom0ell} for $\ell=0$:
\begin{equation}
\frac{\gamma^{\text{CPW}}_{0,0}}2=\frac{\Gamma (\tau ) \Gamma \left(\frac{d-\tau }{2}\right)^2}{\Gamma \left(\frac{d}{2}\right) \Gamma \left(\frac{\tau }{2}\right)^2 \Gamma \left(\frac{d}{2}-\tau \right)}\,,
\end{equation}
where we have divided by the mean field theory coefficient \eqref{mftopeapp} for $n=0$.

It is interesting to contrast with the approach in \cite{Sleight:2018epi}, which obtained both the crossing kernel \eqref{ccscalar} and spinning generalisations in a more streamlined way by using directly the orthogonality properties of CPWs in Mellin space to project onto a given crossed channel conformal block (see e.g. equation \eqref{cpwccint}).\footnote{In particular, this approach uses that the primary operator contribution in a CPW is given by a continuous Hahn polynomial in Mellin space \cite{Costa:2012cb}. See e.g. \eqref{Qpolyapp}. It is instructive to note that continuous Hahn polynomials reappear naturally in the process of evaluating the residues of $6j$ symbols, when collapsing some of the integrals to single out the pole location. These two procedures to extract the OPE data are completely equivalent and encode the same information.} Non-the-less, their Mellin-Barnes integral representation \eqref{6jscalar} is a convenient way to express the $6j$ symbol since it exhibits the poles in $\bar{\tau}$ transparently and is valid for general $d$, which allows the systematic extraction of all residues -- as we have just demonstrated.

\section{Review: Crossing kernels}

\label{sec::crossingkernels}

In order to be as self contained as possible, in this appendix we give some of the expressions derived in \cite{Sleight:2018epi} for crossing kernels of conformal partial waves. In \S \ref{subsec::shadowpro} we also give the resulting expressions upon projecting away the contributions from the shadow conformal multiplet following the approach outlined in \S \ref{subsec::approach}.

\subsection{External scalar operators}

We first give the expressions for identical scalar operators of scaling dimension $\Delta$ in general dimensions $d$. 

The crossing kernel $\mathfrak{J}_{\tau^\prime,\ell^\prime|\ell}^{({\sf t})}\left(t\right)$ of a ${\sf t}$-channel conformal partial wave for the exchange of a spin $\ell^\prime$ primary operator of twist $\tau$ onto leading twist $t=2\Delta$ double-trace operators of spin $\ell$ in the ${\sf s}$-channel reads \cite{Sleight:2018epi}
\begin{align}\label{KernelEqualTau}
    \frac{\mathfrak{J}_{\tau,\ell^\prime|\ell}^{({\sf t})}\left(2\Delta\right)}{ \mathfrak{J}^{({\sf t})}_{\tau,0|0}\left(2\Delta\right)}=\mathcal{Z}_{{\ell^\prime}}\,\sum_{p=0}^{
{\ell^\prime}}\sum_{k=0}^{{\ell^\prime}-p}a^{{\ell^\prime}}_{p,k}\ {}_4F_3\left(\begin{matrix}p-\ell,\ell+p+2 \Delta-1,\tfrac{d-\tau }{2}-k,\tfrac{\tau}{2}+{\ell^\prime}-k\\\tfrac{d}{2},\Delta+p ,\Delta+p \end{matrix};1\right)\,,
\end{align}
where 
\begin{subequations}\label{coeffscc}
\begin{align}
    \mathfrak{J}^{({\sf t})}_{\tau,0|\ell}\left(2\Delta\right)&=\frac{\Gamma (\tau ) \Gamma \left(\frac{d-\tau }{2}\right)^2}{\Gamma \left(\frac{d}{2}\right) \Gamma \left(\frac{\tau }{2}\right)^2 \Gamma \left(\frac{d}{2}-\tau \right)}\,,\\
    \mathcal{Z}_{{\ell^\prime}}&=\frac{2^{\ell^\prime}\left(\frac{\tau +1}{2}\right)_{\ell^\prime} \left(\tau +1-\tfrac{d}{2}\right)_{\ell^\prime}}{\left(\frac{\tau }{2}\right)_{\ell^\prime} (d-{\ell^\prime}-\tau -1)_{\ell^\prime}},\\
    a^{{\ell^\prime}}_{p,k}&=\binom{{\ell^\prime}}{p}\frac{\sqrt{\pi }\,2^{2-\ell-2\Delta}\,\Gamma \left(\ell+\Delta\right) \Gamma (\ell+p+2\Delta-1)}{(\ell-p)! \Gamma \left(\ell+\Delta-\frac{1}{2}\right) \Gamma \left(p+\Delta\right)^2}\,
    \alpha^{\ell^\prime}_{p,k}\,,\\
    \alpha^{\ell^\prime}_{p,k}(t)&=\binom{{\ell^\prime}-p}{k}\,\frac{\left(\tfrac{d+2 p-2}{2}\right)_k}{\left(\tfrac{d+2 {\ell^\prime}-2}{2}-k\right)_k}\,.
\end{align}
\end{subequations}

For identical external scalars the crossing kernels of $\sf{u}$-channel CPWs are proportional to $\sf{t}$-channel crossing kernels \eqref{KernelEqualTau} up to a $(-1)^{\ell+\ell^\prime}$ factor.

\subsection{External spinning operators}
\label{subsec::cccbextspin}

Similar expressions have been obtained for crossing kernels with external spinning legs \cite{Sleight:2018epi}. We restrict to the crossing kernels of CPWs with an exchanged scalar primary operator (so $\ell^\prime=0$) of twist $\tau$.

For four-point correlators of the type \eqref{0O1OOO2} involving two spinning operators of twists $\tau_1$ and $\tau_2$, and spins $J_1$ and $J_2$, the ${\sf t}$-channel and ${\sf u}$-channel CPWs are unique and read \cite{Sleight:2018epi}
\begin{align}\label{CKj1j2t}
     {}^{(\sf{t})}\mathfrak{J}_{\tau,0|\ell}&=\mathcal{C}^{\sf{t}}\frac{(-2)^{-J_1-J_2+\ell} \Gamma (J_1-J_2+\ell+\tau_1)  \Gamma (J_1+J_2+\ell+\tau_1+\tau_2-1)}{ (\ell-J_1-J_2)!\,\Gamma (2 J_1+\tau_1) \Gamma (2 \ell+\tau_1+\tau_2-1) }\\
    &\times\tfrac{\Gamma \left(\frac{2 J_1+\tau +\tau_1-\tau_4}{2}\right) \Gamma \left(\frac{2 J_2+\tau +\tau_2-\tau_3}{2}\right) \Gamma \left(\frac{d+2 J_1-\tau +\tau_1-\tau_4}{2}\right) \Gamma \left(\frac{d+2 J_2-\tau +\tau_2-\tau_3}{2}\right)\Gamma \left(\frac{2 \ell+\tau_1+\tau_2+\tau_3-\tau_4}{2}\right)}{\Gamma \left(\frac{2 J_1+2 J_2+\tau_1+\tau_2+\tau_3-\tau_4}{2}\right) \Gamma \left(\frac{d+2 J_1+2 J_2+\tau_1+\tau_2-\tau_3-\tau_4}{2}\right)}\nonumber\\\nonumber
    &\times \, _4F_3\left(\begin{matrix}J_1+J_2-\ell,J_1+J_2+\ell+\tau_1+\tau_2-1,\frac{d}{2}+J_1-\frac{\tau-\tau_1+\tau_4 }{2},J_1+\frac{\tau+\tau_1-\tau_4 }{2}\\2 J_1+\tau_1,\frac{d+\tau_1+\tau_2-\tau_3-\tau_4}{2}+J_1+J_2,J_1+J_2+\frac{\tau_1+\tau_2+\tau_3-\tau_4}{2}\end{matrix};1\right),
    \end{align}
    and
    \begin{align}\label{CKj1j2u}
     {}^{(\sf{u})}\mathfrak{J}_{\tau,0|\ell}&=\mathcal{C}^{\sf{u}}\frac{2^{-J_1-J_2+\ell} \Gamma (-J_1+J_2+\ell+\tau_2) \Gamma (J_1+J_2+\ell+\tau_1+\tau_2-1)}{ (\ell-J_1-J_2)!\,\Gamma (2 J_2+\tau_2) \Gamma (2 \ell+\tau_1+\tau_2-1)}\\
    &\times \tfrac{\Gamma \left(\frac{2 J_1+\tau +\tau_1-\tau_3}{2}\right) \Gamma \left(\frac{2 J_2+\tau +\tau_2-\tau_4}{2} \right) \Gamma \left(\frac{d+2 J_1-\tau +\tau_1-\tau_3}{2}\right) \Gamma \left(\frac{d+2 J_2-\tau +\tau_2-\tau_4}{2}\right) \Gamma \left(\frac{2 \ell+\tau_1+\tau_2+\tau_3-\tau_4}{2}\right)}{\Gamma \left(\frac{2 J_1+2 J_2+\tau_1+\tau_2+\tau_3-\tau_4}{2}\right) \Gamma \left(\frac{d+2 J_1+2 J_2+\tau_1+\tau_2-\tau_3-\tau_4}{2}\right)}\nonumber\\
    &\times \, _4F_3\left(\begin{matrix}J_1+J_2-\ell,J_1+J_2+\ell+\tau_1+\tau_2-1,\frac{d-\tau+\tau_2-\tau_4}{2}+J_2,J_2+\frac{\tau+\tau_2-\tau_4 }{2}\\2 J_2+\tau_2,\frac{d+\tau_1+\tau_2-\tau_3-\tau_4}{2}+J_1+J_2,J_1+J_2+\frac{\tau_1+\tau_2+\tau_3-\tau_4}{2}\end{matrix};1\right)\,,\nonumber
\end{align}
where we focused on the crossing onto double-trace operators of leading twist $t=\tau_1+\tau_2$ and 
\begin{equation}
     \mathcal{C}^{\sf{t}}=\frac{\kappa_{d-\tau,0}\,\alpha_{J_2,0,0;\tau_2,\tau_3,\tau}\, (-1)^{J_1+J_2}}{\Gamma \left(\frac{\tau }{2}+\frac{\tau_4}{2}-\frac{\tau_1}{2}\right) \Gamma \left(\frac{d}{2}+\frac{\tau_3}{2}-\frac{\tau }{2}-\frac{\tau_2}{2}\right) \Gamma \left(J_1+\frac{\tau }{2}+\frac{\tau_1}{2}-\frac{\tau_4}{2}\right) \Gamma \left(\frac{d}{2}+J_2+\frac{\tau_2}{2}-\frac{\tau }{2}-\frac{\tau_3}{2}\right)},
\end{equation}
and
\begin{equation}
     \mathcal{C}^{\sf{u}}=\frac{\kappa_{d-\tau,0}\alpha_{J_2,0,0;\tau_2,\tau_4,\tau}\, (-1)^{J_2}}{\Gamma \left(\frac{\tau }{2}+\frac{\tau_3}{2}-\frac{\tau_1}{2}\right) \Gamma \left(\frac{d}{2}+\frac{\tau_4}{2}-\frac{\tau }{2}-\frac{\tau_2}{2}\right) \Gamma \left(J_1+\frac{\tau }{2}+\frac{\tau_1}{2}-\frac{\tau_3}{2}\right) \Gamma \left(\frac{d}{2}+J_2+\frac{\tau_2}{2}-\frac{\tau }{2}-\frac{\tau_4}{2}\right)}.
\end{equation}
The above reduces to \eqref{KernelEqualTau} for scalar exchange when $J_1=J_2=0$. In general these expressions give contributions to the weighted average of anomalous dimensions of leading twist double-trace operators $[{\cal O}_{J_1}{\cal O}_{J_2}]$ and $[\mathcal{O}\mathcal{O}]$ whenever $\tau_1+\tau_2=\tau_3+\tau_4$. If $\tau_1+\tau_2\neq\tau_3+\tau_4$ the above crossing kernels encode corrections to the double-trace operator OPE coefficients. 

With similar techniques it is also possible to obtain crossing kernels for correlators of the type \eqref{1o2o} and \eqref{1oo2}, to obtain contributions to the weighted average of anomalous dimensions of double-trace operators $\left[{\cal O}_J{\cal O}\right]$. For instance, for spin $J=1$ we obtained \cite{Sleight:2018epi}:
{\allowdisplaybreaks\begin{align}\label{Jeq1}
   \gamma_{0,\ell}^{\text{CPW}}&=\gamma_{0,1}^{\text{CPW}}\,\frac{2 (d-\tau ) (\tau -\tau_1-\tau_2)}{\tau  (d-\tau +\tau_1+\tau_2)}\Bigg[ {}_4F_3\left(\begin{matrix}1-\ell,\frac{d}{2}-\frac{\tau }{2},\frac{\tau }{2},\ell+\tau_1+\tau_2\\\frac{d}{2}+1,\tau_1+2,\tau_2\end{matrix};1\right)\\\nonumber
   &+\frac{d (\tau_1+\tau_2)}{2 (d-\tau ) (\tau -\tau_1-\tau_2)} \, _4F_3\left(\begin{matrix}1-\ell,\frac{d}{2}-\frac{\tau }{2},\frac{\tau }{2},\ell+\tau_1+\tau_2\\\frac{d}{2},\tau_1+2,\tau_2\end{matrix};1\right)\\\nonumber
   &+\frac{\tau  (d-\tau ) (\ell+\tau_2-1) (\ell+\tau_1+\tau_2)}{4 (d+2) \tau_2 (2 \ell+\tau_1+\tau_2-1)}\, _4F_3\left(\begin{matrix}1-\ell,\frac{d}{2}-\frac{\tau }{2}+1,\frac{\tau }{2}+1,\ell+\tau_1+\tau_2+1\\\frac{d}{2}+2,\tau_1+2,\tau_2+1\end{matrix};1\right)\\\nonumber
   &+\frac{(\ell-1) \tau  (\tau -d) (\ell+\tau_1)}{4 (d+2) \tau_2 (2 \ell+\tau_1+\tau_2-1)}\, _4F_3\left(\begin{matrix}2-\ell,\frac{d}{2}-\frac{\tau }{2}+1,\frac{\tau }{2}+1,\ell+\tau_1+\tau_2\\\frac{d}{2}+2,\tau_1+2,\tau_2+1\end{matrix};1\right)\\\nonumber
   &-\frac{(\tau +2) (d-\tau +2)}{4 (d+2)}\, _4F_3\left(\begin{matrix}1-\ell,\frac{d}{2}-\frac{\tau }{2},\frac{\tau }{2},\ell+\tau_1+\tau_2\\\frac{d}{2}+2,\tau_1+2,\tau_2\end{matrix};1\right)\Bigg]\,,
\end{align}}
where
\begin{align}
    \gamma_{0,1}^{\text{CPW}}&=\frac{2^{\tau -2} \Gamma \left(\frac{\tau +1}{2}\right) (d+2 \Delta -\tau ) \Gamma \left(\frac{d-\tau }{2}\right)^2}{\sqrt{\pi } \Gamma \left(\frac{d}{2}+1\right) (\tau -2 \Delta ) \Gamma \left(\frac{\tau }{2}\right) \Gamma \left(\frac{d}{2}-\tau \right)}\,.
\end{align}

\subsection{Shadow projection}
\label{subsec::shadowpro}

To remove the contributions from the shadow conformal multiplet in conformal partial waves \eqref{cpwdefint} we follow the procedure outlined in \S \ref{subsec::approach}, which employs the Mellin representation of the ${}_4F_3$ hypergeometric functions in the crossing kernels to project away the shadow poles. For the crossing kernels \eqref{CKj1j2t} and \eqref{CKj1j2u}, in this way we obtain

{\footnotesize
\begin{multline}\label{ctapp}
{}^{(\sf{t})}\bar{\mathfrak{I}}_{\tau,0|\ell}=\tfrac{(-1)^{\ell+1} 2^{-J_1-J_2+\ell}\Gamma (\tau ) \Gamma (J_1-J_2+\ell+\tau_1) \Gamma \left(\ell+\frac{\tau_1}{2}+\frac{\tau_2}{2}+\frac{\tau_3}{2}-\frac{\tau_4}{2}\right) \Gamma \left(J_2+\ell+\frac{\tau_1}{2}+\tau_2+\frac{\tau_4}{2}-\frac{\tau }{2}-1\right)}{\Gamma (2 \ell+\tau_1+\tau_2-1) \Gamma \left(\frac{\tau }{2}+\frac{\tau_4}{2}-\frac{\tau_1}{2}\right) \Gamma \left(\frac{\tau }{2}+\frac{\tau_3}{2}-\frac{\tau_2}{2}\right) \Gamma \left(J_1+\frac{\tau_1}{2}+\frac{\tau_4}{2}-\frac{\tau }{2}\right) \Gamma \left(J_2+\frac{\tau_2}{2}+\frac{\tau_3}{2}-\frac{\tau }{2}\right) \Gamma \left(-J_2+\ell+\frac{\tau }{2}+\frac{\tau_1}{2}-\frac{\tau_4}{2}+1\right)}\\
\times \, _4F_3\left(\begin{matrix}-J_2+\frac{\tau }{2}-\frac{\tau_2}{2}-\frac{\tau_3}{2}+1,-\frac{d}{2}-J_2+\frac{\tau }{2}-\frac{\tau_2}{2}+\frac{\tau_3}{2}+1,-J_1+\frac{\tau }{2}-\frac{\tau_1}{2}-\frac{\tau_4}{2}+1,J_1+\frac{\tau }{2}+\frac{\tau_1}{2}-\frac{\tau_4}{2}\\-\frac{d}{2}+\tau +1,-J_2+\ell+\frac{\tau }{2}+\frac{\tau_1}{2}-\frac{\tau_4}{2}+1,-J_2-\ell+\frac{\tau }{2}-\frac{\tau_1}{2}-\tau_2-\frac{\tau_4}{2}+2\end{matrix};1\right),
\end{multline}}
and
{\footnotesize
\begin{multline}\label{cuapp}
    {}^{(\sf{u})}\bar{\mathfrak{I}}_{\tau,0|\ell}=\tfrac{(-1)^{\ell+1-J_1}(-2)^{-J_1-J_2+\ell}\Gamma (\tau ) \Gamma (-J_1+J_2+\ell+\tau_2) \Gamma \left(\ell+\frac{\tau_1}{2}+\frac{\tau_2}{2}+\frac{\tau_3}{2}-\frac{\tau_4}{2}\right) \Gamma \left(J_1+\ell+\tau_1+\frac{\tau_2}{2}+\frac{\tau_4}{2}-\frac{\tau }{2}-1\right)}{\Gamma (2 \ell+\tau_1+\tau_2-1) \Gamma \left(\frac{\tau }{2}+\frac{\tau_3}{2}-\frac{\tau_1}{2}\right) \Gamma \left(\frac{\tau }{2}+\frac{\tau_4}{2}-\frac{\tau_2}{2}\right) \Gamma \left(J_1+\frac{\tau_1}{2}+\frac{\tau_3}{2}-\frac{\tau }{2}\right) \Gamma \left(J_2+\frac{\tau_2}{2}+\frac{\tau_4}{2}-\frac{\tau }{2}\right) \Gamma \left(-J_1+\ell+\frac{\tau }{2}+\frac{\tau_2}{2}-\frac{\tau_4}{2}+1\right)}\\\times
    \, _4F_3\left(\begin{matrix}-J_1+\frac{\tau }{2}-\frac{\tau_1}{2}-\frac{\tau_3}{2}+1,-\frac{d}{2}-J_1+\frac{\tau }{2}-\frac{\tau_1}{2}+\frac{\tau_3}{2}+1,-J_2+\frac{\tau }{2}-\frac{\tau_2}{2}-\frac{\tau_4}{2}+1,J_2+\frac{\tau }{2}+\frac{\tau_2}{2}-\frac{\tau_4}{2}\\-\frac{d}{2}+\tau +1,-J_1-\ell+\frac{\tau }{2}-\tau_1-\frac{\tau_2}{2}-\frac{\tau_4}{2}+2,-J_1+\ell+\frac{\tau }{2}+\frac{\tau_2}{2}-\frac{\tau_4}{2}+1\end{matrix};1\right).
\end{multline}}

\subsection{Expanding crossing kernels in terms of Wilson polynomials}\label{CrossingToWilson}

While expanding crossing kernels in a basis of Wilson polynomials is a straightforward task, it can be quite cumbersome in general. In this section we present a general method to work out such an expansion for crossing kernels expressed in terms of the hypergeometric functions, considering explicitly the example of crossing kernels for external scalars reviewed in \eqref{KernelEqualTau}.\footnote{For the case of external spinning operators, the spinning crossing kernels we consider in this work (which are reviewed in \S \ref{subsec::cccbextspin}) are for an exchanged scalar ($\ell^\prime=0$) in the crossed channel. It has already been observed in \cite{Sleight:2018epi} that the latter crossing kernels are proportional to a Wilson polynomial, and the result is recalled in equation \eqref{0spinningdecomp}.} To this end, it is sufficient to obtain such a decomposition of the following type of hypergeometric function:
\begin{multline}
    {}_4F_3\left(\begin{matrix}p-\ell,\ell+p+2\Delta-1,\tfrac{d}4-m+\tfrac{\ell^\prime}2-\tfrac{i\nu}2,\tfrac{d}4-m+\tfrac{\ell^\prime}2+\tfrac{i\nu}2\\\tfrac{d}2,\Delta+p,\Delta+p\end{matrix};1\right)\\=\sum_{j}\bar{\beta}_{\ell,j}^{(\ell^\prime)}\mathcal{W}_{\ell-j}(\nu^2;\tfrac{d+2\ell^\prime}4,\tfrac{d+2\ell^\prime}4,-\tfrac{d-2\ell^\prime}4+\Delta,-\tfrac{d-2\ell^\prime}4+\Delta),
\end{multline}
or more generally
\begin{multline}
    \mathcal{H}^{(\ell)}_{\ell^\prime,p,m}(\nu^2;a_i)\equiv{}_4F_3\left(\begin{matrix}p-\ell,a_1+a_2+a_3+a_4-2 \ell^\prime+\ell+p-1,a_1-m-\frac{i\nu }{2},a_1-m+\frac{i\nu }{2}\\a_1+a_2-\ell^\prime,a_1+a_3-\ell^\prime+p,a_1+a_4-\ell^\prime+p\end{matrix};1\right)\\=\sum_{j}\bar{\beta}_{\ell,j}^{(\ell^\prime)}\mathcal{W}_{\ell-j}(\nu^2;a_i)\,.
\end{multline}
The coefficients $\bar{\beta}_{\ell,j}^{(\ell^\prime)}$ can then be extracted using the orthogonality of Wilson polynomials:
\begin{equation}\label{intrepbeta}
    \bar{\beta}_{\ell,j}^{(\ell^\prime)}=\frac1{\mathcal{N}_{\ell-j}}\int_{-\infty}^{+\infty}\frac{d\nu}{2\pi}\,w(a_i)\,\mathcal{H}^{(\ell)}_{\ell^\prime,p,m}(\nu^2;a_i)\mathcal{W}_{\ell-j}(\nu^2;a_i)\,,
\end{equation}
where the normalisation factor $\mathcal{N}_{\ell}$ is defined as
\begin{multline}
    \mathcal{N}_{\ell}\equiv\int_{-\infty}^{+\infty}\frac{d\nu}{2\pi}\,w(a_i)\,\mathcal{W}_{\ell}(\nu^2;a_i)\mathcal{W}_{\ell}(\nu^2;a_i)\\=\frac{\ell! (a_2+a_3)_{\ell} (a_2+a_4)_{\ell} (a_3+a_4)_{\ell} (a_1+a_2+a_3+a_4+{\ell}-1)_{\ell}}{(a_1+a_2)_{\ell} (a_1+a_3)_{\ell} (a_1+a_4)_{\ell}(a_1+a_2+a_3+a_4)_{2 \ell}}\\\times\,\frac{\Gamma (a_1+a_2) \Gamma (a_1+a_3) \Gamma (a_1+a_4) \Gamma (a_2+a_3) \Gamma (a_2+a_4) \Gamma (a_3+a_4)}{\Gamma (a_1+a_2+a_3+a_4)}\,.
\end{multline}
In order to evaluate the above integral we simply applied a case of Barnes' Lemma (which at the level of the ${}_7F_6$ is equivalent to Dougall's Theorem).

Going back to the integral representation \eqref{intrepbeta} for the coefficients $\bar{\beta}_{\ell,j}^{(\ell^\prime)}$, one effective strategy is to consider an expansion of the functions $\mathcal{H}^{(\ell)}_{\ell^\prime,p,m}(\nu^2;a_i)$ in terms of telescopic Pochhammer symbols $(a_1+\tfrac{i\nu}2)_{k_1}(a_1-\tfrac{i\nu}2)_{k_1}$, while at the same time expanding the Wilson polynomial in $(a_2+\tfrac{i\nu}2)_{k_2}(a_2-\tfrac{i\nu}2)_{k_2}$. Such an expansion takes the form:
\begin{align}
    \mathcal{H}^{(\ell)}_{\ell^\prime,p,m}(\nu^2;a_i)&=\sum_{k=0}^\ell \sum_{q=0}^{k}h_{k,q}\,(a_1+\tfrac{i\nu}2)_{q}(a_1-\tfrac{i\nu}2)_{q}\,,
\end{align}
with
\begin{align}
    h_{k,q}&=\tfrac{\Gamma (m+1) (-1)^{q-k} \Gamma (k+1) \Gamma (2 a_1-m+k)}{\Gamma (q+1) \Gamma (k-q+1) \Gamma (2 a_1-m+q) \Gamma (m-k+q+1)}\\\nonumber
    &\times\tfrac{(-1)^k \Gamma (\ell-p+1) \Gamma (a_1+a_2-J) \Gamma (a_1+a_3-J+p) \Gamma (a_1+a_4-J+p) \Gamma (a_1+a_2+a_3+a_4-2 J+k+\ell+p-1)}{\Gamma (k+1) \Gamma (-k+\ell-p+1) \Gamma (a_1+a_2-J+k) \Gamma (a_1+a_3-J+k+p) \Gamma (a_1+a_4-J+k+p) \Gamma (a_1+a_2+a_3+a_4-2 J+\ell+p-1)}\,.
\end{align}
In this way, the integral \eqref{intrepbeta} for the coefficients $\bar{\beta}_{\ell,j}^{(\ell^\prime)}$ reduces to a sum of spectral integrals of  Wilson polynomials, which can be evaluated explicitly:
\begin{multline}
    \bar{\beta}_{\ell,j}^{(\ell^\prime)}=\sum_{k,q=0}^{j}\,h_{\ell-k,\ell-k-q}\,\\ \times \underbrace{\frac{1}{\mathcal{N}_{\ell-j}}\int_{-\infty}^{+\infty}\frac{d\nu}{2\pi}\,w(a_1+q,a_{i>1})\,\mathcal{W}_{\ell-j}(\nu^2;a_i)}_{\tfrac{(-1)^{j-\ell} \Gamma (q+1) \Gamma (a_1+a_2+q) \Gamma (a_1+a_3+q) \Gamma (a_1+a_4+q) (a_1+a_2+a_3+a_4-2 j+2 \ell-1) \Gamma (a_1+a_2+a_3+a_4-j+\ell-1)}{\Gamma (a_1+a_2) \Gamma (a_1+a_3) \Gamma (a_1+a_4) \Gamma (-j+\ell+1) \Gamma (j-\ell+q+1) \Gamma (a_1+a_2+a_3+a_4-j+\ell+q)}}\,,
\end{multline}
where in order to bound the summation at $k=j$ and $q=j$ we have taken into account that any polynomial of degree lower than the degree of the Wilson polynomial cancels in the above sum. The above coefficients can be combined with eq.~\eqref{KernelEqualTau} to give the full decomposition of the crossing kernels in terms of Wilson polynomials:
\begin{equation}
    \widehat{\mathfrak{J}}_{\frac{d}{2}+i\nu,\ell^\prime|\ell}^{({\sf t})}\left(2\Delta\right)=\sum_{j=0}^{2\ell^\prime}\beta_{\ell,j}^{(\ell^\prime)}\mathcal{W}_{\ell-j}(\nu^2;\tfrac{d+2 \ell^\prime}{4},\tfrac{d+2 \ell^\prime}{4},-\tfrac{d}{4}+\Delta +\tfrac{\ell^\prime}{2},-\tfrac{d}{4}+\Delta +\tfrac{\ell^\prime}{2}),
\end{equation}
with
\begin{equation}
    \beta_{\ell,j}^{(\ell^\prime)}=\frac{\pi ^{\frac{d-1}{2}} (-1)^{\ell^\prime}\,2^{2 \Delta -\ell^\prime+\ell-2} \Gamma (\Delta )^2 \Gamma (\ell+1) \Gamma \left(\ell+\Delta -\tfrac{1}{2}\right)}{\Gamma \left(\tfrac{d}{2}\right) \Gamma (\ell+\Delta ) \Gamma (\ell+2 \Delta -1)}\sum_{p=0}^{\ell^\prime}\sum_{i=0}^{\ell^\prime-p}a_{p,i}^{\ell^\prime}\bar{\beta}_{\ell,j}^{(\ell^\prime)}\,,
\end{equation}
which can be expressed as the double-sum of a product of two hypergeometric functions:

\begin{empheq}[box=\fbox]{align}
   \label{betaCoeffs}
    \beta_{\ell,j}^{(\ell^\prime)}&=\tfrac{\pi ^{d/2}(-1)^{j+\ell^\prime+\ell} 2^{-\ell^\prime} J(2 \Delta -2 j+2 \ell^\prime+2 \ell-1)\,\ell!\,\Gamma (\Delta )^2 \Gamma (\ell^\prime) \Gamma (2 \ell^\prime+\ell-j+2 \Delta -1)}{(\ell-j)!\Gamma \left(\frac{d}{2}-1\right)\Gamma \left(\frac{d}{2}+\ell^\prime-1\right) \Gamma \left(\frac{d}{2}+\ell^\prime\right)\Gamma (\ell^\prime+\Delta )^2 \Gamma (\ell+2 \Delta -1)}\\\nonumber
    &\times\sum_{i=0}^{\ell^\prime}\sum_{k=0}^j\tfrac{ (-1)^{-k}  \Gamma \left(\frac{d}{2}+i-1\right) \Gamma \left(\frac{d}{2}-i+\ell^\prime-1\right)  \Gamma (2 \ell-k+2 \Delta -1) \Gamma \left(\frac{d}{2}+\ell^\prime+\ell-k\right)  \Gamma (\ell^\prime+\ell-k+\Delta )^2}{i! k! (\ell^\prime-i)! (j-k)!  \Gamma \left(\frac{d}{2}-k+\ell\right) \Gamma (\ell-k+\Delta )^2 \Gamma (2 \ell^\prime-j-k+2 (\ell+\Delta ))}\\\nonumber
    &\times
    \, _4F_3\left(\begin{matrix}-k,\frac{d}{2}+i-1,i-\ell^\prime,-k+2 \ell+2 \Delta -1\\\frac{d}{2}-1,\ell-k+\Delta ,\ell-k+\Delta\end{matrix} ;1\right) \\
    &\times\, _4F_3\left(\begin{matrix}-i,k-j,-\frac{d}{2}+i-\ell^\prime+k-\ell+1,j-2 \ell^\prime+k-2\ell-2 \Delta +1\\-\frac{d}{2}-\ell^\prime+k-\ell+1,k-\ell^\prime-\ell-\Delta +1,k-\ell^\prime-\ell-\Delta +1\end{matrix};1\right)\,.\nonumber
\end{empheq}

Below we list some simplified expressions for specific (low) values of $\ell^\prime$:

\paragraph{\underline{$\ell^\prime=0$}:}
\begin{align}
\beta_{\ell,0}^{(0)}=\frac{\pi ^{d/2}}{\Gamma \left(\frac{d}{2}\right)},
\end{align}

{\footnotesize \paragraph{\underline{$\ell^\prime=1$}:}
\begin{subequations}
\begin{align}
\beta_{\ell,0}^{(1)}&=-\frac{\pi ^{d/2} (d+2 \ell) (\Delta +\ell) (2 \Delta +\ell-1) (2 \Delta +\ell)}{4 \Delta ^2 \Gamma \left(\frac{d}{2}+1\right) (2 \Delta +2 \ell-1)}\,,\\
\beta_{\ell,1}^{(1)}&=\frac{\pi ^{d/2} \ell (-d+2 \Delta +2) (2 \Delta +\ell-1)}{4 \Delta ^2 \Gamma \left(\frac{d}{2}+1\right)}\,,\\
\beta_{\ell,2}^{(1)}&=\frac{\pi ^{d/2} (\ell-1) \ell (\Delta +\ell-1) (-d+4 \Delta +2 \ell-2)}{4 \Delta ^2 \Gamma \left(\frac{d}{2}+1\right) (2 \Delta +2 \ell-1)}\,.
\end{align}
\end{subequations}}

\paragraph{\underline{$\ell^\prime=2$}:}

{\footnotesize \begin{subequations}
\begin{align}
\beta_{\ell,0}^{(2)}&=\tfrac{(d-1) \pi ^{d/2} (d+2 \ell) (d+2 \ell+2) (\Delta +\ell) (\Delta +\ell+1) (2 \Delta +\ell-1) (2 \Delta +\ell) (2 \Delta +\ell+1) (2 \Delta +\ell+2)}{16 d \Delta ^2 (\Delta +1)^2 \Gamma \left(\frac{d}{2}+2\right) (2 \Delta +2 \ell-1) (2 \Delta +2 \ell+1)}\,,\\
\beta_{\ell,1}^{(2)}&=\tfrac{\pi ^{d/2} \ell (d+2 \ell) (\Delta +\ell) (2 \Delta +\ell-1) (2 \Delta +\ell) (2 \Delta +\ell+1) (d (d-2 \Delta -3)+2 (\Delta +\ell))}{8 d \Delta ^2 (\Delta +1)^2 \Gamma \left(\frac{d}{2}+2\right) (2 \Delta +2 \ell-1)}\,,\\
\beta_{\ell,2}^{(2)}&=\tfrac{\pi ^{d/2} (\ell-1) \ell (2 \Delta +\ell-1) (2 \Delta +\ell)}{8 d \Delta ^2 (\Delta +1)^2 \Gamma \left(\frac{d}{2}+2\right) (2 \Delta +2 \ell-3) (2 \Delta +2 \ell+1)} \Big(d^3 \left(3 \Delta -3 \left(\Delta ^2+2 \Delta  \ell+(\ell-1) \ell\right)+2\right)\\\nonumber
&+d^2 \left(12 \Delta ^3-19 \Delta +\Delta ^2 (24 \ell-1)+2 \Delta  \ell (6 \ell+5)+11 (\ell-1) \ell-8\right)\\\nonumber
&+2 d \Big(\Delta  (15-4 \Delta  (\Delta  (\Delta +2)-2))+2 \ell^4+(8 \Delta -4) \ell^3+\left(6 \Delta ^2-22 \Delta -6\right) \ell^2\\\nonumber
&-2 (2 \Delta -1) (\Delta  (\Delta +7)+4) \ell+5\Big)+4 \Big((\Delta -1)^3 (2 \Delta +1)+3 \ell^4+6 (2 \Delta -1) \ell^3\\\nonumber
&+(\Delta  (17 \Delta -18)+2) \ell^2+\Delta  (\Delta  (10 \Delta -17)+4) \ell+\ell\Big)\Big),\\
\beta_{\ell,3}^{(2)}&=\tfrac{\pi ^{d/2} (\ell-2) (\ell-1) \ell (\Delta +\ell-1) (2 \Delta +\ell-1) (d-4 \Delta -2 \ell+2) (d (d-2 \Delta -3)-2 (\Delta +\ell-1))}{8 d \Delta ^2 (\Delta +1)^2 \Gamma \left(\frac{d}{2}+2\right) (2 \Delta +2 \ell-1)}\,,\\
\beta_{\ell,4}^{(2)}&=\tfrac{(d-1) \pi ^{d/2} (\ell-3) (\ell-2) (\ell-1) \ell (\Delta +\ell-2) (\Delta +\ell-1) (d-4 \Delta -2 \ell+2) (d-4 \Delta -2 \ell+4)}{16 d \Delta ^2 (\Delta +1)^2 \Gamma \left(\frac{d}{2}+2\right) (2 \Delta +2 \ell-3) (2 \Delta +2 \ell-1)}\,.
\end{align}
\end{subequations}}

\subsection{Sub-leading twist $D_j$}\label{Dcoeffs}

Here we list the coefficients $D_j$ defined in equation \eqref{gamma0000nl} for few values of $n$. Note that in \eqref{gamma0000nl} we take $\tau=\frac{d}{2}+i\nu$.\\

\noindent\underline{$n=0$:}
{\footnotesize 
\begin{equation}
D_0=\frac12
\end{equation}}\\

\noindent\underline{$n=1$}
{\footnotesize\begin{subequations}
\begin{align}
    D_0=&\frac{(d-2 (\Delta +1)) (\tau -2 \Delta )^2}{2 (d-2 \tau )}\left[d-2 \Delta -(\tau -2) \tau -2\right]\,,\\
    D_1=&\frac{(d-2 (\Delta +1)) (-d+2 \Delta +\tau )^2}{2 (d-2 \tau )}\,\left[ (d-3) d+2 (\Delta +\tau +1)+\tau(\tau-2 d)\right]\,,
\end{align}
\end{subequations}}
\\

\noindent\underline{$n=2$}
{\allowdisplaybreaks\small
\begin{subequations}
\begin{align}
    &D_0= \frac{(d-2 (\Delta +2)) (-2 \Delta +\tau -2)^2 (\tau -2 \Delta )^2}{4 (d-2 \tau ) (d-2 \tau +2)} \Big(\tau ^4 (d-2 \Delta -3)-4 \tau ^3 (d-2 \Delta -3)-4 \tau ^2 (d-2 \Delta -3)^2\nonumber\\
    &\hspace{60pt}+8 \tau  (d-2 \Delta -3) (d-2 (\Delta +1))+2 (d-2 (\Delta +1))^2 (d-2 (\Delta +2))\Big)\\
    &D_1= \frac{(d-2 (\Delta +2)) (\tau -2 \Delta )^2 (-d+2 \Delta +\tau )^2}{2 (d-2 \tau +2) (d-2 (\tau +1))} \Big(-\tau ^2 (d-2 \Delta -3) \left(d^2-4 d+8 \Delta +8\right)\nonumber\\
    &\hspace{60pt}+2 (d-2 (\Delta +1)) \left(d^3-2 d^2 (\Delta +2)+d (4 \Delta +2)-4 \left(\Delta ^2+\Delta -1\right)\right)\nonumber\\
    &\hspace{60pt}+\tau ^4 (-d+2 \Delta +3)+2 d \tau ^3 (d-2 \Delta -3)-4 d \tau  (d-2 \Delta -3) (d-2 (\Delta +1))\Big)\\
    &D_2= \frac{(d-2 (\Delta +2)) (-d+2 \Delta +\tau )^2 (-d+2 \Delta +\tau +2)^2}{4 (d-2 \tau ) (d-2 (\tau +1))} \Big(d^5-2 d^4 \Delta -11 d^4+24 d^3 \Delta +46 d^3-16 d^2 \Delta ^2\nonumber\\
    &\hspace{60pt}-92 d^2 \Delta -92 d^2+56 d \Delta ^2+\tau ^4 (d-2 \Delta -3)-4 (d-1) \tau ^3 (d-2 \Delta -3)\nonumber\\
    &\hspace{60pt}+2 \tau ^2 (d (3 d-8)+4 \Delta +6) (d-2 \Delta -3)-4 (d-1) \tau  (d-2 \Delta -3) \left((d-2)^2+4 \Delta \right)\nonumber\\
    &\hspace{60pt}+144 d \Delta +88 d-16 \Delta ^3-64 \Delta ^2-80 \Delta -32\Big)
\end{align}
\end{subequations}}

\noindent\underline{$n=3$}
{\allowdisplaybreaks\footnotesize
\begin{subequations}
\begin{align}
    &D_0=\frac{(d-2 (\Delta +2)) (d-2 (\Delta +3)) (-2 \Delta +\tau -4)^2 (-2 \Delta +\tau -2)^2 (\tau -2 \Delta )^2}{12 (d-2 \tau ) (d-2 \tau +2) (d-2 \tau +4)}\nonumber\\
    &\times \Big(12 \tau  (d-2 (\Delta +1)) \left(3 d^2-12 d (\Delta +2)+12 \Delta  (\Delta +4)+44\right)\nonumber\\
    &+2 \tau ^2 \left(-9 d^3+54 d^2 (\Delta +2)-2 d (54 \Delta  (\Delta +4)+205)+4 \Delta  (18 \Delta  (\Delta +6)+205)+484\right)\nonumber\\
    &+\tau ^6 (-d+2 \Delta +5)+6 \tau ^5 (d-2 \Delta -5)+\tau ^4 (d-2 \Delta -5) (9 d-18 \Delta -34)\nonumber\\
    &-12 \tau ^3 (3 d-6 \Delta -8) (d-2 \Delta -5)+6 (d-2 (\Delta +1))^2 (d-2 (\Delta +2)) (d-2 (\Delta +3))\Big)\\
    &D_1=\frac{(d-2 (\Delta +2)) (d-2 (\Delta +3)) (-2 \Delta +\tau -2)^2 (\tau -2 \Delta )^2 (-d+2 \Delta +\tau )^2}{4 (d-2 \tau ) (d-2 \tau +4) (d-2 (\tau +1))}\nonumber\\
    &\times \Big(-4 (d+1) \tau  (d-2 (\Delta +1)) \left(3 d^2-12 d (\Delta +2)+12 \Delta  (\Delta +4)+44\right)\nonumber\\
    &-2 \tau ^2 \left(12 ((d-7) d+28) \Delta ^2-4 d (d (3 d-19)+61) \Delta +d (d (d (3 d-26)+78)-146)+72 \Delta ^3+452 \Delta +164\right)\nonumber\\
    &+2 (d-2 (\Delta +1)) \left(12 ((d-1) d+4) \Delta ^2+6 d (d (7-2 d)+8) \Delta +d (d (3 (d-7) d+8)+148)+24 \Delta ^3-8 (15 \Delta +26)\right)\nonumber\\
    &+\tau ^6 (d-2 \Delta -5)-2 (d+1) \tau ^5 (d-2 \Delta -5)+\tau ^4 ((d-7) d+18 \Delta +26) (d-2 \Delta -5)\nonumber\\
    &+4 (d+1) \tau ^3 (d-2 \Delta -5) (3 d-6 \Delta -8)\Big)\\
    &D_2=\frac{(d-2 (\Delta +2)) (d-2 (\Delta +3)) (\tau -2 \Delta )^2 (-d+2 \Delta +\tau )^2 (-d+2 \Delta +\tau +2)^2}{4 (d-2 \tau ) (d-2 \tau +2) (d-2 (\tau +2))}\nonumber\\
    &\times \Big(3 d^6-d^5 (12 \Delta +\tau  (\tau +14)+37)+d^4 \left(12 \Delta ^2+2 \Delta  (\tau  (\tau +26)+61)+\tau  (\tau  (4 \tau +37)+146)+164\right)\nonumber\\
    &-2 d^3 \left(24 \Delta ^2 (\tau +3)+4 \Delta  \left(\tau  (\tau +7)^2+60\right)+\tau  (\tau  (\tau  (3 \tau +28)+139)+260)+152\right)\nonumber\\
    &+d^2 (96 \Delta ^3+48 \Delta ^2 (\tau  (2 \tau +7)+12)+4 \Delta  (\tau  (\tau  (3 \tau  (\tau +10)+149)+280)+188)\nonumber\\
    &+\tau  (\tau  (\tau  (\tau  (4 \tau +47)+264)+732)+816)+80)+d(-192 \Delta ^3 (\tau +1)-24 \Delta ^2 (\tau  (\tau  (4 \tau +13)+42)+18)\nonumber\\
    &-4 \Delta  (\tau  (\tau  (\tau  (\tau  (2 \tau +13)+108)+260)+388)-72)-\tau  (\tau  (\tau  (\tau  (\tau  (\tau +22)+111)+452)+772)+672)+656)\nonumber\\
    &+(2 \Delta +5) \tau ^6+2 (2 \Delta +5) \tau ^5+2 (2 \Delta +5) (9 \Delta +13) \tau ^4+8 (2 \Delta +5) (3 \Delta +4) \tau ^3\nonumber\\
    &+8 (\Delta  (6 \Delta  (3 \Delta +14)+113)+41) \tau ^2+32 (\Delta +1) (3 \Delta  (\Delta +4)+11) \tau +32 (\Delta +1) (3 \Delta  (\Delta  (\Delta +2)-5)-26)\Big)\\
    &D_3=\frac{(d-2 (\Delta +2)) (d-2 (\Delta +3)) (-d+2 \Delta +\tau )^2 (-d+2 \Delta +\tau +2)^2 (-d+2 \Delta +\tau +4)^2}{12 (d-2 \tau ) (d-2 (\tau +1)) (d-2 (\tau +2))}\nonumber\\
    &\times\Big(d^7-2 d^6 (\Delta +3 \tau +10)+d^5 (12 \Delta  (\tau +4)+3 \tau  (5 \tau +32)+163)\nonumber\\
    &-d^4 \left(36 \Delta ^2+\Delta  (6 \tau  (5 \tau +34)+410)+\tau  (\tau  (20 \tau +189)+610)+704\right)\nonumber\\
    &+d^3 \left(72 \Delta ^2 (2 \tau +5)+8 \Delta  (\tau  (\tau  (5 \tau +42)+160)+210)+\tau  (\tau  (\tau  (15 \tau +196)+900)+1976)+1744\right)\nonumber\\
    &-2 d^2 (72 \Delta ^3+36 \Delta ^2 (3 \tau  (\tau +4)+20)+\Delta  (3 \tau  (\tau  (\tau  (5 \tau +44)+248)+600)+1792)\nonumber\\
    &+\tau  (\tau  (\tau  (3 \tau  (\tau +19)+326)+1032)+1720)+1240)+d (96 \Delta ^3 (3 \tau +5)+72 \Delta ^2 (\tau  (\tau  (2 \tau +9)+30)+34)\nonumber\\
    &+4 \Delta  (\tau  (\tau  (\tau  (3 \tau  (\tau +8)+194)+630)+1180)+960)+\tau  (\tau  (\tau  (\tau  (\tau  (\tau +36)+229)+956)+2260)+3040)+1872)\nonumber\\
    &-96 \Delta ^4-48 \Delta ^3 (3 \tau  (\tau +2)+14)-12 \Delta ^2 (3 \tau  (\tau +2) (\tau  (\tau +2)+20)+136)\nonumber\\
    &-2 \Delta  (\tau  (\tau +2) (\tau  (\tau +2) (\tau  (\tau +2)+67)+552)+816)-\tau  (\tau +2) (5 \tau  (\tau +2) (\tau  (\tau +2)+22)+528)-576\Big)
\end{align}
\end{subequations}}

\section{Expansion in inverse powers of conformal spin}\label{sec:LargeSpin}
The large spin expansion naturally arranges in terms of the conformal spin \cite{Alday:2015eya} 
\begin{equation}
    \mathfrak{J}^2=\left(\ell+\tfrac{\tau_{n,\ell}}2\right)\left(\ell+\tfrac{\tau_{n,\ell}}2-1\right).
\end{equation}
In this appendix we give further details on how to perform the large spin expansion at the level of the crossing kernels via the Mellin representation \eqref{mellinhyper} of the hypergeometric function ${}_4F_3$ which encodes their dependence on $\ell$.

In particular, this dependence is generally encoded in the following ratios of Gamma functions at the level of the Mellin integrand:
\begin{equation}
    \frac{\Gamma (\ell+1) \Gamma (\ell-s+\tau_{n,\ell}-1)}{\Gamma (\ell+s+1) \Gamma (\ell+\tau_{n,\ell}-1)}=\frac{\Gamma \left(\lambda-\frac{\tau_{n,\ell}}{2} +1\right) \Gamma \left(\lambda-s+\frac{\tau_{n,\ell}}{2} -1\right)}{\Gamma \left(\lambda+\frac{\tau_{n,\ell}}{2} -1\right) \Gamma \left(\lambda+s -\frac{\tau_{n,\ell}}{2}+1\right)}\,,
\end{equation}
where, to make the expansion in $1/\mathfrak{J}$ manifest, it is useful to express the dependence on $\ell$ in terms of $\lambda=\ell+\tfrac{\tau_{n,\ell}}{2}$, which is done in the second equality. This has the general structure
\begin{equation}
    \frac{\Gamma(\lambda-\alpha)}{\Gamma(\lambda+\alpha)}\frac{\Gamma(\lambda-\beta)}{\Gamma(\lambda+\beta)}\,.
\end{equation}

In this way we can use the following asymptotic expansion for simple ratios of Gamma functions in $\lambda$ \cite{Fields,Dey:2017fab}:
\begin{multline}\label{asymratgamma}
     \frac{\Gamma(\lambda-\alpha)}{\Gamma(\lambda+\alpha)}=\sum_{j=0}^\infty\frac{\Gamma(2j+2\alpha)}{\Gamma(2\alpha)(2j)!}\,B_{2j}^{(1-2\alpha)}\left(\tfrac{1-2\alpha}2\right)(\lambda-\tfrac12)^{2\alpha-2j}\\=\sum_{j=0}^\infty\frac{\Gamma(2j+2\alpha)}{\Gamma(2\alpha)(2j)!}\,B_{2j}^{(1-2\alpha)}\left(\tfrac{1-2\alpha}2\right)\Big(\underbrace{\lambda(\lambda+1)}_{\mathfrak{J}^2}+\tfrac14\Big)^{-\alpha-j},
\end{multline}
in terms of generalised Bernoulli polynomials $B_{2j}^{(n)}$. Binomially expanding the factor encoding the dependence on $\mathfrak{J}$, we obtain the following expansion in $1/\mathfrak{J}^2$:
\begin{equation}\label{prodgammaexp}
    \frac{\Gamma(\lambda-\alpha)}{\Gamma(\lambda+\alpha)}=\sum_{i=0}^\infty\sum_{j=0}^\infty\frac{\Gamma(2j+2\alpha)}{\Gamma(2\alpha)(2j)!}\,B_{2j}^{(1-2\alpha)}\left(\tfrac{1-2\alpha}2\right)\,4^{-i}\tbinom{-\alpha -j}{i}\,\mathfrak{J}^{-2 (\alpha +j+i)}
\end{equation}
This then gives
\begin{align} \nonumber
    \frac{\Gamma(\lambda-\alpha)}{\Gamma(\lambda+\alpha)}\frac{\Gamma(\lambda-\beta)}{\Gamma(\lambda+\beta)}=&\sum_{i_a=0}^\infty\sum_{j_a=0}^\infty\frac{\Gamma(2j_1+2\alpha)}{\Gamma(2\alpha)(2j_1)!}\frac{\Gamma(2j_2+2\beta)}{\Gamma(2\beta)(2j_2)!}\,B_{2j_1}^{(1-2\alpha)}\left(\tfrac{1-2\alpha}2\right)B_{2j_2}^{(1-2\beta)}\left(\tfrac{1-2\beta}2\right)\\ \nonumber &\times 4^{-i_1-i_2} \binom{-\alpha -j_1}{i_1}\binom{-\beta -j_2}{i_2}\,\mathfrak{J}^{-2 (\alpha +\beta+j_1+i_1 +j_2+i_2)}\\ 
    &\equiv\sum_{k=0}^\infty d_{\alpha,\beta}^k\mathfrak{J}^{-2(\alpha+\beta+k)},
\end{align}
where for convenience we have defined the coefficients:
\begin{multline}\label{dbern}
    d_{\alpha,\beta}^k=\sum_{i_1+i_2+j_1+j_2=k}\frac{\Gamma(2j_1+2\alpha)}{\Gamma(2\alpha)(2j_1)!}\frac{\Gamma(2j_2+2\beta)}{\Gamma(2\beta)(2j_2)!}\,B_{2j_1}^{(1-2\alpha)}\left(\tfrac{1+2\alpha}2\right)B_{2j_2}^{(1-2\beta)}\left(\tfrac{1+2\beta}2\right)\\\times 4^{-i_1-i_2} \binom{-\alpha -j_1}{i_1}\binom{-\beta -j_2}{i_2},
\end{multline}
which are polynomials in $\alpha$ and $\beta$.

\end{appendix}

\bibliography{refs}

\providecommand{\href}[2]{#2}\begingroup\raggedright\begin{thebibliography}{10}

\bibitem{Komargodski:2012ek}
Z.~Komargodski and A.~Zhiboedov, \emph{{Convexity and Liberation at Large
  Spin}}, \href{http://dx.doi.org/10.1007/JHEP11(2013)140}{\emph{JHEP}
  {\bfseries 11} (2013) 140},
  [\href{https://arxiv.org/abs/1212.4103}{{\ttfamily 1212.4103}}].

\bibitem{Fitzpatrick:2012yx}
A.~L. Fitzpatrick, J.~Kaplan, D.~Poland and D.~Simmons-Duffin, \emph{{The
  Analytic Bootstrap and AdS Superhorizon Locality}},
  \href{http://dx.doi.org/10.1007/JHEP12(2013)004}{\emph{JHEP} {\bfseries 12}
  (2013) 004}, [\href{https://arxiv.org/abs/1212.3616}{{\ttfamily 1212.3616}}].

\bibitem{PhysRevD.8.4383}
C.~G. Callan and D.~J. Gross, \emph{Bjorken scaling in quantum field theory},
  \href{http://dx.doi.org/10.1103/PhysRevD.8.4383}{\emph{Phys. Rev. D}
  {\bfseries 8} (Dec, 1973) 4383--4394}.

\bibitem{Korchemsky:1992xv}
G.~P. Korchemsky and G.~Marchesini, \emph{{Structure function for large x and
  renormalization of Wilson loop}},
  \href{http://dx.doi.org/10.1016/0550-3213(93)90167-N}{\emph{Nucl. Phys.}
  {\bfseries B406} (1993) 225--258},
  [\href{https://arxiv.org/abs/hep-ph/9210281}{{\ttfamily hep-ph/9210281}}].

\bibitem{Belitsky:2003ys}
A.~V. Belitsky, A.~S. Gorsky and G.~P. Korchemsky, \emph{{Gauge / string
  duality for QCD conformal operators}},
  \href{http://dx.doi.org/10.1016/S0550-3213(03)00542-X}{\emph{Nucl. Phys.}
  {\bfseries B667} (2003) 3--54},
  [\href{https://arxiv.org/abs/hep-th/0304028}{{\ttfamily hep-th/0304028}}].

\bibitem{Alday:2007mf}
L.~F. Alday and J.~M. Maldacena, \emph{{Comments on operators with large
  spin}}, \href{http://dx.doi.org/10.1088/1126-6708/2007/11/019}{\emph{JHEP}
  {\bfseries 11} (2007) 019},
  [\href{https://arxiv.org/abs/0708.0672}{{\ttfamily 0708.0672}}].

\bibitem{Simmons-Duffin:2016wlq}
D.~Simmons-Duffin, \emph{{The Lightcone Bootstrap and the Spectrum of the 3d
  Ising CFT}}, \href{http://dx.doi.org/10.1007/JHEP03(2017)086}{\emph{JHEP}
  {\bfseries 03} (2017) 086},
  [\href{https://arxiv.org/abs/1612.08471}{{\ttfamily 1612.08471}}].

\bibitem{Simmons-Duffin:2016gjk}
D.~Simmons-Duffin, \emph{{The Conformal Bootstrap}},  in \emph{{Proceedings,
  Theoretical Advanced Study Institute in Elementary Particle Physics: New
  Frontiers in Fields and Strings (TASI 2015): Boulder, CO, USA, June 1-26,
  2015}}, pp.~1--74, 2017.
\newblock \href{https://arxiv.org/abs/1602.07982}{{\ttfamily 1602.07982}}.
\newblock \href{http://dx.doi.org/10.1142/9789813149441_0001}{DOI}.

\bibitem{Poland:2018epd}
D.~Poland, S.~Rychkov and A.~Vichi, \emph{{The Conformal Bootstrap: Theory,
  Numerical Techniques, and Applications}},
  \href{https://arxiv.org/abs/1805.04405}{{\ttfamily 1805.04405}}.

\bibitem{Hartman:2015lfa}
T.~Hartman, S.~Jain and S.~Kundu, \emph{{Causality Constraints in Conformal
  Field Theory}}, \href{http://dx.doi.org/10.1007/JHEP05(2016)099}{\emph{JHEP}
  {\bfseries 05} (2016) 099},
  [\href{https://arxiv.org/abs/1509.00014}{{\ttfamily 1509.00014}}].

\bibitem{Caron-Huot:2017vep}
S.~Caron-Huot, \emph{{Analyticity in Spin in Conformal Theories}},
  \href{http://dx.doi.org/10.1007/JHEP09(2017)078}{\emph{JHEP} {\bfseries 09}
  (2017) 078}, [\href{https://arxiv.org/abs/1703.00278}{{\ttfamily
  1703.00278}}].

\bibitem{Sleight:2018epi}
C.~Sleight and M.~Taronna, \emph{{Spinning Mellin Bootstrap: Conformal Partial
  Waves, Crossing Kernels and Applications}},
  \href{https://arxiv.org/abs/1804.09334}{{\ttfamily 1804.09334}}.

\bibitem{Krasnov:2005fu}
K.~Krasnov and J.~Louko, \emph{{SO(1,d+1) Racah coefficients: Type I
  representations}}, \href{http://dx.doi.org/10.1063/1.2180626}{\emph{J. Math.
  Phys.} {\bfseries 47} (2006) 033513},
  [\href{https://arxiv.org/abs/math-ph/0502017}{{\ttfamily math-ph/0502017}}].

\bibitem{Gadde:2017sjg}
A.~Gadde, \emph{{In search of conformal theories}},
  \href{https://arxiv.org/abs/1702.07362}{{\ttfamily 1702.07362}}.

\bibitem{Hogervorst:2017sfd}
M.~Hogervorst and B.~C. van Rees, \emph{{Crossing symmetry in alpha space}},
  \href{http://dx.doi.org/10.1007/JHEP11(2017)193}{\emph{JHEP} {\bfseries 11}
  (2017) 193}, [\href{https://arxiv.org/abs/1702.08471}{{\ttfamily
  1702.08471}}].

\bibitem{Hogervorst:2017kbj}
M.~Hogervorst, \emph{{Crossing Kernels for Boundary and Crosscap CFTs}},
  \href{https://arxiv.org/abs/1703.08159}{{\ttfamily 1703.08159}}.

\bibitem{Gopakumar:2018xqi}
R.~Gopakumar and A.~Sinha, \emph{{On the Polyakov-Mellin bootstrap}},
  \href{https://arxiv.org/abs/1809.10975}{{\ttfamily 1809.10975}}.

\bibitem{Dolan:2000ut}
F.~A. Dolan and H.~Osborn, \emph{{Conformal four point functions and the
  operator product expansion}},
  \href{http://dx.doi.org/10.1016/S0550-3213(01)00013-X}{\emph{Nucl. Phys.}
  {\bfseries B599} (2001) 459--496},
  [\href{https://arxiv.org/abs/hep-th/0011040}{{\ttfamily hep-th/0011040}}].

\bibitem{Heemskerk:2009pn}
I.~Heemskerk, J.~Penedones, J.~Polchinski and J.~Sully, \emph{{Holography from
  Conformal Field Theory}},
  \href{http://dx.doi.org/10.1088/1126-6708/2009/10/079}{\emph{JHEP} {\bfseries
  10} (2009) 079}, [\href{https://arxiv.org/abs/0907.0151}{{\ttfamily
  0907.0151}}].

\bibitem{Fitzpatrick:2011dm}
A.~L. Fitzpatrick and J.~Kaplan, \emph{{Unitarity and the Holographic
  S-Matrix}}, \href{http://dx.doi.org/10.1007/JHEP10(2012)032}{\emph{JHEP}
  {\bfseries 10} (2012) 032},
  [\href{https://arxiv.org/abs/1112.4845}{{\ttfamily 1112.4845}}].

\bibitem{Alday:2015eya}
L.~F. Alday, A.~Bissi and T.~Lukowski, \emph{{Large spin systematics in CFT}},
  \href{http://dx.doi.org/10.1007/JHEP11(2015)101}{\emph{JHEP} {\bfseries 11}
  (2015) 101}, [\href{https://arxiv.org/abs/1502.07707}{{\ttfamily
  1502.07707}}].

\bibitem{Vos:2014pqa}
G.~Vos, \emph{{Generalized Additivity in Unitary Conformal Field Theories}},
  \href{http://dx.doi.org/10.1016/j.nuclphysb.2015.07.013}{\emph{Nucl. Phys.}
  {\bfseries B899} (2015) 91--111},
  [\href{https://arxiv.org/abs/1411.7941}{{\ttfamily 1411.7941}}].

\bibitem{Alday:2015ewa}
L.~F. Alday and A.~Zhiboedov, \emph{{An Algebraic Approach to the Analytic
  Bootstrap}}, \href{http://dx.doi.org/10.1007/JHEP04(2017)157}{\emph{JHEP}
  {\bfseries 04} (2017) 157},
  [\href{https://arxiv.org/abs/1510.08091}{{\ttfamily 1510.08091}}].

\bibitem{Dey:2017fab}
P.~Dey, K.~Ghosh and A.~Sinha, \emph{{Simplifying large spin bootstrap in
  Mellin space}}, \href{http://dx.doi.org/10.1007/JHEP01(2018)152}{\emph{JHEP}
  {\bfseries 01} (2018) 152},
  [\href{https://arxiv.org/abs/1709.06110}{{\ttfamily 1709.06110}}].

\bibitem{Kaviraj:2015cxa}
A.~Kaviraj, K.~Sen and A.~Sinha, \emph{{Analytic bootstrap at large spin}},
  \href{http://dx.doi.org/10.1007/JHEP11(2015)083}{\emph{JHEP} {\bfseries 11}
  (2015) 083}, [\href{https://arxiv.org/abs/1502.01437}{{\ttfamily
  1502.01437}}].

\bibitem{Alday:2017gde}
L.~F. Alday, A.~Bissi and E.~Perlmutter, \emph{{Holographic Reconstruction of
  AdS Exchanges from Crossing Symmetry}},
  \href{http://dx.doi.org/10.1007/JHEP08(2017)147}{\emph{JHEP} {\bfseries 08}
  (2017) 147}, [\href{https://arxiv.org/abs/1705.02318}{{\ttfamily
  1705.02318}}].

\bibitem{Aharony:2018npf}
O.~Aharony, L.~F. Alday, A.~Bissi and R.~Yacoby, \emph{{The Analytic Bootstrap
  for Large $N$ Chern-Simons Vector Models}},
  \href{https://arxiv.org/abs/1805.04377}{{\ttfamily 1805.04377}}.

\bibitem{Kaviraj:2015xsa}
A.~Kaviraj, K.~Sen and A.~Sinha, \emph{{Universal anomalous dimensions at large
  spin and large twist}},
  \href{http://dx.doi.org/10.1007/JHEP07(2015)026}{\emph{JHEP} {\bfseries 07}
  (2015) 026}, [\href{https://arxiv.org/abs/1504.00772}{{\ttfamily
  1504.00772}}].

\bibitem{Dey:2016zbg}
P.~Dey, A.~Kaviraj and K.~Sen, \emph{{More on analytic bootstrap for O(N)
  models}}, \href{http://dx.doi.org/10.1007/JHEP06(2016)136}{\emph{JHEP}
  {\bfseries 06} (2016) 136},
  [\href{https://arxiv.org/abs/1602.04928}{{\ttfamily 1602.04928}}].

\bibitem{Li:2015itl}
D.~Li, D.~Meltzer and D.~Poland, \emph{{Conformal Collider Physics from the
  Lightcone Bootstrap}},
  \href{http://dx.doi.org/10.1007/JHEP02(2016)143}{\emph{JHEP} {\bfseries 02}
  (2016) 143}, [\href{https://arxiv.org/abs/1511.08025}{{\ttfamily
  1511.08025}}].

\bibitem{Elkhidir:2017iov}
E.~Elkhidir and D.~Karateev, \emph{{Scalar-Fermion Analytic Bootstrap in 4D}},
  \href{https://arxiv.org/abs/1712.01554}{{\ttfamily 1712.01554}}.

\bibitem{Alday:2017vkk}
L.~F. Alday and S.~Caron-Huot, \emph{{Gravitational S-matrix from CFT
  dispersion relations}},  \href{https://arxiv.org/abs/1711.02031}{{\ttfamily
  1711.02031}}.

\bibitem{Giombi:2018vtc}
S.~Giombi, V.~Kirilin and E.~Perlmutter, \emph{{Double-Trace Deformations of
  Conformal Correlations}},
  \href{http://dx.doi.org/10.1007/JHEP02(2018)175}{\emph{JHEP} {\bfseries 02}
  (2018) 175}, [\href{https://arxiv.org/abs/1801.01477}{{\ttfamily
  1801.01477}}].

\bibitem{ElShowk:2011ag}
S.~El-Showk and K.~Papadodimas, \emph{{Emergent Spacetime and Holographic
  CFTs}}, \href{http://dx.doi.org/10.1007/JHEP10(2012)106}{\emph{JHEP}
  {\bfseries 10} (2012) 106},
  [\href{https://arxiv.org/abs/1101.4163}{{\ttfamily 1101.4163}}].

\bibitem{Alday:2010zy}
L.~F. Alday, B.~Eden, G.~P. Korchemsky, J.~Maldacena and E.~Sokatchev,
  \emph{{From correlation functions to Wilson loops}},
  \href{http://dx.doi.org/10.1007/JHEP09(2011)123}{\emph{JHEP} {\bfseries 09}
  (2011) 123}, [\href{https://arxiv.org/abs/1007.3243}{{\ttfamily 1007.3243}}].

\bibitem{Mack:2009mi}
G.~Mack, \emph{{D-independent representation of Conformal Field Theories in D
  dimensions via transformation to auxiliary Dual Resonance Models. Scalar
  amplitudes}},  \href{https://arxiv.org/abs/0907.2407}{{\ttfamily 0907.2407}}.

\bibitem{Sen:2015doa}
K.~Sen and A.~Sinha, \emph{{On critical exponents without Feynman diagrams}},
  \href{http://dx.doi.org/10.1088/1751-8113/49/44/445401}{\emph{J. Phys.}
  {\bfseries A49} (2016) 445401},
  [\href{https://arxiv.org/abs/1510.07770}{{\ttfamily 1510.07770}}].

\bibitem{Gopakumar:2016wkt}
R.~Gopakumar, A.~Kaviraj, K.~Sen and A.~Sinha, \emph{{Conformal Bootstrap in
  Mellin Space}},
  \href{http://dx.doi.org/10.1103/PhysRevLett.118.081601}{\emph{Phys. Rev.
  Lett.} {\bfseries 118} (2017) 081601},
  [\href{https://arxiv.org/abs/1609.00572}{{\ttfamily 1609.00572}}].

\bibitem{Gopakumar:2016cpb}
R.~Gopakumar, A.~Kaviraj, K.~Sen and A.~Sinha, \emph{{A Mellin space approach
  to the conformal bootstrap}},
  \href{http://dx.doi.org/10.1007/JHEP05(2017)027}{\emph{JHEP} {\bfseries 05}
  (2017) 027}, [\href{https://arxiv.org/abs/1611.08407}{{\ttfamily
  1611.08407}}].

\bibitem{Costa:2012cb}
M.~S. Costa, V.~Goncalves and J.~Penedones, \emph{{Conformal Regge theory}},
  \href{http://dx.doi.org/10.1007/JHEP12(2012)091}{\emph{JHEP} {\bfseries 12}
  (2012) 091}, [\href{https://arxiv.org/abs/1209.4355}{{\ttfamily 1209.4355}}].

\bibitem{Korchemsky:1994um}
G.~P. Korchemsky, \emph{{Bethe ansatz for QCD pomeron}},
  \href{http://dx.doi.org/10.1016/0550-3213(95)00099-E}{\emph{Nucl. Phys.}
  {\bfseries B443} (1995) 255--304},
  [\href{https://arxiv.org/abs/hep-ph/9501232}{{\ttfamily hep-ph/9501232}}].

\bibitem{andrews_askey_roy_1999}
G.~E. Andrews, R.~Askey and R.~Roy, \emph{Special Functions}.
\newblock Encyclopedia of Mathematics and its Applications. Cambridge
  University Press, 1999,
  \href{http://dx.doi.org/10.1017/CBO9781107325937}{10.1017/CBO9781107325937}.

\bibitem{Mack1974}
G.~Mack, \emph{Group Theoretical Approach to Conformal Invariant Quantum Field
  Theory}, pp.~123--157.
\newblock Springer US, Boston, MA, 1974.
\newblock 10.1007/978-1-4615-8909-9-7.

\bibitem{Mack:1974sa}
G.~Mack, \emph{{Osterwalder-Schrader Positivity in Conformal Invariant Quantum
  Field Theory}}, \href{http://dx.doi.org/10.1007/3-540-07160-1-3}{\emph{Lect.
  Notes Phys.} {\bfseries 37} (1975) 66--91}.

\bibitem{Dobrev:1975ru}
V.~K. Dobrev, V.~B. Petkova, S.~G. Petrova and I.~T. Todorov, \emph{{Dynamical
  Derivation of Vacuum Operator Product Expansion in Euclidean Conformal
  Quantum Field Theory}},
  \href{http://dx.doi.org/10.1103/PhysRevD.13.887}{\emph{Phys. Rev.} {\bfseries
  D13} (1976) 887}.

\bibitem{Dolan:2011dv}
F.~A. Dolan and H.~Osborn, \emph{{Conformal Partial Waves: Further Mathematical
  Results}},  \href{https://arxiv.org/abs/1108.6194}{{\ttfamily 1108.6194}}.

\bibitem{Polyakov:1974gs}
A.~M. Polyakov, \emph{{Nonhamiltonian approach to conformal quantum field
  theory}}, {\emph{Zh. Eksp. Teor. Fiz.} {\bfseries 66} (1974) 23--42}.

\bibitem{Costa:2014kfa}
M.~S. Costa, V.~Gonçalves and J.~Penedones, \emph{{Spinning AdS Propagators}},
  \href{http://dx.doi.org/10.1007/JHEP09(2014)064}{\emph{JHEP} {\bfseries 09}
  (2014) 064}, [\href{https://arxiv.org/abs/1404.5625}{{\ttfamily 1404.5625}}].

\bibitem{Sleight:2017fpc}
C.~Sleight and M.~Taronna, \emph{{Spinning Witten Diagrams}},
  \href{http://dx.doi.org/10.1007/JHEP06(2017)100}{\emph{JHEP} {\bfseries 06}
  (2017) 100}, [\href{https://arxiv.org/abs/1702.08619}{{\ttfamily
  1702.08619}}].

\bibitem{Wilson1980}
J.~A. Wilson, \emph{Some hypergeometric orthogonal polynomials},
  \href{http://dx.doi.org/10.1137/0511064}{\emph{SIAM Journal on Mathematical
  Analysis} {\bfseries 11} (1980) 690--701},
  [\href{https://arxiv.org/abs/https://doi.org/10.1137/0511064}{{\ttfamily
  https://doi.org/10.1137/0511064}}].

\bibitem{Slater}
L.~J. Slater, \emph{{Generalized Hypergeometric Functions}}.
\newblock Cambridge University Press, 1966.

\bibitem{Groenevelt2006}
W.~Groenevelt, \emph{Wilson function transforms related to racah coefficients},
  \href{http://dx.doi.org/10.1007/s10440-006-9024-7}{\emph{Acta Applicandae
  Mathematica} {\bfseries 91} (Apr, 2006) 133--191}.

\bibitem{doi:10.1063/1.524047}
J.~Raynal, \emph{On the definition and properties of generalized 6‐j
  symbols}, \href{http://dx.doi.org/10.1063/1.524047}{\emph{Journal of
  Mathematical Physics} {\bfseries 20} (1979) 2398--2415},
  [\href{https://arxiv.org/abs/https://doi.org/10.1063/1.524047}{{\ttfamily
  https://doi.org/10.1063/1.524047}}].

\bibitem{Hogervorst:2016hal}
M.~Hogervorst, \emph{{Dimensional Reduction for Conformal Blocks}},
  \href{http://dx.doi.org/10.1007/JHEP09(2016)017}{\emph{JHEP} {\bfseries 09}
  (2016) 017}, [\href{https://arxiv.org/abs/1604.08913}{{\ttfamily
  1604.08913}}].

\bibitem{Fields}
J.~L. Fields, \emph{{A note on the asymptotic expansion of a ratio of gamma
  functions}}, {\emph{Proc. Edinburgh Math. Soc.} {\bfseries 15} (1964) 43,45}.

\bibitem{Groenevelt2003}
W.~{Groenevelt}, \emph{{The Wilson function transform}}, {\emph{ArXiv
  Mathematics e-prints} (June, 2003) },
  [\href{https://arxiv.org/abs/math/0306424}{{\ttfamily math/0306424}}].

\bibitem{Joung:2012rv}
E.~Joung, L.~Lopez and M.~Taronna, \emph{{On the cubic interactions of massive
  and partially-massless higher spins in (A)dS}},
  \href{http://dx.doi.org/10.1007/JHEP07(2012)041}{\emph{JHEP} {\bfseries 07}
  (2012) 041}, [\href{https://arxiv.org/abs/1203.6578}{{\ttfamily 1203.6578}}].

\bibitem{Bailey}
W.~N. Bailey, \emph{{Generalized Hypergeometric Series}}.
\newblock Cambridge University Press, 1935.

\bibitem{Liu:2018jhs}
J.~Liu, E.~Perlmutter, V.~Rosenhaus and D.~Simmons-Duffin,
  \emph{{$d$-dimensional SYK, AdS Loops, and $6j$ Symbols}},
  \href{https://arxiv.org/abs/1808.00612}{{\ttfamily 1808.00612}}.

\bibitem{HypZ=1}
W.~Bühring, \emph{Generalized hypergeometric functions at unit argument},
  {\emph{Proceedings of the American Mathematical Society} {\bfseries 114}
  (1992) 145--153}.

\bibitem{NACHTMANN1973237}
O.~Nachtmann, \emph{Positivity constraints for anomalous dimensions},
  \href{http://dx.doi.org/https://doi.org/10.1016/0550-3213(73)90144-2}{\emph{Nuclear
  Physics B} {\bfseries 63} (1973) 237 -- 247}.

\bibitem{Sleight:2017pcz}
C.~Sleight and M.~Taronna, \emph{{Higher-Spin Gauge Theories and Bulk
  Locality}},
  \href{http://dx.doi.org/10.1103/PhysRevLett.121.171604}{\emph{Phys. Rev.
  Lett.} {\bfseries 121} (2018) 171604},
  [\href{https://arxiv.org/abs/1704.07859}{{\ttfamily 1704.07859}}].

\bibitem{Fradkin:1986ka}
E.~S. Fradkin and M.~A. Vasiliev, \emph{{Candidate to the Role of Higher Spin
  Symmetry}},
  \href{http://dx.doi.org/10.1016/S0003-4916(87)80025-8}{\emph{Annals Phys.}
  {\bfseries 177} (1987) 63}.

\bibitem{Mikhailov:2002bp}
A.~Mikhailov, \emph{{Notes on higher spin symmetries}},
  \href{https://arxiv.org/abs/hep-th/0201019}{{\ttfamily hep-th/0201019}}.

\bibitem{Eastwood:2002su}
M.~G. Eastwood, \emph{{Higher symmetries of the Laplacian}},
  \href{http://dx.doi.org/10.4007/annals.2005.161.1645}{\emph{Annals Math.}
  {\bfseries 161} (2005) 1645--1665},
  [\href{https://arxiv.org/abs/hep-th/0206233}{{\ttfamily hep-th/0206233}}].

\bibitem{Vasiliev:2003ev}
M.~A. Vasiliev, \emph{{Nonlinear equations for symmetric massless higher spin
  fields in (A)dS(d)}},
  \href{http://dx.doi.org/10.1016/S0370-2693(03)00872-4}{\emph{Phys. Lett.}
  {\bfseries B567} (2003) 139--151},
  [\href{https://arxiv.org/abs/hep-th/0304049}{{\ttfamily hep-th/0304049}}].

\bibitem{Joung:2014qya}
E.~Joung and K.~Mkrtchyan, \emph{{Notes on higher-spin algebras: minimal
  representations and structure constants}},
  \href{http://dx.doi.org/10.1007/JHEP05(2014)103}{\emph{JHEP} {\bfseries 05}
  (2014) 103}, [\href{https://arxiv.org/abs/1401.7977}{{\ttfamily 1401.7977}}].

\bibitem{Lang:1992zw}
K.~Lang and W.~Ruhl, \emph{{The Critical O(N) sigma model at dimensions 2 < d <
  4: Fusion coefficients and anomalous dimensions}},
  \href{http://dx.doi.org/10.1016/0550-3213(93)90417-N}{\emph{Nucl. Phys.}
  {\bfseries B400} (1993) 597--623}.

\bibitem{Leonhardt:2003du}
T.~Leonhardt and W.~Ruhl, \emph{{The Minimal conformal O(N) vector sigma model
  at d = 3}}, \href{http://dx.doi.org/10.1088/0305-4470/37/4/023}{\emph{J.
  Phys.} {\bfseries A37} (2004) 1403--1413},
  [\href{https://arxiv.org/abs/hep-th/0308111}{{\ttfamily hep-th/0308111}}].

\bibitem{Alday:2015ota}
L.~F. Alday and A.~Zhiboedov, \emph{{Conformal Bootstrap With Slightly Broken
  Higher Spin Symmetry}},
  \href{http://dx.doi.org/10.1007/JHEP06(2016)091}{\emph{JHEP} {\bfseries 06}
  (2016) 091}, [\href{https://arxiv.org/abs/1506.04659}{{\ttfamily
  1506.04659}}].

\bibitem{Giombi:2016zwa}
S.~Giombi, V.~Gurucharan, V.~Kirilin, S.~Prakash and E.~Skvortsov, \emph{{On
  the Higher-Spin Spectrum in Large N Chern-Simons Vector Models}},
  \href{http://dx.doi.org/10.1007/JHEP01(2017)058}{\emph{JHEP} {\bfseries 01}
  (2017) 058}, [\href{https://arxiv.org/abs/1610.08472}{{\ttfamily
  1610.08472}}].

\bibitem{Alday:2016njk}
L.~F. Alday, \emph{{Large Spin Perturbation Theory for Conformal Field
  Theories}},
  \href{http://dx.doi.org/10.1103/PhysRevLett.119.111601}{\emph{Phys. Rev.
  Lett.} {\bfseries 119} (2017) 111601},
  [\href{https://arxiv.org/abs/1611.01500}{{\ttfamily 1611.01500}}].

\bibitem{Giombi:2017rhm}
S.~Giombi, V.~Kirilin and E.~Skvortsov, \emph{{Notes on Spinning Operators in
  Fermionic CFT}}, \href{http://dx.doi.org/10.1007/JHEP05(2017)041}{\emph{JHEP}
  {\bfseries 05} (2017) 041},
  [\href{https://arxiv.org/abs/1701.06997}{{\ttfamily 1701.06997}}].

\bibitem{Turiaci:2018nua}
G.~J. Turiaci and A.~Zhiboedov, \emph{{Veneziano Amplitude of Vasiliev
  Theory}},  \href{https://arxiv.org/abs/1802.04390}{{\ttfamily 1802.04390}}.

\bibitem{Maldacena:2012sf}
J.~Maldacena and A.~Zhiboedov, \emph{{Constraining conformal field theories
  with a slightly broken higher spin symmetry}},
  \href{http://dx.doi.org/10.1088/0264-9381/30/10/104003}{\emph{Class. Quant.
  Grav.} {\bfseries 30} (2013) 104003},
  [\href{https://arxiv.org/abs/1204.3882}{{\ttfamily 1204.3882}}].

\bibitem{Giombi:2011kc}
S.~Giombi, S.~Minwalla, S.~Prakash, S.~P. Trivedi, S.~R. Wadia and X.~Yin,
  \emph{{Chern-Simons Theory with Vector Fermion Matter}},
  \href{http://dx.doi.org/10.1140/epjc/s10052-012-2112-0}{\emph{Eur. Phys. J.}
  {\bfseries C72} (2012) 2112},
  [\href{https://arxiv.org/abs/1110.4386}{{\ttfamily 1110.4386}}].

\bibitem{Aharony:2011jz}
O.~Aharony, G.~Gur-Ari and R.~Yacoby, \emph{{d=3 Bosonic Vector Models Coupled
  to Chern-Simons Gauge Theories}},
  \href{http://dx.doi.org/10.1007/JHEP03(2012)037}{\emph{JHEP} {\bfseries 03}
  (2012) 037}, [\href{https://arxiv.org/abs/1110.4382}{{\ttfamily 1110.4382}}].

\bibitem{Aharony:2012nh}
O.~Aharony, G.~Gur-Ari and R.~Yacoby, \emph{{Correlation Functions of Large N
  Chern-Simons-Matter Theories and Bosonization in Three Dimensions}},
  \href{http://dx.doi.org/10.1007/JHEP12(2012)028}{\emph{JHEP} {\bfseries 12}
  (2012) 028}, [\href{https://arxiv.org/abs/1207.4593}{{\ttfamily 1207.4593}}].

\bibitem{Diaz:2006nm}
D.~E. Diaz and H.~Dorn, \emph{{On the AdS higher spin / O(N) vector model
  correspondence: Degeneracy of the holographic image}},
  \href{http://dx.doi.org/10.1088/1126-6708/2006/07/022}{\emph{JHEP} {\bfseries
  07} (2006) 022}, [\href{https://arxiv.org/abs/hep-th/0603084}{{\ttfamily
  hep-th/0603084}}].

\bibitem{Sleight:2016dba}
C.~Sleight and M.~Taronna, \emph{{Higher Spin Interactions from Conformal Field
  Theory: The Complete Cubic Couplings}},
  \href{http://dx.doi.org/10.1103/PhysRevLett.116.181602}{\emph{Phys. Rev.
  Lett.} {\bfseries 116} (2016) 181602},
  [\href{https://arxiv.org/abs/1603.00022}{{\ttfamily 1603.00022}}].

\bibitem{Giombi:2017hpr}
S.~Giombi, C.~Sleight and M.~Taronna, \emph{{Spinning AdS Loop Diagrams: Two
  Point Functions}},
  \href{http://dx.doi.org/10.1007/JHEP06(2018)030}{\emph{JHEP} {\bfseries 06}
  (2018) 030}, [\href{https://arxiv.org/abs/1708.08404}{{\ttfamily
  1708.08404}}].

\bibitem{Bekaert:2016ezc}
X.~Bekaert, J.~Erdmenger, D.~Ponomarev and C.~Sleight, \emph{{Bulk quartic
  vertices from boundary four-point correlators}},  in \emph{{Proceedings,
  International Workshop on Higher Spin Gauge Theories: Singapore, Singapore,
  November 4-6, 2015}}, pp.~291--303, 2017.
\newblock \href{https://arxiv.org/abs/1602.08570}{{\ttfamily 1602.08570}}.
\newblock \href{http://dx.doi.org/10.1142/9789813144101_0015}{DOI}.

\bibitem{Taronna:2016ats}
M.~Taronna, \emph{{Pseudo-local Theories: A Functional Class Proposal}},  in
  \emph{{Proceedings, International Workshop on Higher Spin Gauge Theories:
  Singapore, Singapore, November 4-6, 2015}}, pp.~59--84, 2017.
\newblock \href{https://arxiv.org/abs/1602.08566}{{\ttfamily 1602.08566}}.
\newblock \href{http://dx.doi.org/10.1142/9789813144101_0006}{DOI}.

\end{thebibliography}\endgroup
\bibliographystyle{JHEP}

\end{document}